%% file: Hicks_etal_2009_v2.tex
\documentclass[numberedappendix]{emulateapj}

\submitted{Accepted by ApJ}

\shorttitle{Molecular Gas in AGN}
\shortauthors{E. K. S. Hicks et al.}

\def \as{\hbox{$^{\prime\prime}$}}
\def \Msun{M$_{\odot}$}
\def \Lsun{L$_{\odot}$}
\def \tMsun{$\times$ 10$^7$ \Msun}

\def \mbh{$M_{BH}$}
\def \mdyn{$M_{dyn}$}
\def \mgas{$M_{gas}$}
\def \mic{$\mu$m}
\def \lam{$\lambda$}
\def \sig{$\sigma$}
\def \it{\textit{}}
\def \htwo{H$_{2}$}
\def \htwos{H$_{2}$ 1-0~S(1)}
\def \kms{km s$^{-1}$}
\def \sm{$\sim$}
\def \kb{{\em K}-band}
\def \deg{\,\hbox{$^\circ$}}
\def \inc{{\em i}}
\def \vs{$V_{rot}$/\sig}
\def \vrot{$V_{rot}$}
\def \nh{{\em N$_{H}$}}
\def \tanh{$\times$ 10$^{23}$ cm$^{-2}$}
\def \tbnh{$\times$ 10$^{24}$ cm$^{-2}$}
\def \tcnh{$\times$ 10$^{22}$ cm$^{-2}$}

\def \n{{\em n}}
\def \z{{\em z}$_{o}$}
\def \fg{{\em f$_{g}$}}
\def \h2o{H$_{2}$O}

\begin{document}

\title{The Role of Molecular Gas in Obscuring Seyfert Active Galactic Nuclei\altaffilmark{*,\dag}}

\author{E. K. S. Hicks\altaffilmark{1}, R. I. Davies\altaffilmark{1}, M. A. Malkan\altaffilmark{2}, R. Genzel\altaffilmark{1,3}, L. J. Tacconi\altaffilmark{1}, F. M{\"u}ller S\'anchez\altaffilmark{1}, A. Sternberg\altaffilmark{4}} 

\altaffiltext{*}{Based on observations at the ESO Very Large Telescope (60.A-9235, 070.B-0649, 070.B-0664, 074.B-9012, 076.B-0098, 076.B-0117, 077.B-0514).}
\altaffiltext{\dag}{Based on observations at the W. M. Keck Observatory, which is operated as a scientific partnership among the California Institute of Technology, the University of California and the National Aeronautics and Space Administration. The Observatory was made possible by the generous financial support of the W.M. Keck Foundation.}
\altaffiltext{1}{Max Planck Institut f\"ur extraterrestrische Physik, Postfach 1312, 85741, Garching, Germany}
\altaffiltext{2}{Department of Physics and Astronomy, University of California, Los Angeles, CA, 90095-1562, United States}
\altaffiltext{3}{Department of Physics, 366 Le Conte Hall, University of California, Berkeley, CA, 94720-7300, United States}
\altaffiltext{4}{School of Physics and Astronomy, Tel Aviv University, Tel Aviv 69978, Israel}

\begin{abstract}
In a sample of local active galactic nuclei (AGNs) studied at a spatial resolution on the order of 10 pc, we show that the interstellar medium traced by the molecular hydrogen $\nu$=1-0 S(1) line at 2.1 \mic\ forms a geometrically thick, clumpy disk.  The kinematics of the molecular gas reveals general rotation, although an additional significant component of random bulk motion is required by the high local velocity dispersion.  The size scale of the typical gas disk is found to have a radius of $\sim$30 pc with a comparable vertical height.  Within this radius the average gas mass is estimated to be $\sim$10$^7$ \Msun\ based on a typical gas mass fraction of 10\%, which suggests column densities of \nh\ $\sim$ 5 \tanh.  Extinction of the stellar continuum within this same region suggests lower column densities of \nh\ $\sim$ 2 $\times$ 10$^{22}$ cm$^{-2}$, indicating that the gas distribution on these scales is dominated by dense clumps.  In half of the observed Seyfert galaxies this lower column density is still great enough to obscure the AGN at optical/infrared wavelengths.  We conclude, based on the spatial distribution, kinematics, and column densities that the molecular gas observed is spatially mixed with the nuclear stellar population and is likely to be associated with the outer extent of any smaller scale nuclear obscuring structure.  Furthermore, we find that the velocity dispersion of the molecular gas is correlated with the star formation rate per unit area, suggesting a link between the two phenomena, and that the gas surface density follows known ``Schmidt-Kennicutt'' relations.  The molecular/dusty structure on these scales may be dynamic since it is possible that the velocity dispersion of the gas, and hence the vertical disk height, is maintained by a short, massive inflow of material into the nuclear region and/or by intense, short-lived nuclear star formation.  
\end{abstract}


\section{Introduction}
\label{sec:intro}

\input{tab1.tex}


Although there is a great deal of evidence supporting the widely accepted active galactic nuclei (AGNs) unification model (e.g. \citealt{ant93}, \citealt{urry95}), characterization, or even proof, of the key model component, an obscuring medium, has been difficult to attain.  A popular version of the unification model proposes the existence of an optically and geometrically thick component of gas/dust that obscures the optical and ultraviolet (UV) AGN emission along some fraction of the lines of sight such that, depending on the viewing angle, the observed characteristics differ, resulting in type 1 and 2 AGNs.  In this scenario, the obscuring medium, which is classically thought of as a static canonical ``torus" surrounding the AGN accretion disk, is thought to begin at the dust sublimation radius and extend outward.  

Thermal dust emission has been spatially resolved around AGNs by infrared (IR) interferometric techniques in the AGN NGC 1068 (\citealt{jaffe04}, \citealt{weigelt04}, \citealt{poncelet06}, \citealt{poncelet07}) and Circinus (\citealt{tristram07}).  This emission is interpreted as reradiated emission from dust heated by the optical/ultraviolet emission originating in the accretion disk.  The data suggest a two-component structure in both galaxies consisting of a subparsec, warm, disklike component that is surrounded by a slightly lower temperature, geometrically thick, torus-like component that extends out to a radius of a few parsecs.  In several nearby AGNs \h2o\ masers are found to trace a subparsec warped disk (e.g. \citealt{herrnstein96}, \citealt{greenhill96}, \citealt{greenhill03}), and in NGC 1068 and Circinus these warped disks have spatial scales and orientations that are consistent with the warm disklike component detected in the IR observations.  The compact size of the thermal dust emission implies that this obscuring medium is in a clumpy distribution, rather than smoothly distributed as assumed in early torus models (\citealt{elitzur06b}).  In obscured nuclei, X-ray observations imply column densities on these parsec scales of \nh\ $\sim$ 10$^{22}$ cm$^{-2}$ to greater than \nh\ $\sim$ 10$^{24}$ cm$^{-2}$ (\citealt{risaliti99}, \citealt{treister04}), which is indeed high enough to obscure the optical and ultraviolet emission from the accretion disk ($\gtrsim$10$^{22}$ cm$^{-2}$ for Seyfert galaxies; \citealt{treister04}). 

Models of clumpy tori predict that the obscuring region extends from subparsec radii out to sizes on the order of tens of parsecs in typical Seyfert galaxies (e.g. \citealt{cameron93}, \citealt{schartmann05}, \citealt{honig06}, \citealt {fritz06}, \citealt{schartmann08a}, \citealt{schartmann08b}).  Little is known about the characteristics of the gas and dust on these scales beyond the relatively small-scale structures inferred from the IR interferometry and \h2o\ masers measurements.  Observations of the thermal dust emission are temperature dependent and the inferred radii are thus temperature-weighted.  These measurements therefore represent only the warm dust and do not trace any potentially cooler extended structures in the nuclear region.  Due to the glare from the AGN, studies probing the obscuring region predicted by models are challenging, and those that have been carried out focus primarily on Seyfert 2 galaxies to reduce the impact of the AGN emission (e.g. \citealt{fernandes04}, \citealt{silge05}).  However, with adaptive optics (AO) the AGN emission can now be confined to a smaller region, making possible a detailed study of the nuclear region (central $\sim$100 pc) in even bright Seyfert 1 galaxies.  This provides a more direct view of the nuclear region and eliminates potential biases imposed by the greater amount of obscuration present in Seyfert 2s.  It is therefore now possible to establish the respective roles of the smaller and larger scale components in fulfilling the functions, such as obscuration of the AGN, of a unifying torus. 

In an effort to better characterize the obscuring medium present out to the several tens of parsec scales predicted by models, and investigate its association with the nuclear starburst activity and fueling of the AGN, we have observed a sample of nine local AGNs using \kb\ AO assisted integral field spectroscopy.  The bright molecular hydrogen emission from the 1-0 S(1) rovibrational transition at 2.1218 \mic\ is used to measure, on scales of $\sim$10 pc, the two-dimensional (2-D) distribution and kinematics of the nuclear gas within the central $\sim$100 pc.  As reported by \citet{christopher05} to be the case in the Galactic Center, this warm molecular gas traces the cold/dense gas within the nuclear region.  Sample selection, observations, and data reduction are discussed in $\S$ \ref{sec:obs}, and the general properties of the molecular gas are described in $\S$ \ref{sec:prop}.  Presented in $\S$ \ref{sec:sb-agn} are correlations between the observed properties of the molecular gas and the inferred nuclear star formation.  Possible scenarios for maintaining the estimated vertical structure of the gas disk are investigated in $\S$ \ref{sec:disp}, and in $\S$ \ref{sec:AGN} the role of the molecular hydrogen on these scales in obscuring the AGN and its connection with the nuclear starburst activity are discussed.  The overall conclusions of the study are outlined in $\S$ \ref{sec:conc}, while a more detailed discussion of each of the observed galaxies can be found in the Appendix.

\section{Observations and Data Reduction}
\label{sec:obs}

\subsection{Sample Selection}
The nine galaxies were observed at the ESO Very Large Telescope (VLT) and the W. M. Keck Observatory.  Although the AGNs observed were not selected to comprise a complete sample, both Seyfert 1s and 2s, as well as a LINER (low-ionization nuclear emission-line region), are included.  Table \ref{tab:obj} lists the full sample of galaxies selected.

The criteria for target selection of the galaxies observed at the VLT were (1) the nucleus is bright enough to use for AO correction, (2) the galaxy is close enough that small spatial scales can be resolved ($\sim$20 pc), and (3) the galaxies are well studied such that complementary data can be found in the literature (e.g. radio, submillimeter observations).  Although, the first criterion biases the sample to type 1 Seyferts, two of the six galaxies observed are Seyfert 2s.

The sample observed at the Keck Observatory was selected from all AGNs with a black hole mass estimate from reverberation mapping and a redshift such that the 2.1218 \mic\ molecular hydrogen emission and stellar 2.3 \mic\ CO bandheads are observable in the {\em K} band.  This sample is therefore composed of relatively nearby Seyfert 1 galaxies, all of which have a nucleus that is bright enough for either a correction with the natural guide star (NGS) AO system or as a tip-tilt guide to be used in conjunction with the laser guide star (LGS) AO system.  Both the OSIRIS and SINFONI subsamples contain NGC 3227 and NGC 7469, and a comparison of the data from each instrument is presented in the Appendix.

\subsection{Integral Field Spectra from SINFONI}
\label{sec:obs-sinfoni}
Integral field spectroscopy was obtained for several AGN at the VLT UT4 with the AO near-IR integral field spectrograph SINFONI (\citealt{eisenhauer03}, \citealt{bonnet04}).  The pixel scale of the {\em R} \sm\ 4300 \kb\ spectra (approximately 1.95-2.45 \mic) is either 0\as.125$\times$0\as.25 or 0\as.05$\times$0\as.1, which results in a field of view (FOV) of 0\as.8$\times$0\as.8 or 3.2\as$\times$3.2\as, respectively.  For each galaxy, the nucleus was used as the reference for a near-diffraction-limited correction by the AO module (\citealt{bonnet03}).  The exposure times for Circinus, NGC 3783, and NGC 1068 were 30, 40, and 110 minutes, respectively, and for the remaining galaxies 60 minutes of on-source data were obtained.  A summary of the data, including the specific spatial resolution achieved for each galaxy, is given in Table \ref{tab:obj}.

The data were reduced using the SINFONI custom reduction package SPRED (\citealt{abuter06}), which includes all of the typical reduction steps applied to near-IR spectra with the additional routines necessary to reconstruct the data cube.  After background subtraction, which is performed with sky frames obtained interspersed with the on-source exposures, the data are flat-fielded and corrected for dead/hot pixels.  Telluric correction and flux calibration are carried out using A- and B-type stars.  Residuals from the OH line emission are minimized using the methods outlined in \citet{davies07a}.  The flux calibration is confirmed by comparison to 3\as\ aperture Two Micron All Sky Survey and 1\as-3\as\ aperture NACO or {\em Hubble Space Telescope} ({\em HST}) NICMOS images and was found to be consistent to within 20\%.

The spatial resolution of the SINFONI data (Table \ref{tab:obj}) is determined from the data themselves rather than from additional point-spread function (PSF) calibration frames.  This is done because of the uncertainty involved in using a standard star to recreate the AO performance achieved with an AGN which has spatially extended background galaxy light, as well as to avoid mismatches due to the variability of the ambient seeing.  Two methods, outlined in \citealt{davies04a}, are used to determine the PSF directly from the data, which also has the advantage of including all effects associated with the construction of the data cube.  The first method makes use of the broad Br$\gamma$ (\lam 2.1661 \mic) emission assumed to be coming from the broad-line region (BLR), which is expected to be unresolved for the galaxies in this sample (\citealt{peterson04}).  The second method applies the fact that the intrinsic equivalent width of the CO 2-0 2.29 \mic\ bandhead is expected to fall within a limited range (\citealt{oliva95}, \citealt{forster00}, Davies et al. 2007, hereafter D07), and therefore the observed $W_{CO2-0}$ can be used to determine the dilution of the stellar feature due to the AGN continuum emission, which, at this resolution, is also a point source (predicted sizes are less than 2 pc; \citealt{jaffe04}, \citealt{tristram07}).

An analysis of the nuclear star formation properties (star formation rates (SFRs), time since the last star-forming episode, stellar distribution, etc.) in each of the galaxies in this subsample is presented in D07.  Detailed studies of a few of the galaxies in the subsample can also be found in the literature:  Circinus (\citealt{mueller06}), NGC 3227 (\citealt{davies06}), and NGC 1068 (\citealt{mueller08}).

\subsection{Integral Field Spectra from OSIRIS}
At the W. M. Keck Observatory, the integral field spectrometer OSIRIS (\citealt{larkin06}) on Keck II was used with both the NGS and LGS AO systems (\citealt{vandam04}, \citealt{vandam06}, \citealt{wizinowich06}).  \kb\ {\em R} $\sim$ 3000 spectra, covering 1.965-2.381 \mic, were obtained with a pixel scale of 0\as.035 pixel$^{-1}$ over an FOV of 0\as.56 $\times$ 2\as.24 for each of the galaxies.  Typically 10 minute exposures were taken with off-source sky exposures obtained interspersed with the on-source frames.  Total on-source exposures vary from 40 minutes for NGC 4051, to 60 minutes for NGC 4151, and 70 minutes for the remaining three galaxies.  Of the 70 minutes of on-source exposure for NGC 7469, 20 minutes of this was obtained with the LGS AO system, while the rest of the data for this galaxy, and the others, was obtained with a correction from the NGS AO system using the AGN nucleus as the reference.  The spatial resolution achieved for each galaxy, via the methods discussed in the previous section, are given in Table \ref{tab:obj}.  As a consistency check, the PSF full-width at half-maxima (FWHMs) estimated using these methods were compared to that of the PSFs determined from observations of standard stars (using the same AO setting as used for the galaxies), and they are found to be consistent to within the uncertainties discussed in $\S$ \ref{sec:obs-sinfoni} of standard star PSF measurements (e.g. changes in seeing and difficulty in reproducing the AO performance of the galaxy observations).
 
Reduction of the OSIRIS data was carried out with the OSIRIS final data reduction pipeline (DRP), which performs reduction steps typically applied to near-IR spectra as well as reconstruction of the data cube.  These steps include background subtraction, flat fielding, and correction for dead/hot pixels, as well as correction of detector nonlinearity, removal of detector cross talk, and wavelength calibration.  Telluric correction was done using A-type stars and was applied to the data in the DRP.  Flux calibration of the OSIRIS data is expected to be accurate to about 30\%.

\subsection{Wide FOV Long Single Slit Spectra from ISAAC}
To assess the kinematic properties of the galaxies on a larger scale than is measured here, long-slit data were also obtained with ISAAC on the VLT UT1 for three of the galaxies measured with SINFONI.  Seeing-limited data were obtained as part of a service observing program carried out in 2006.  The short wavelength arm of ISAAC was used in the medium resolution mode ({\em R} \sm\ 3000) with a 0.122 \mic\ narrowband filter centered at 2.144 \mic\ to obtain spectra of the \htwos\ emission line.  Two orthogonal slit positions were obtained for each of the galaxies (see Table \ref{tab:obj} for details) with a 2\hbox{$^{\prime}$}\ long and 0\as.3 wide slit.  The orientations of the two slit positions for each object were selected to align with the major and minor axes as determined from 2.2 \mic\ {\em HST} NICMOS images.  Exposure times totaled 40 minutes per slit position, and the target was nodded along the slit to perform sky subtraction.  The ISAAC data were reduced using standard near-IR reduction techniques, and were carried out with the ESO pipeline software.  

\clearpage
\section{General Properties of the Molecular Gas}
\label{sec:prop}

\begin{figure*}[!ht]	
\epsscale{0.85}
\plotone{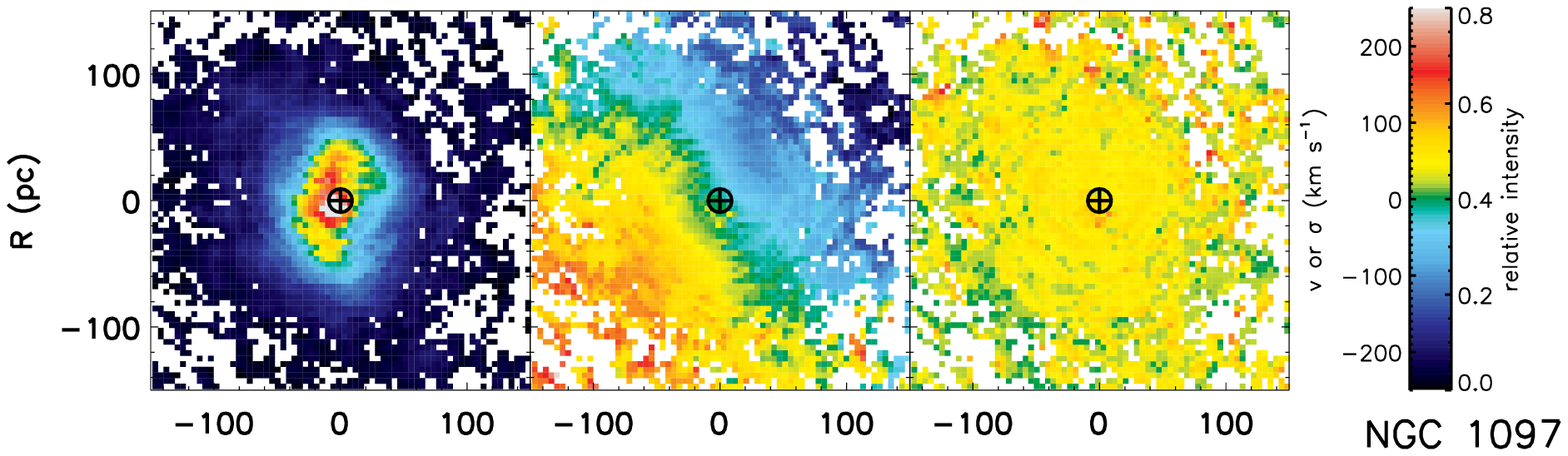}
\plotone{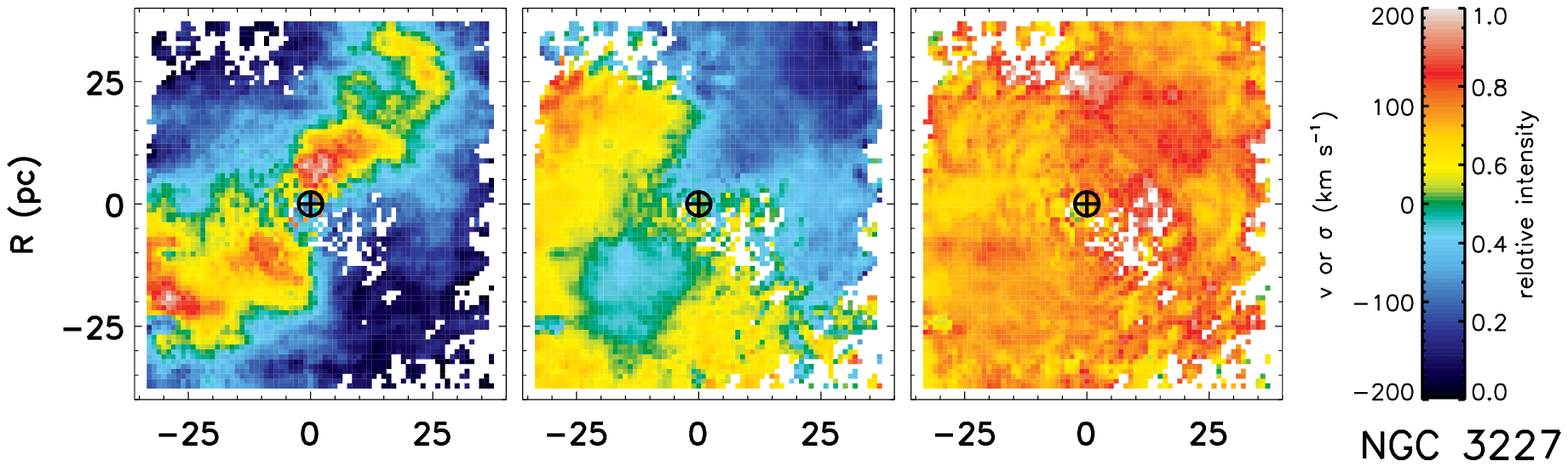}
\plotone{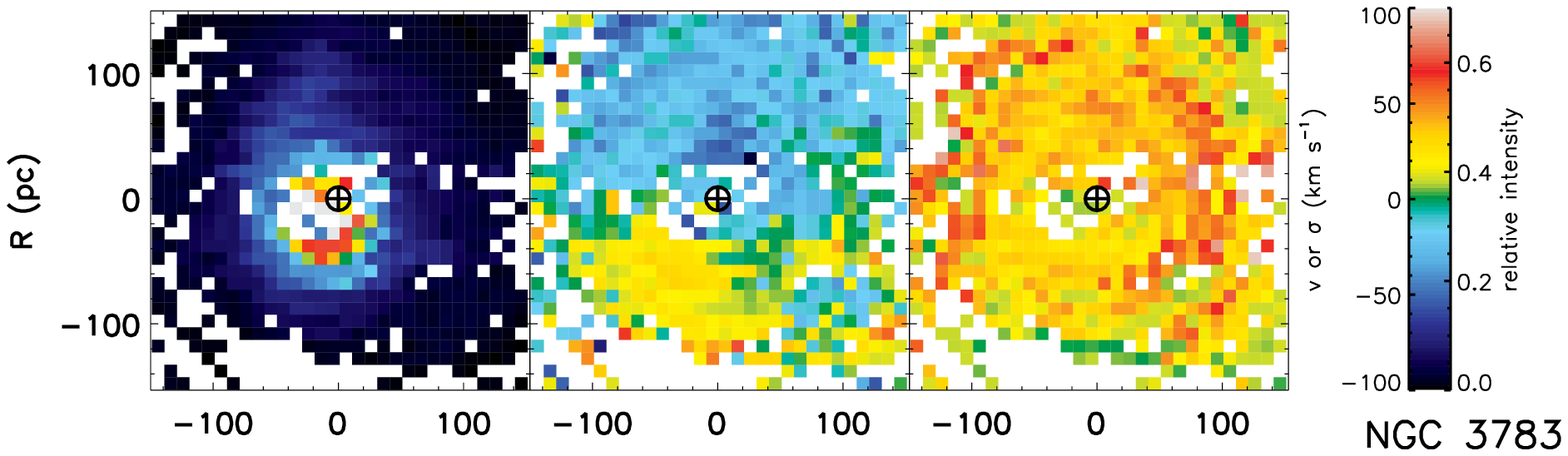}
\plotone{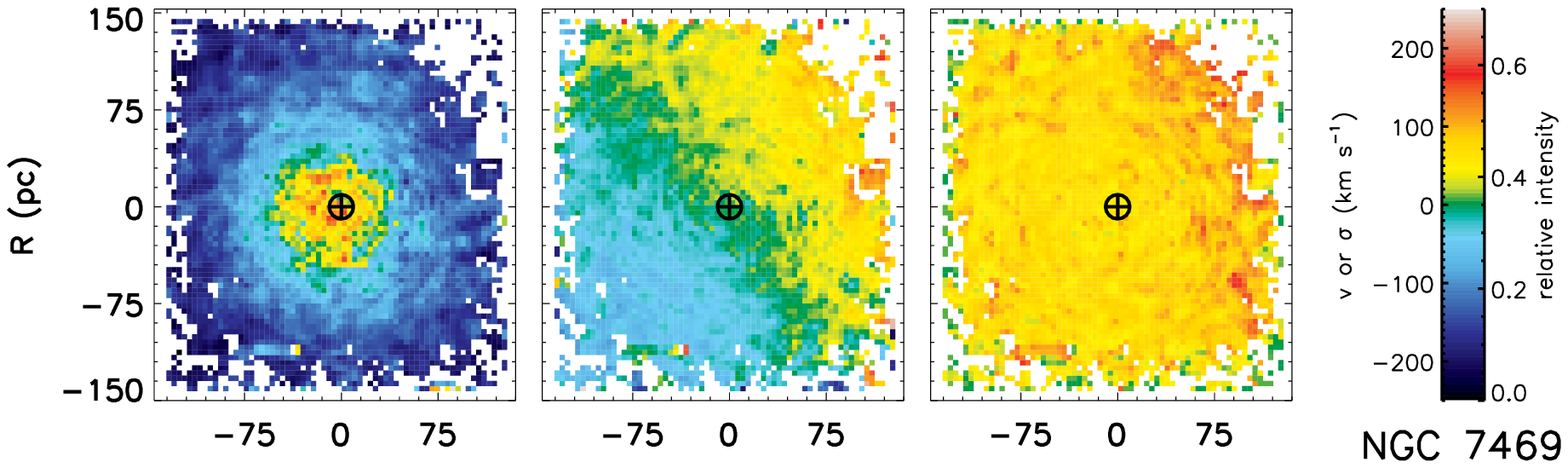}
\plotone{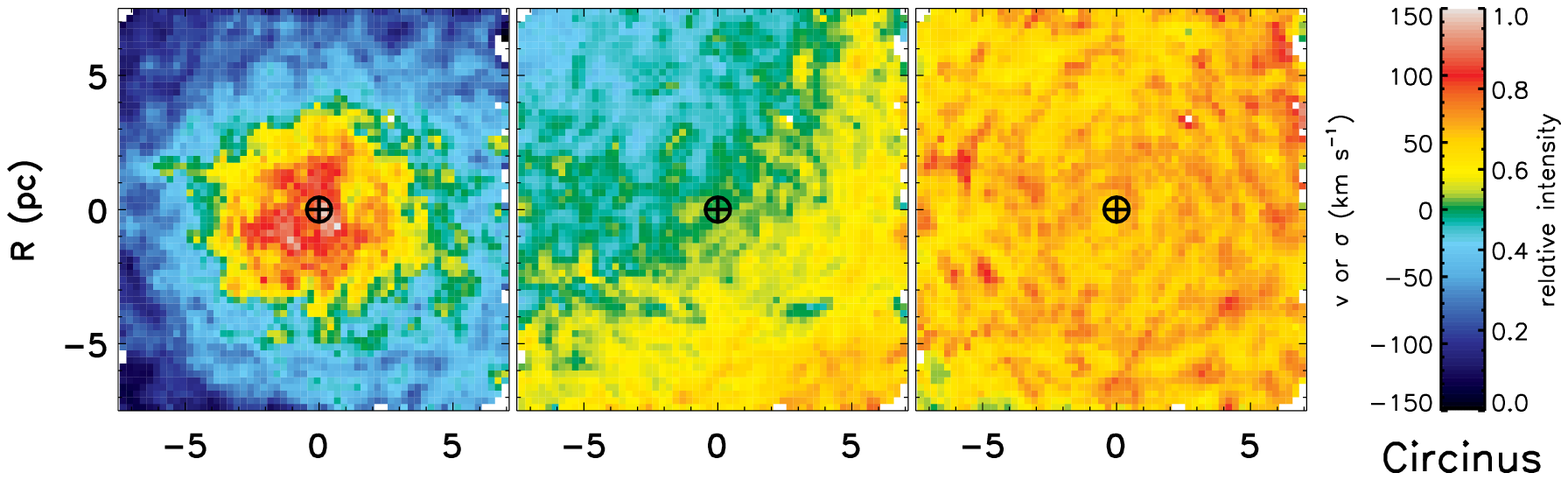}
\plotone{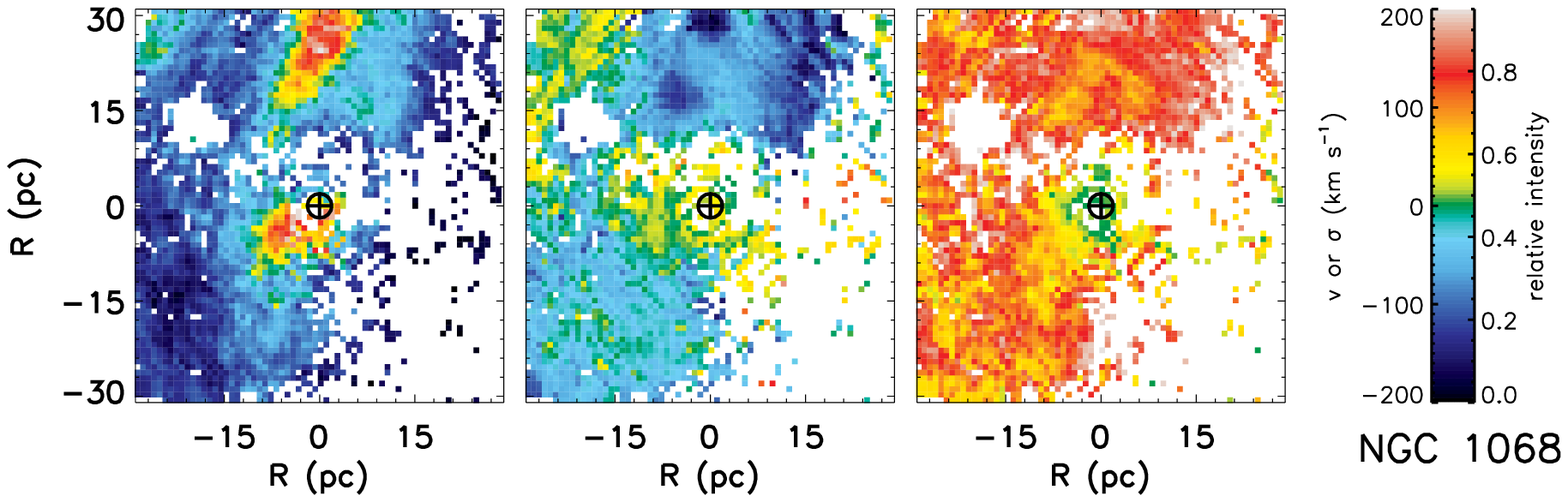}
\caption[]{2-D maps of the \htwo\ flux distribution, velocity, and velocity dispersion, from left to right, for each Seyfert galaxy observed with SINFONI.  In all maps of galaxies observed with SINFONI, North is alighted with the y-axis and East is to the left.  The circled cross marks the location of the nonstellar continuum. \label{fig:maps-sinfoni}}
\end{figure*} 

\begin{figure*}[!ht]	
\epsscale{0.85}
\plotone{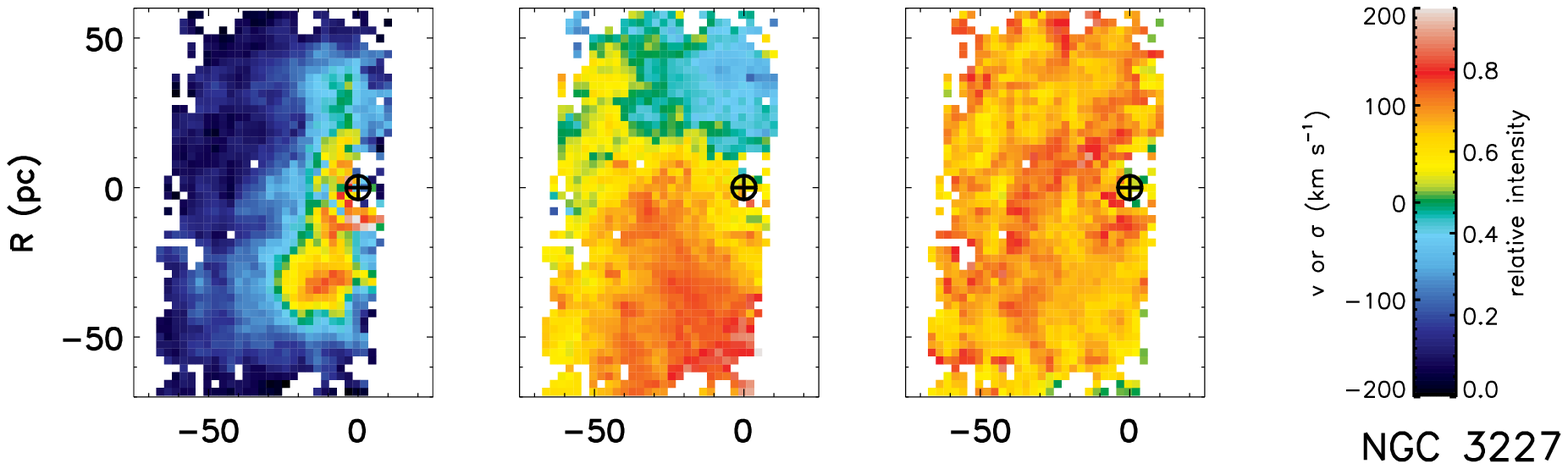}
\plotone{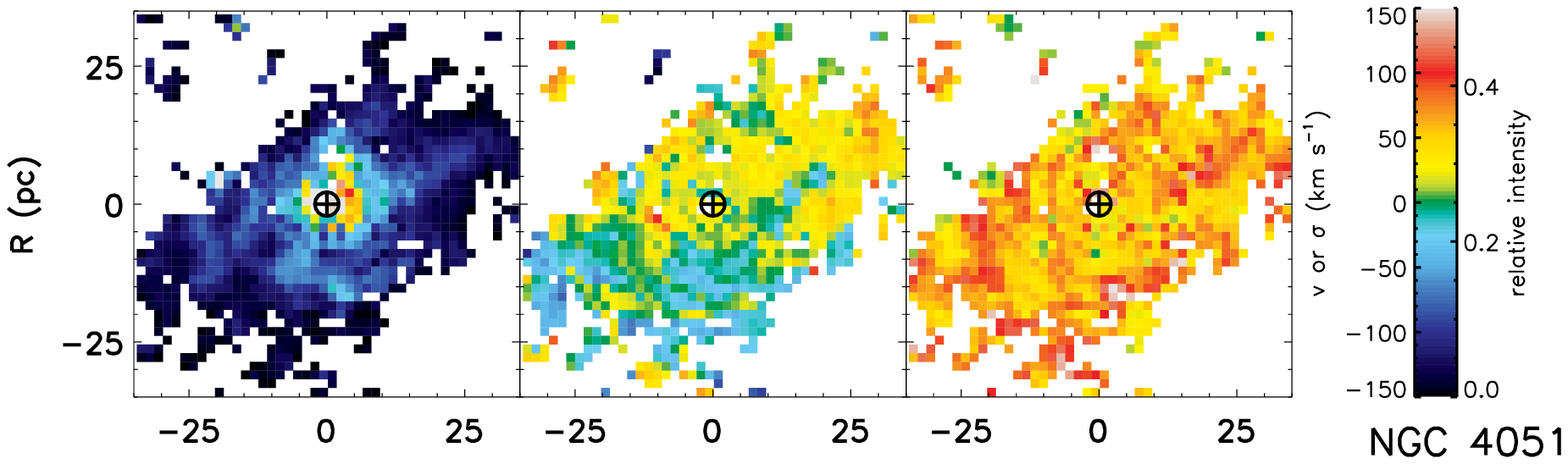}
\plotone{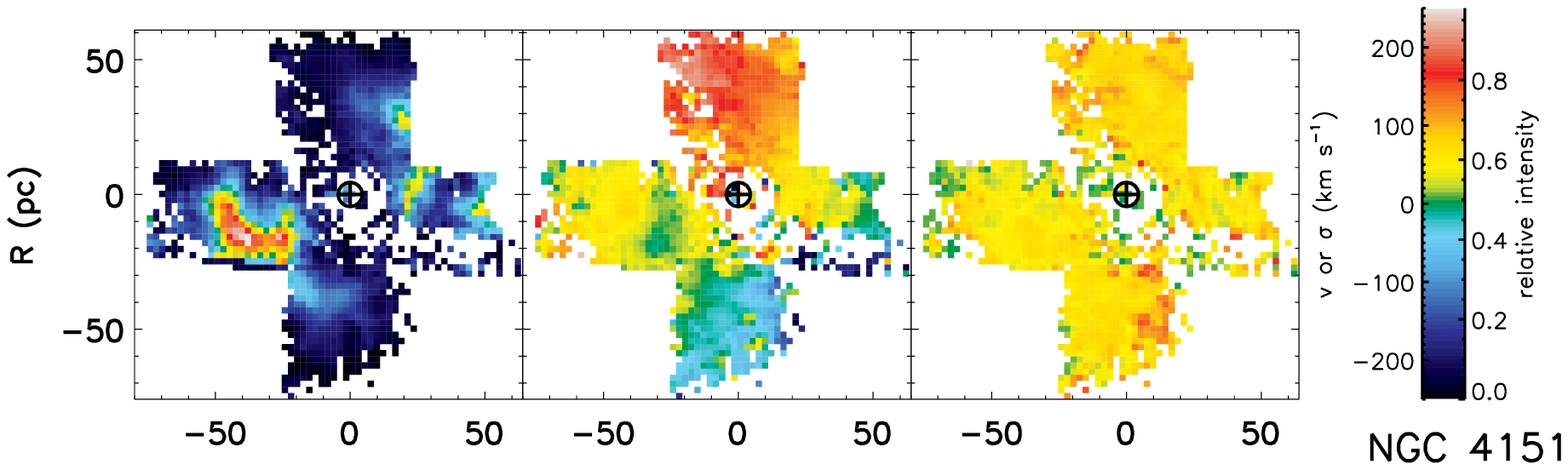}
\plotone{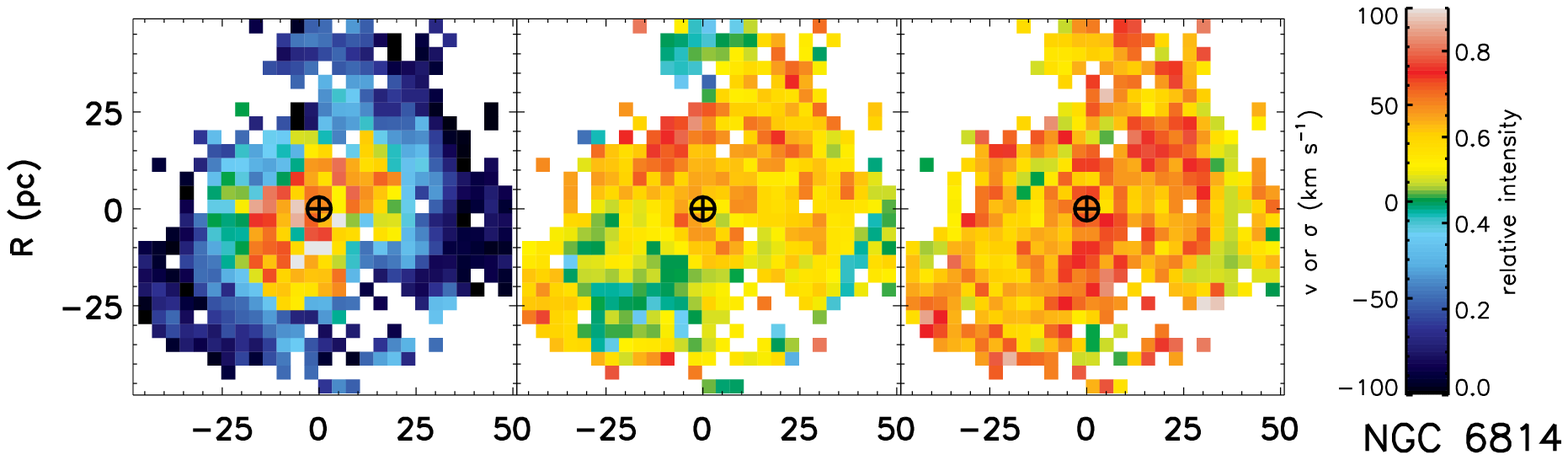}
\plotone{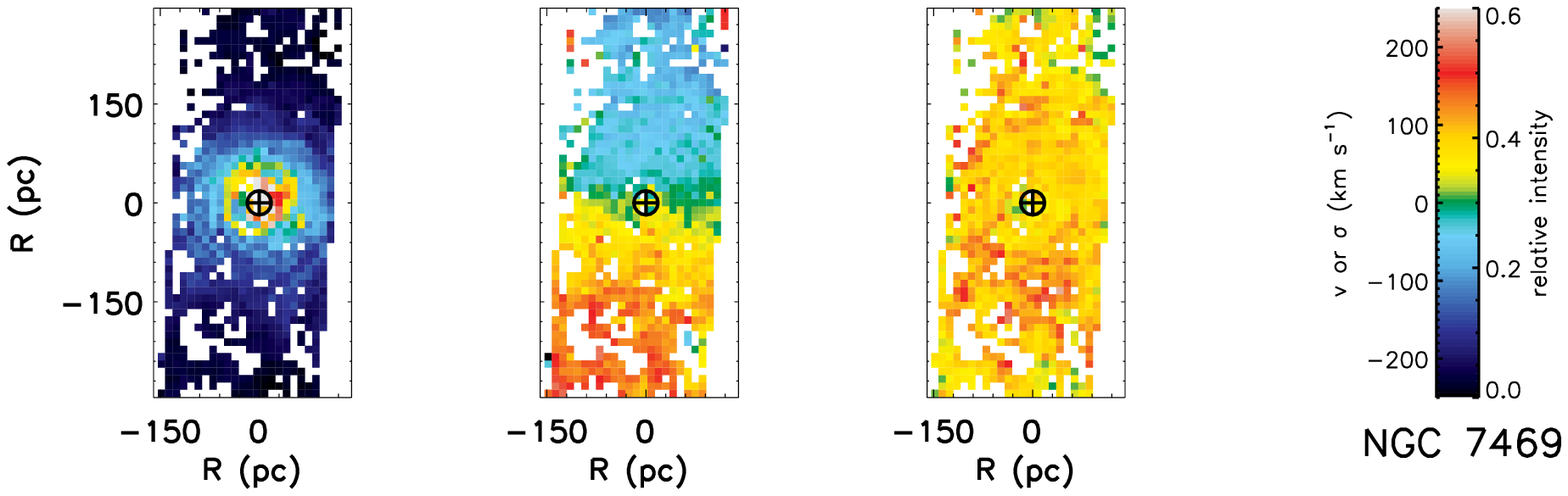}
\caption[]{2-D maps of the \htwo\ flux distribution, velocity, and velocity dispersion, from left to right, for each Seyfert galaxy observed with OSIRIS.  The y-axes in maps of NGC 3227 are aligned along a PA of -45\deg, and those of NGC 7469 are alighted along a PA of -48\deg.  In all other maps North is aligned with the y-axis and East is to the left.  The circled cross marks the location of the nonstellar continuum.  \label{fig:maps-osiris}}
\end{figure*} 

In order to draw general conclusions about the properties of molecular hydrogen in AGNs, all of the galaxies in the sample are evaluated using a consistent method to minimize differences due to spatial resolution.  The distribution and kinematics of the molecular gas are measured by fitting the spectrum of the \htwos\ emission line with an unresolved line profile (e.g. a sky line) convolved with a Gaussian, as well as a linear function to the line free continuum, at each spaxel.  This fitting process gives the intrinsic velocity dispersion (i.e. instrumental broadening has been accounted for), and it is this value that is used in the subsequent analysis, figures, and tables.  The uncertainty of the fit is estimated using Monte Carlo techniques by refitting the best-fit Gaussian with added noise of the same statistics as the data.  This is done 100 times, and the standard deviations of the best-fit parameters are used as the uncertainties.  Typically the uncertainties in velocity and velocity dispersion (\sig) are on the order of 5-15 \kms.  The 2-D maps of the \htwo\ flux distribution, velocity, and \sig\ for each of the observed galaxies are shown in Figures \ref{fig:maps-sinfoni} and \ref{fig:maps-osiris}. 

\begin{figure*}[!ht]	
\epsscale{0.4}
\plotone{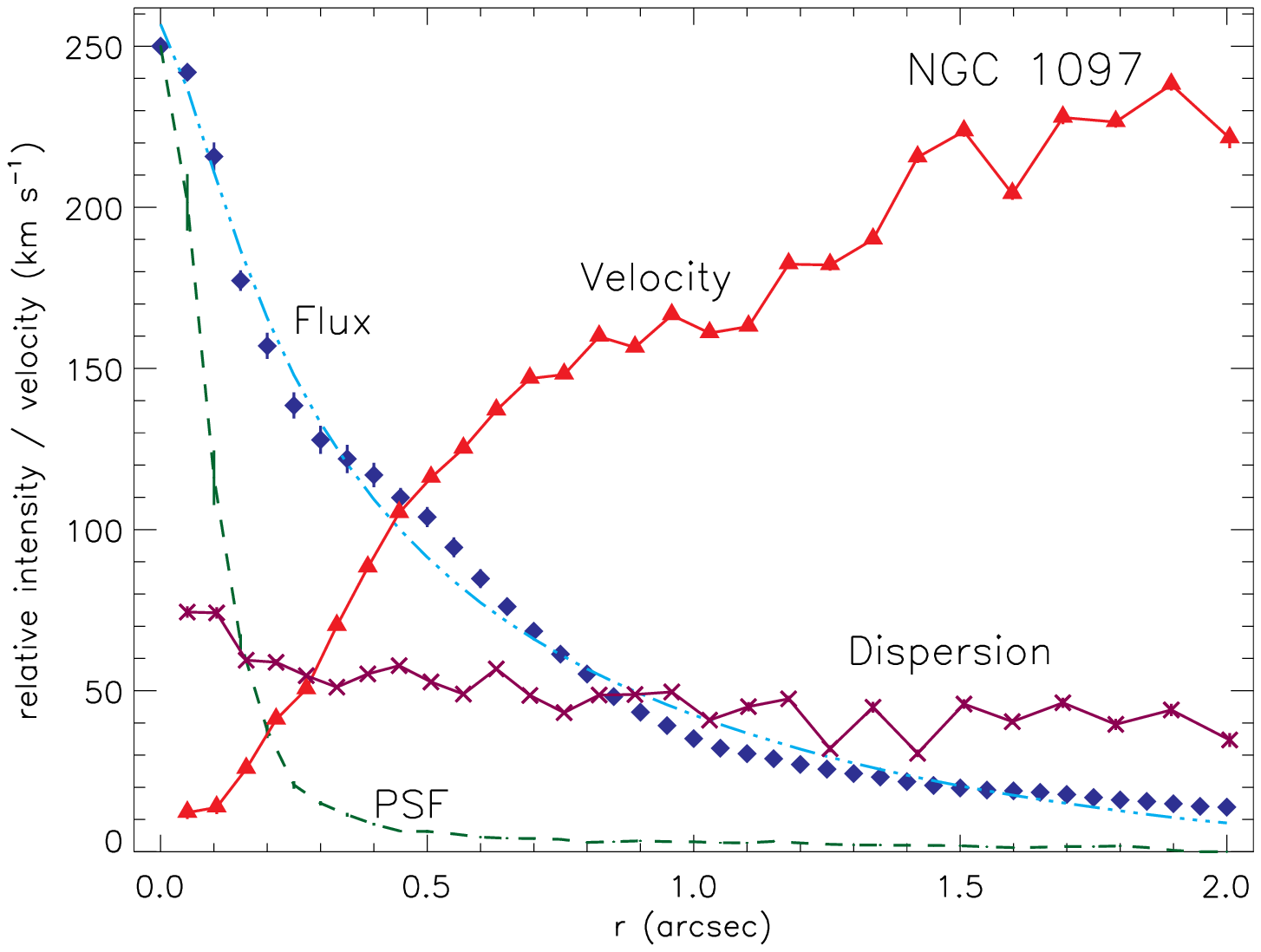}
\plotone{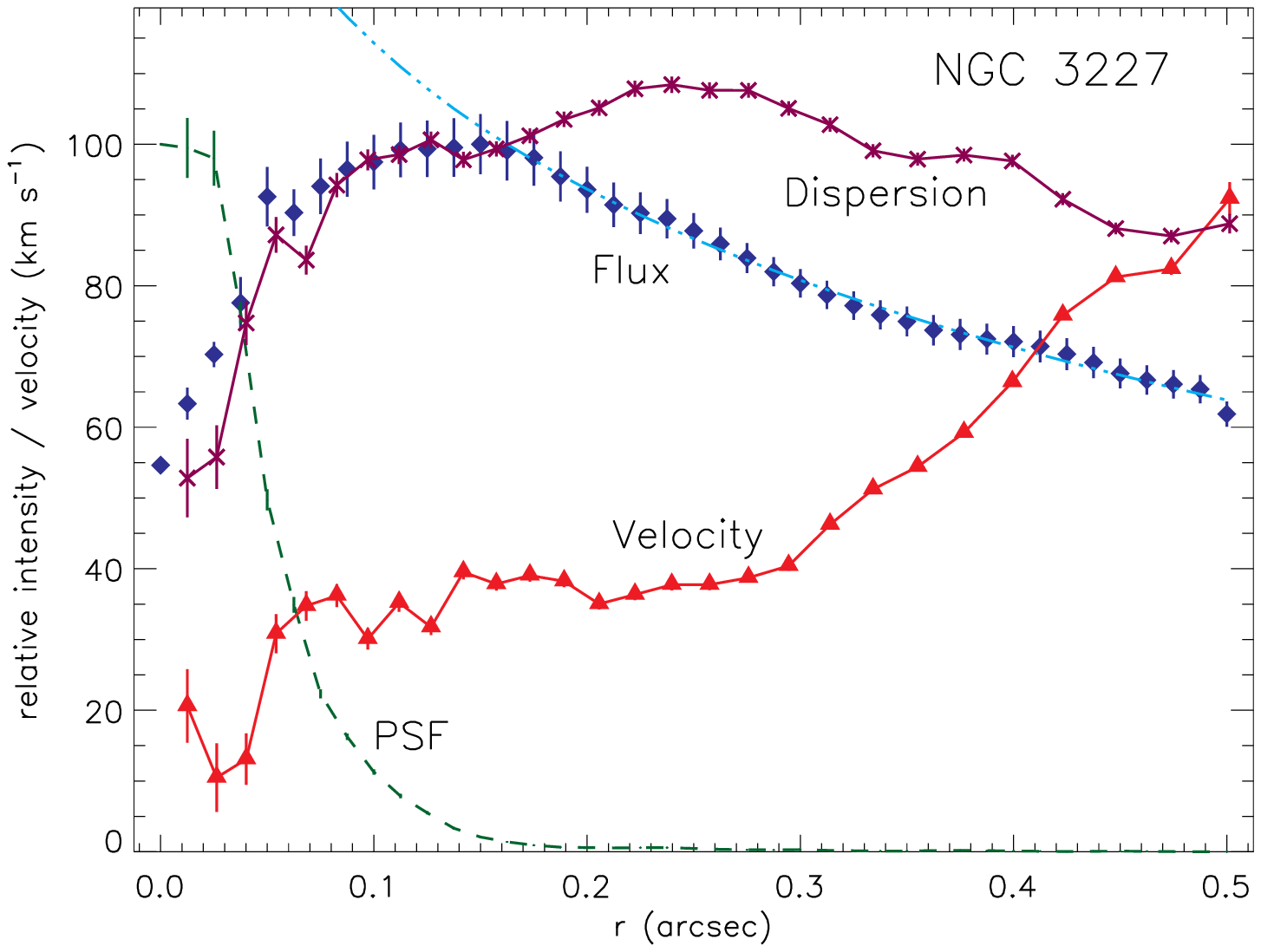}
\plotone{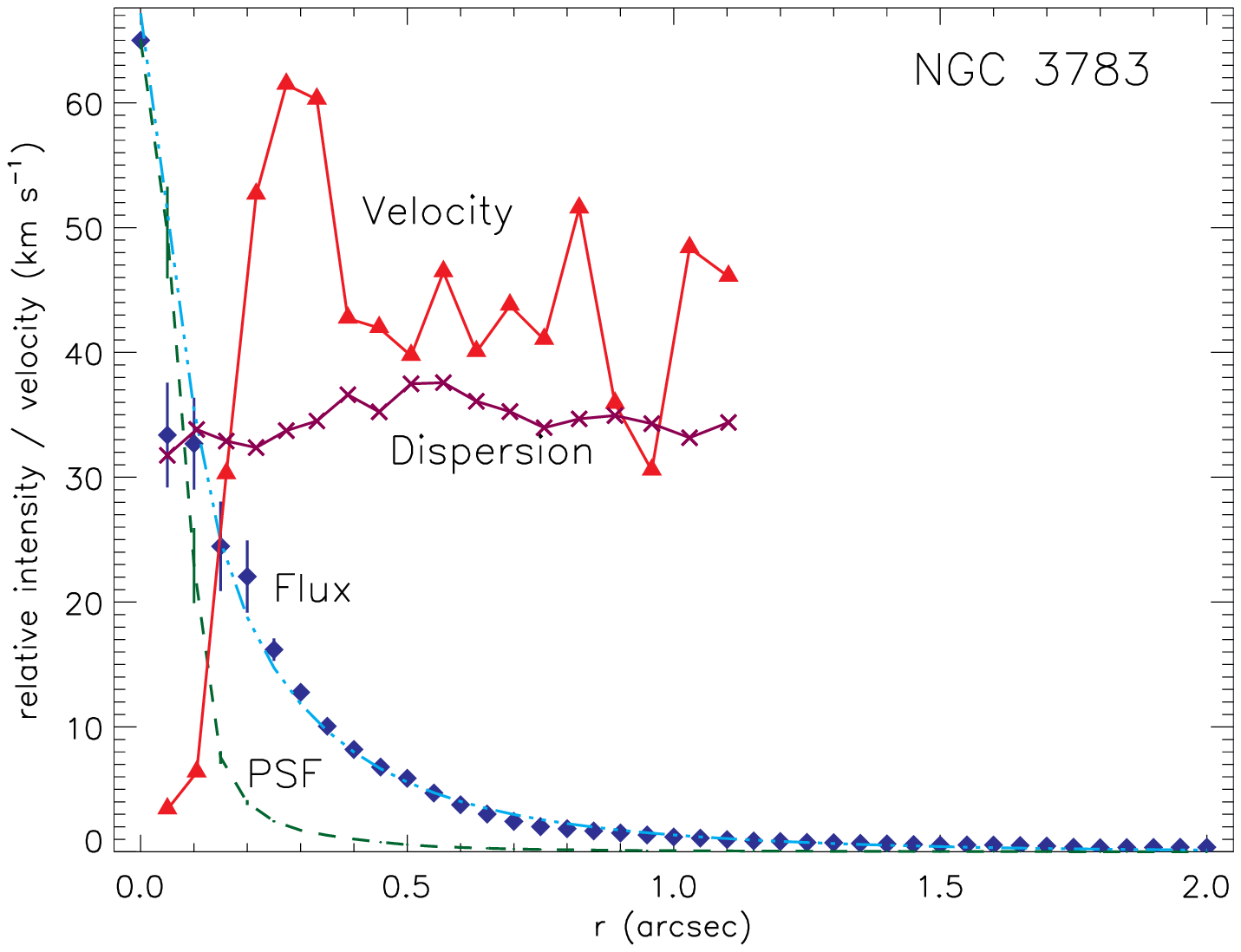}
\plotone{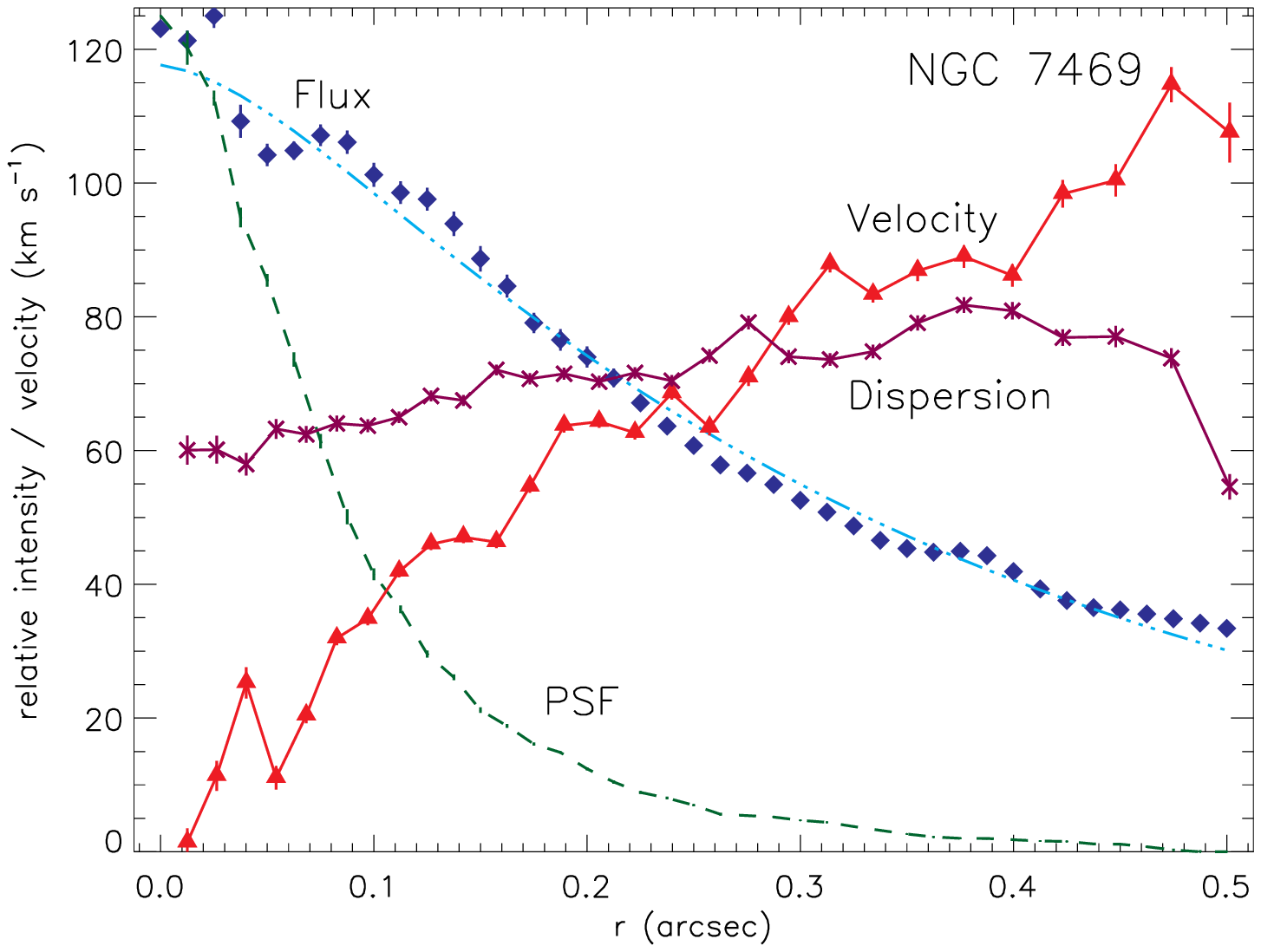}
\plotone{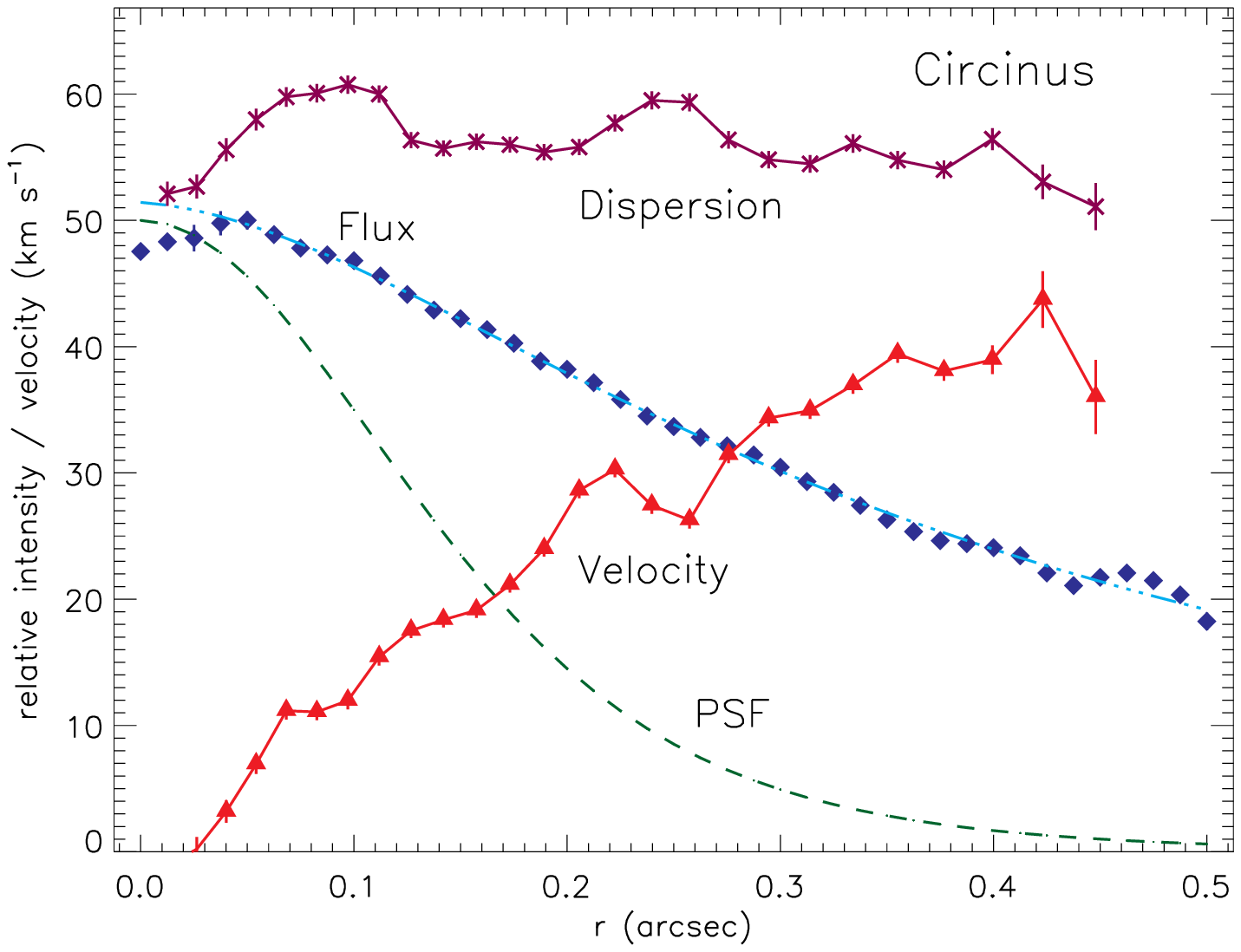}
\plotone{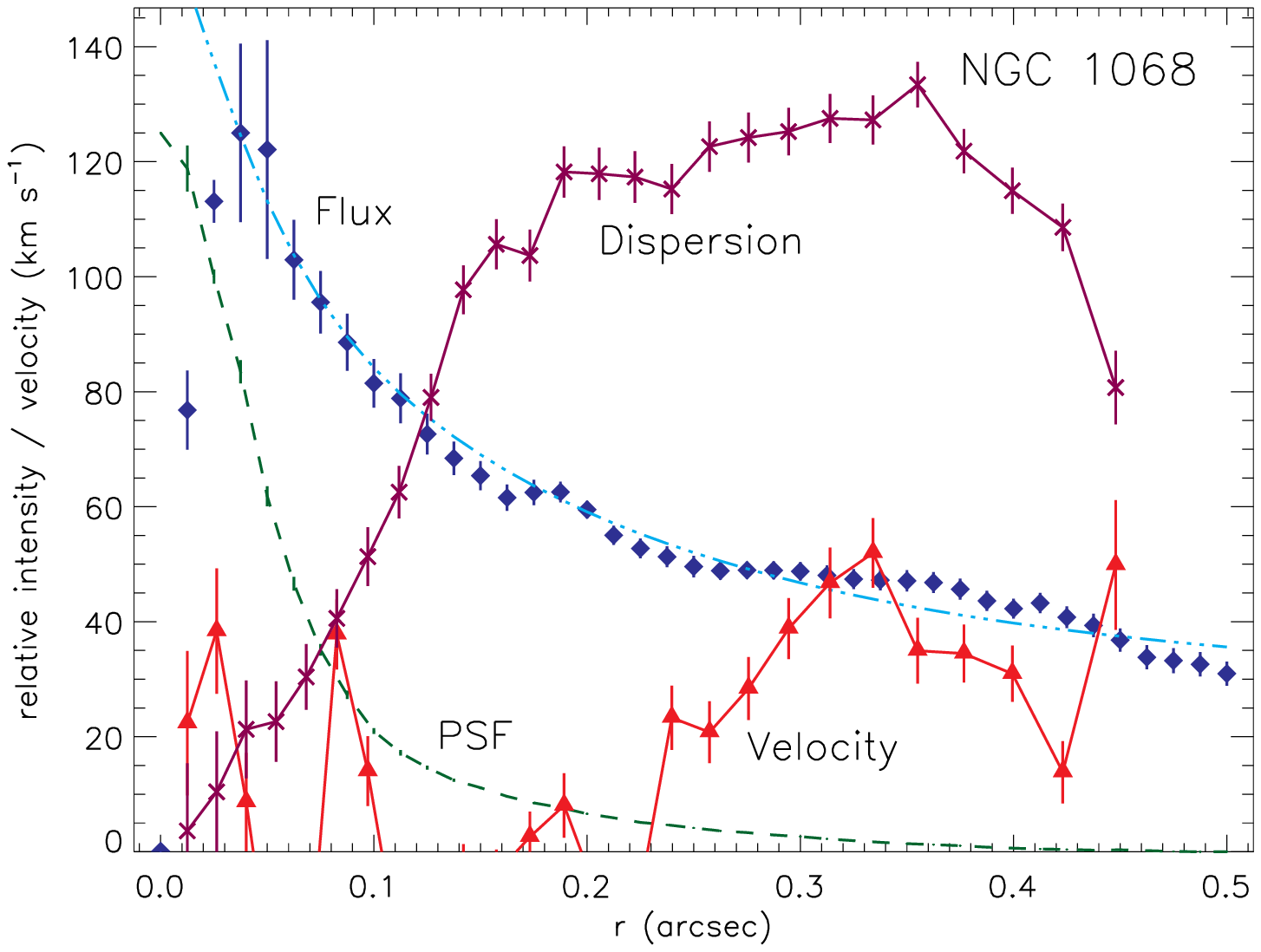}
\caption[]{Azimuthal averages based on 2-D measurements made with SINFONI.  As suggested by the labels, the plots show the \htwo\ flux distribution (diamonds), the S\'{e}rsic fit to this distribution (dash-dotted curve), the PSF (dashed curve), the inclination corrected rotational velocity (solid curve with triangle data points), and the velocity dispersion (solid curve with ``x" data points).  The galaxy name is given in the upper right of each plot. \label{fig:rad-sinfoni}}
\end{figure*}

\begin{figure*}[!ht]	
\epsscale{0.4}
\plotone{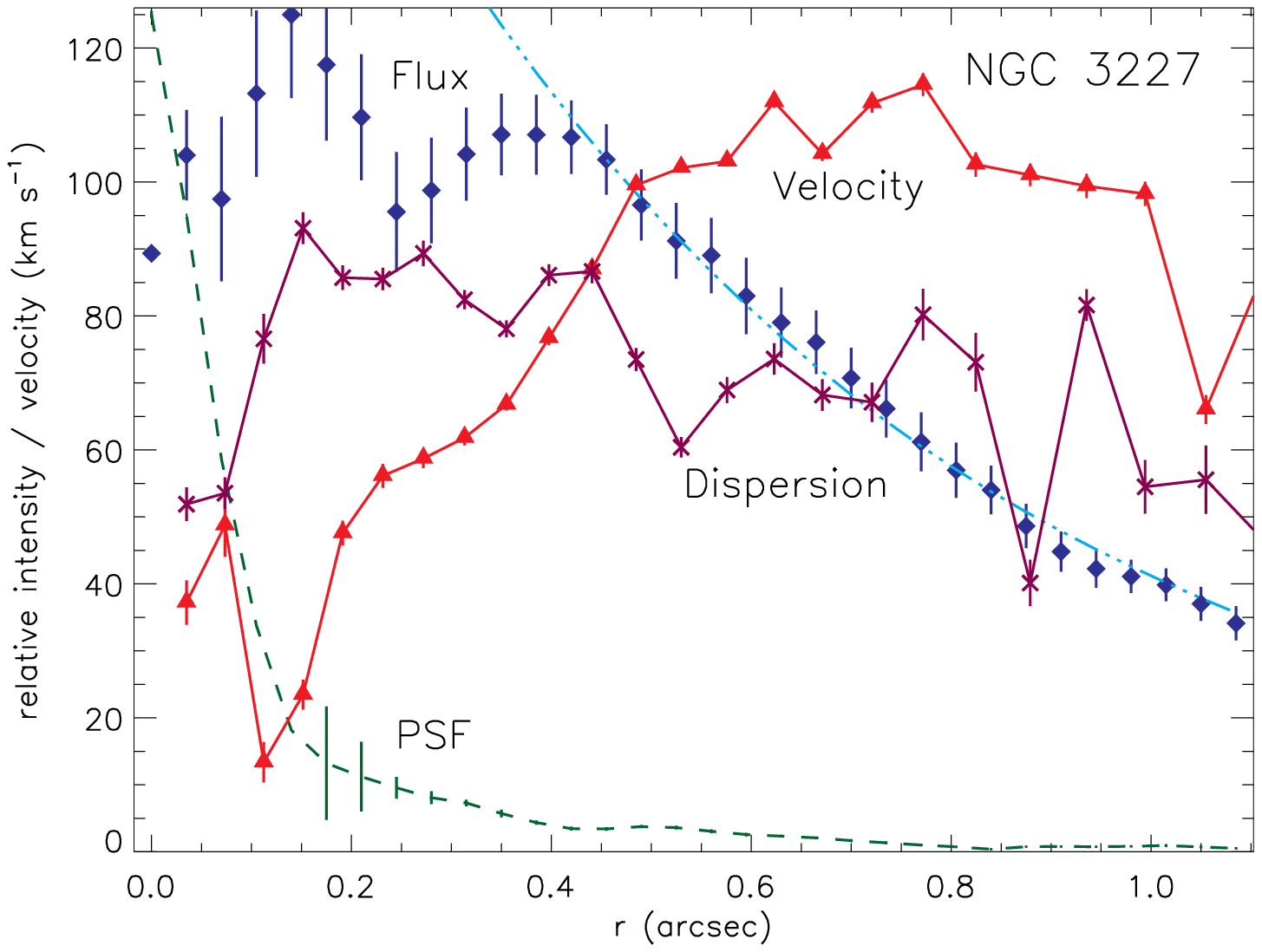}
\plotone{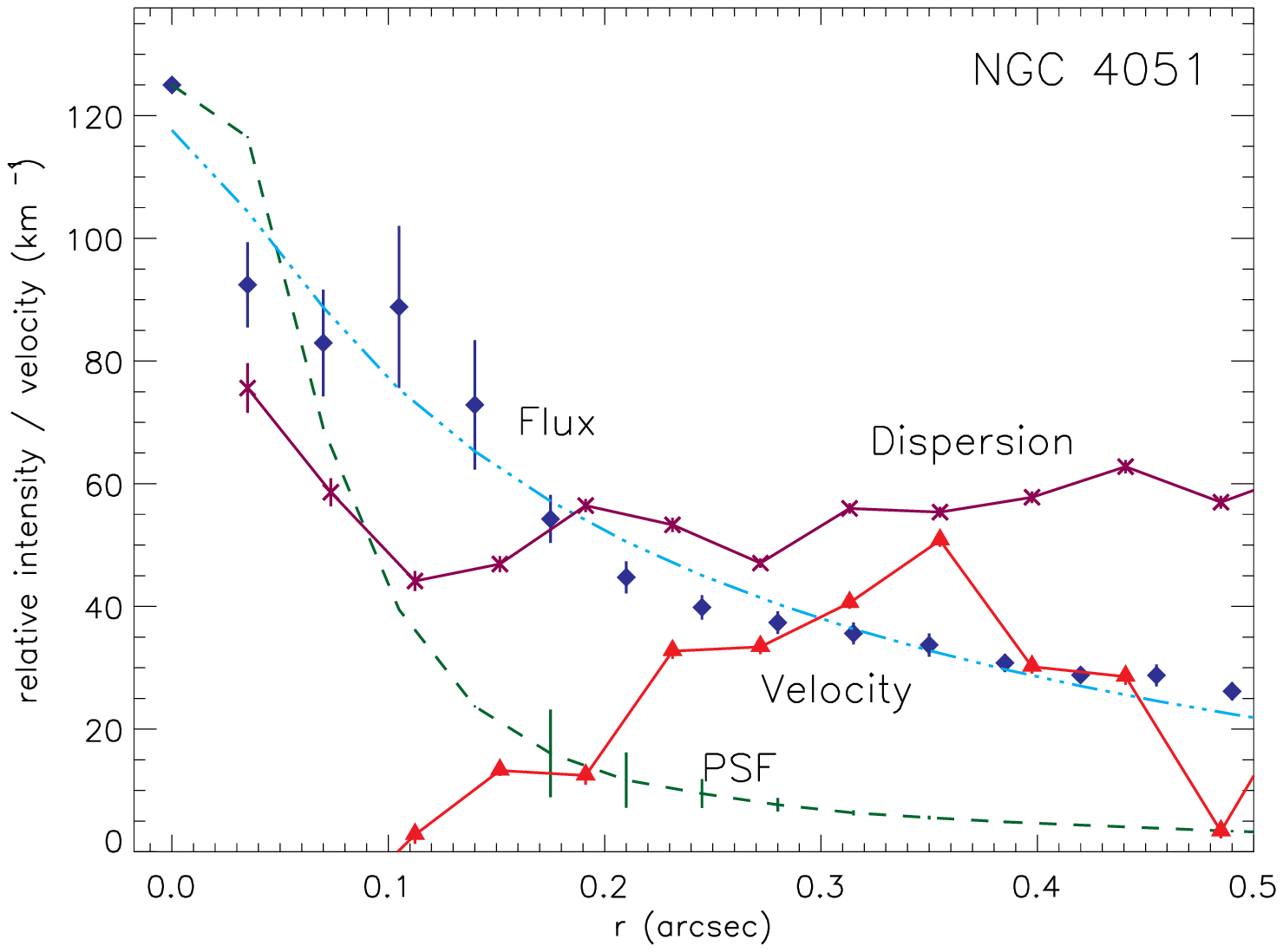}
\plotone{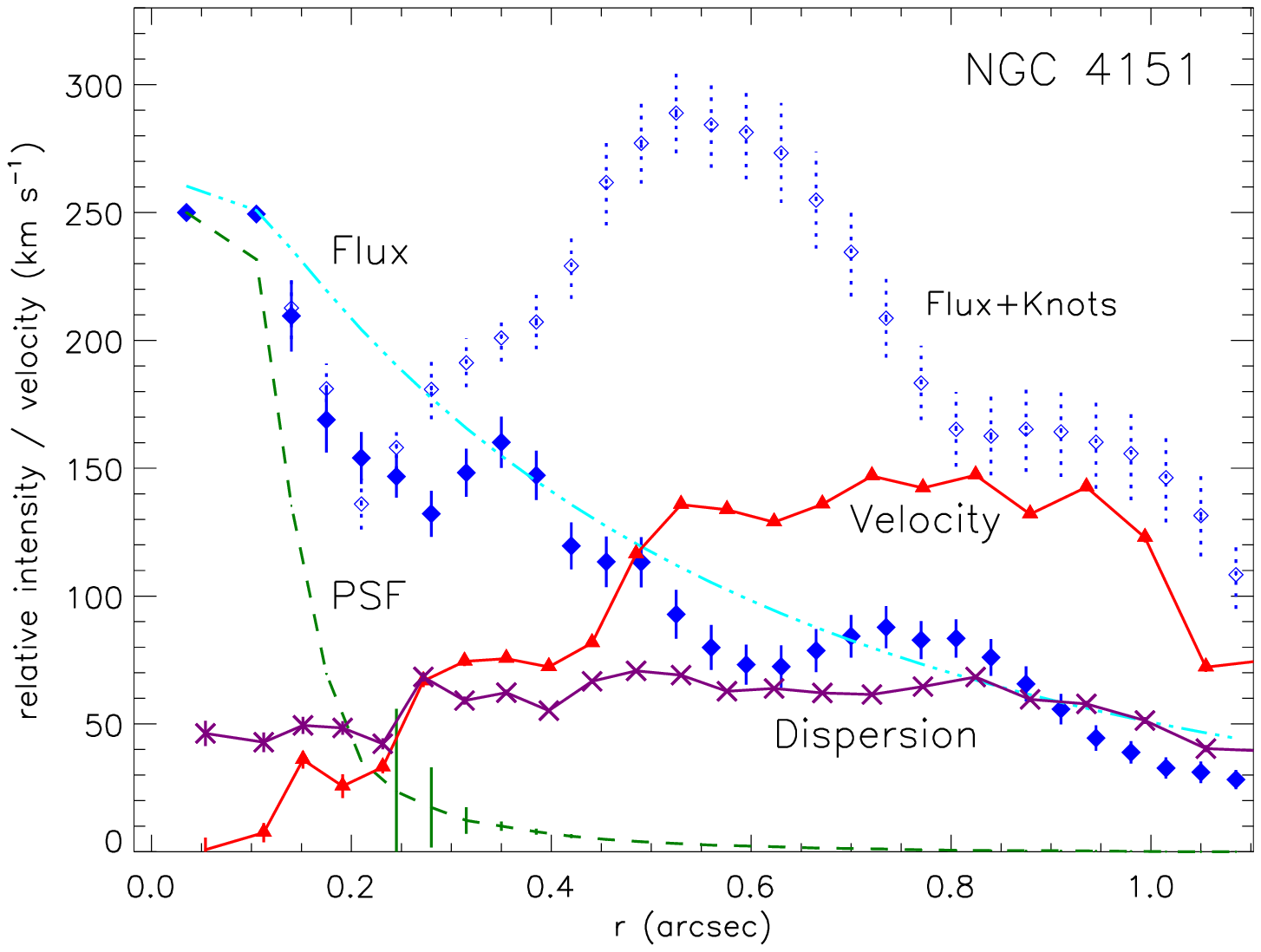}
\plotone{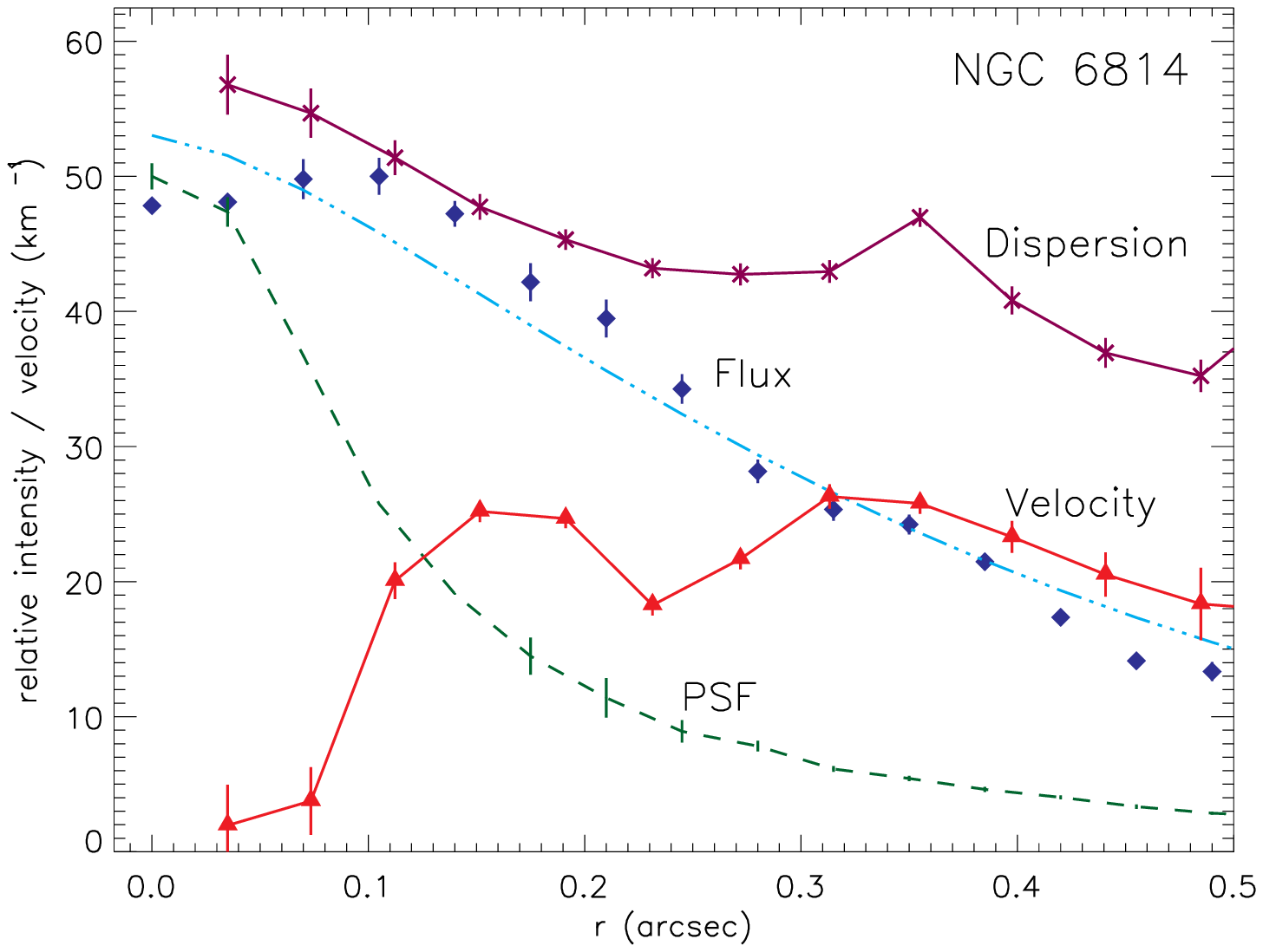}
\plotone{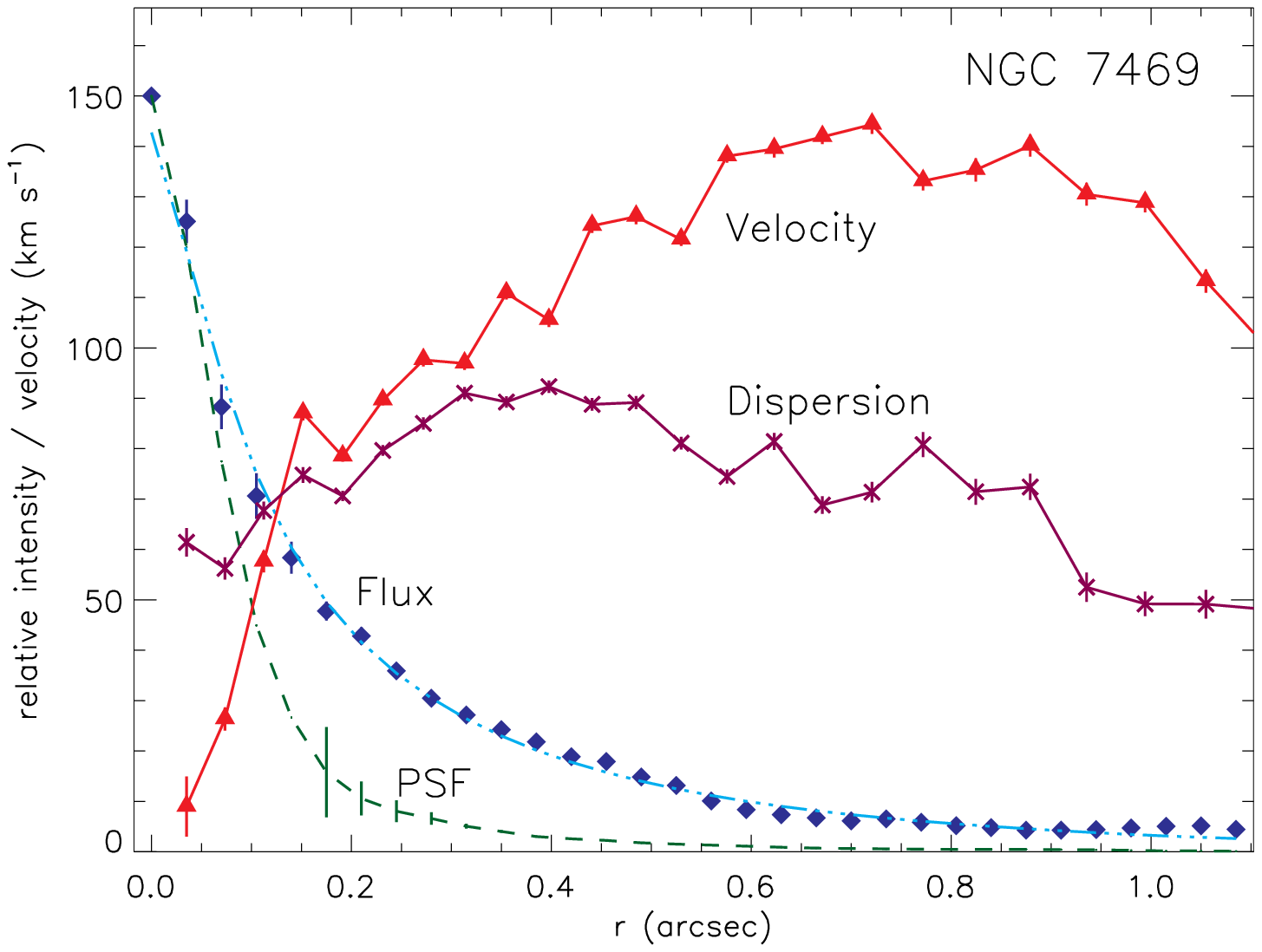}
\caption[]{Azimuthal averages based on 2-D measurements made with OSIRIS.  As suggested by the labels, the plots show the \htwo\ flux distribution (diamonds), the S\'{e}rsic fit to this distribution (dash-dotted curve), the PSF (dashed curve), the inclination corrected rotational velocity (solid curve with triangle data points), and the velocity dispersion (solid curve with ``x" data points).  The galaxy name is given in the upper right of each plot.  For NGC 4151, the azimuthal average of the measured flux distribution, including the bright knots of emission, is indicated by the open diamonds with dashed error bars, while the azimuthal average of just the extended emission is shown by the solid diamonds with solid error bars (see \ref{sec:obj-4151} for details).  \label{fig:rad-osiris}}
\end{figure*}

The gas kinematics is characterized using the method of kinemetry, which is a generalization of surface photometry to the higher order moments of the velocity distribution (\citealt{krajnovic06}).  The kinematic center is assumed to be coincident with the nucleus defined as the peak in the nonstellar emission.  As justified in $\S$ \ref{sec:kin_mass}, the kinematics are ultimately fit holding the position angle (PA) and inclination angle constant since neither is found to vary significantly in the region measured.  Furthermore, the best fit PA and inclination angle of the \htwo\ kinematics is consistent with that fit of the stellar kinematics and we conservatively assume the best-fit values to the stellar kinematics when possible.  The azimuthal average of the flux distribution and kinematics in each galaxy are also computed, correcting for these best-fit PA and inclination angles, to further investigate the general properties of \htwo\ in the observed AGN (Figures \ref{fig:rad-sinfoni} and \ref{fig:rad-osiris}).  

\begin{figure*}[!ht]	
\epsscale{0.4}
\plotone{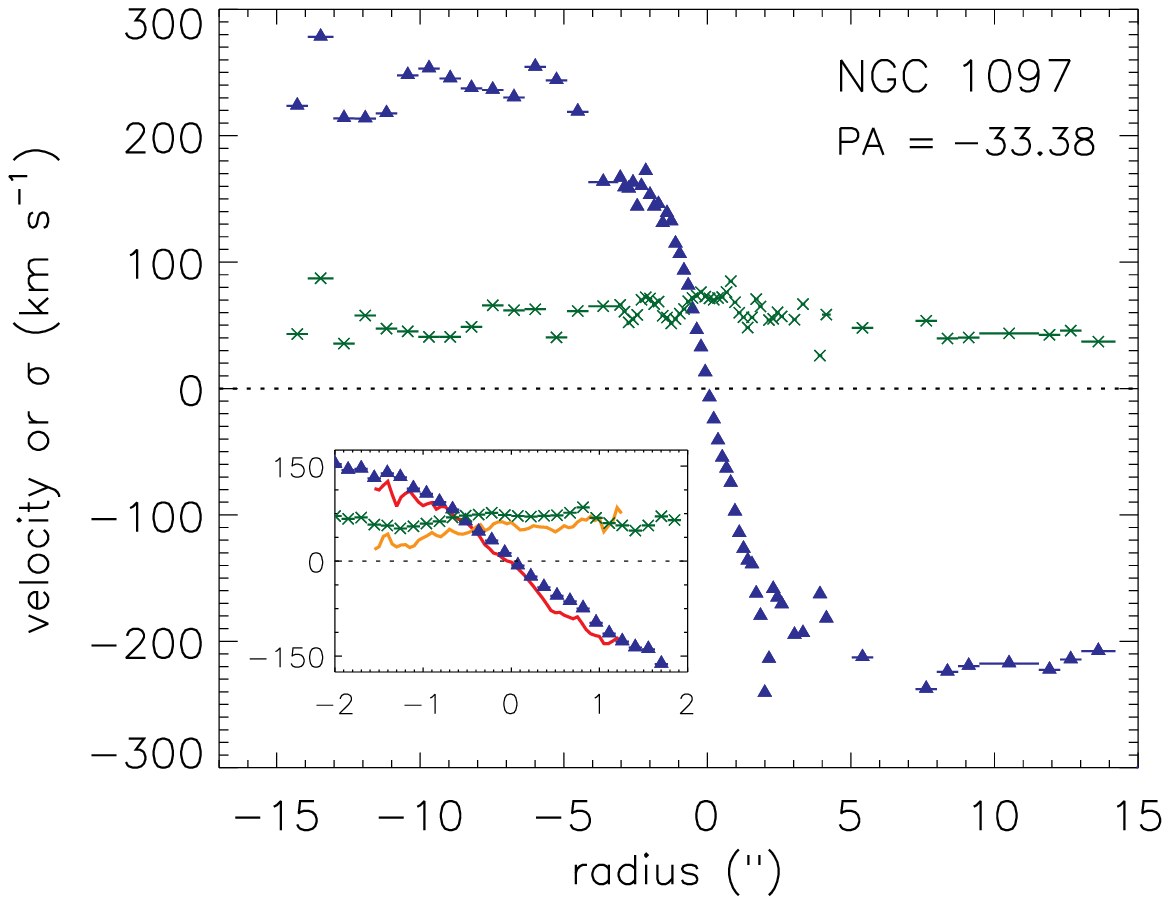}
\plotone{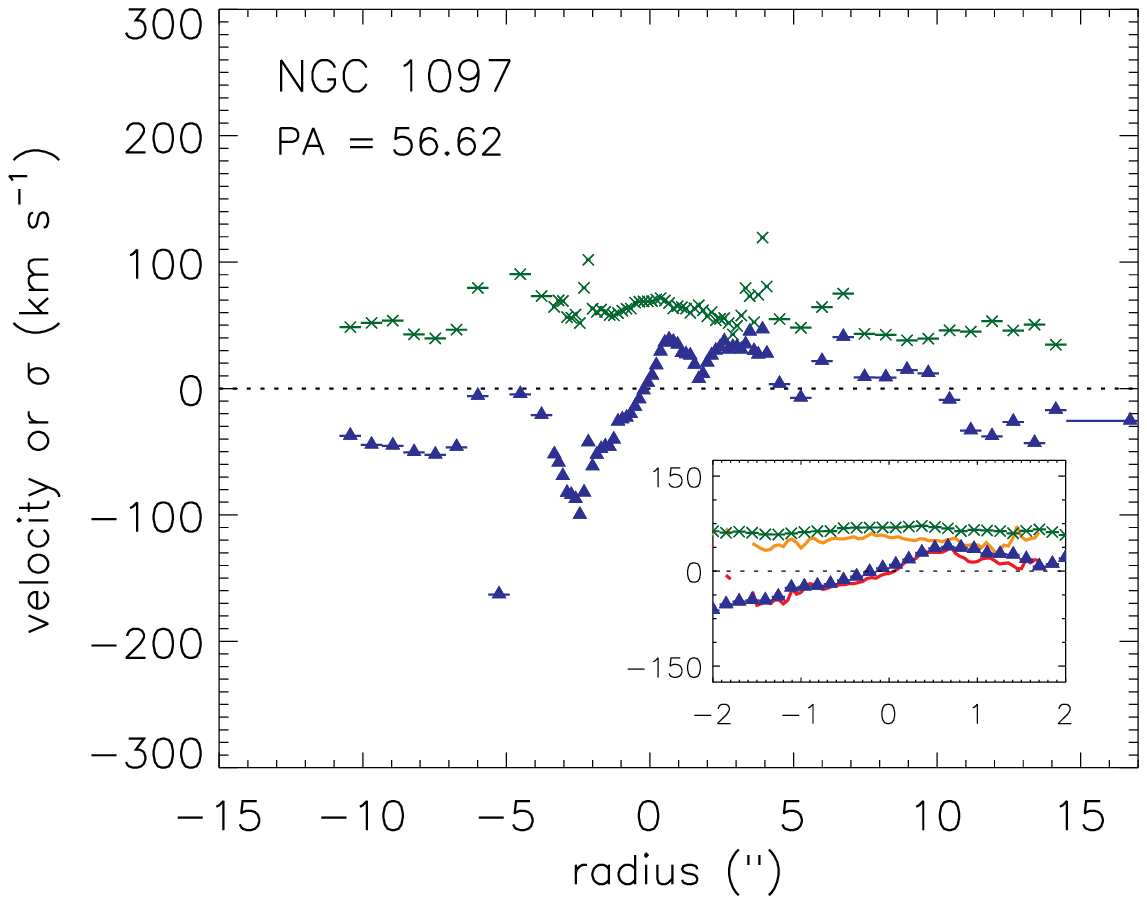} 
\plotone{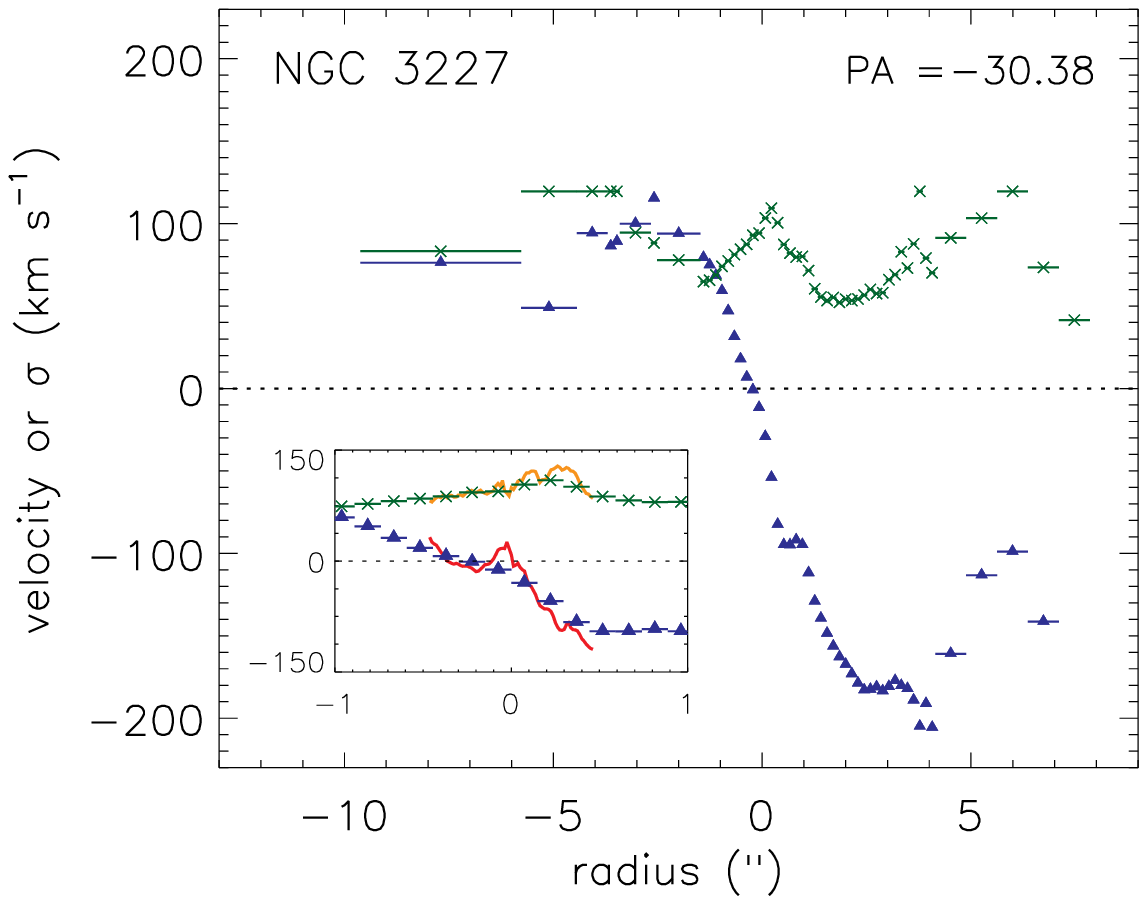} 
\plotone{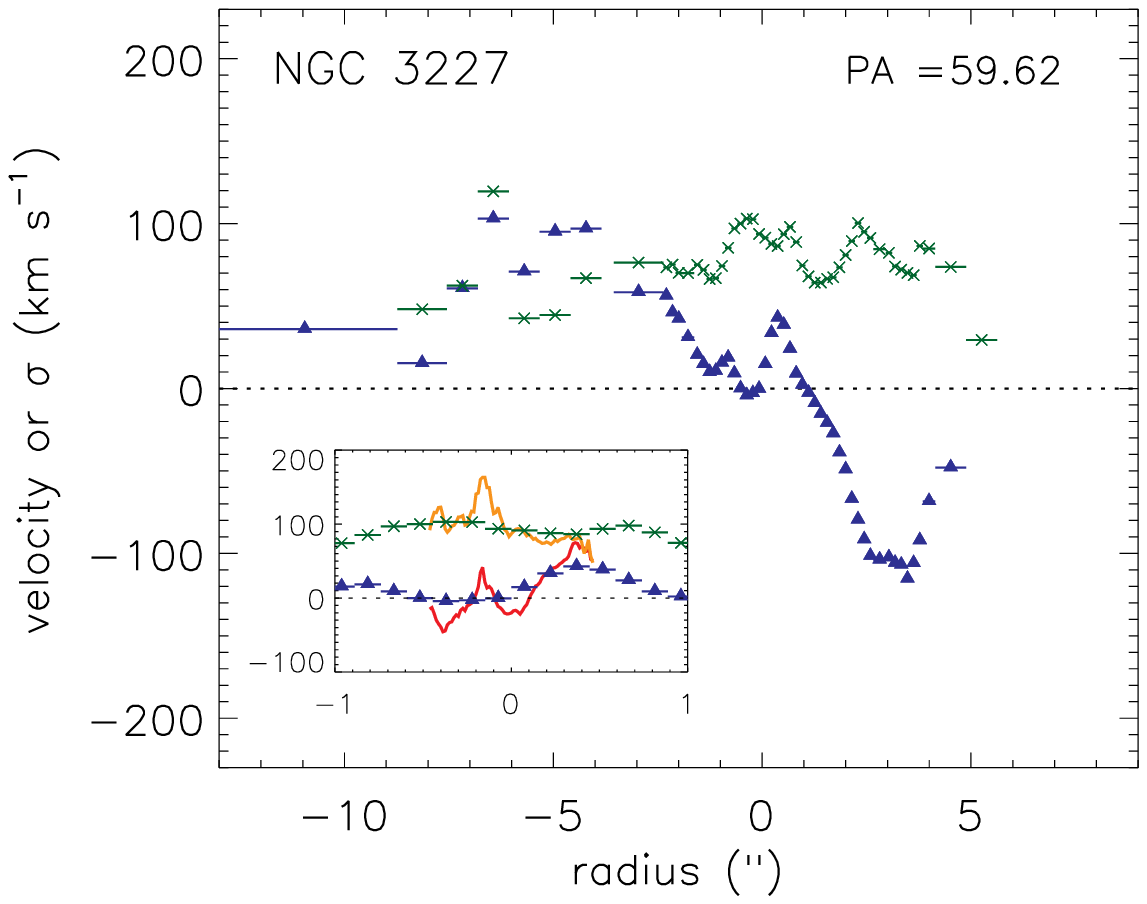}
\plotone{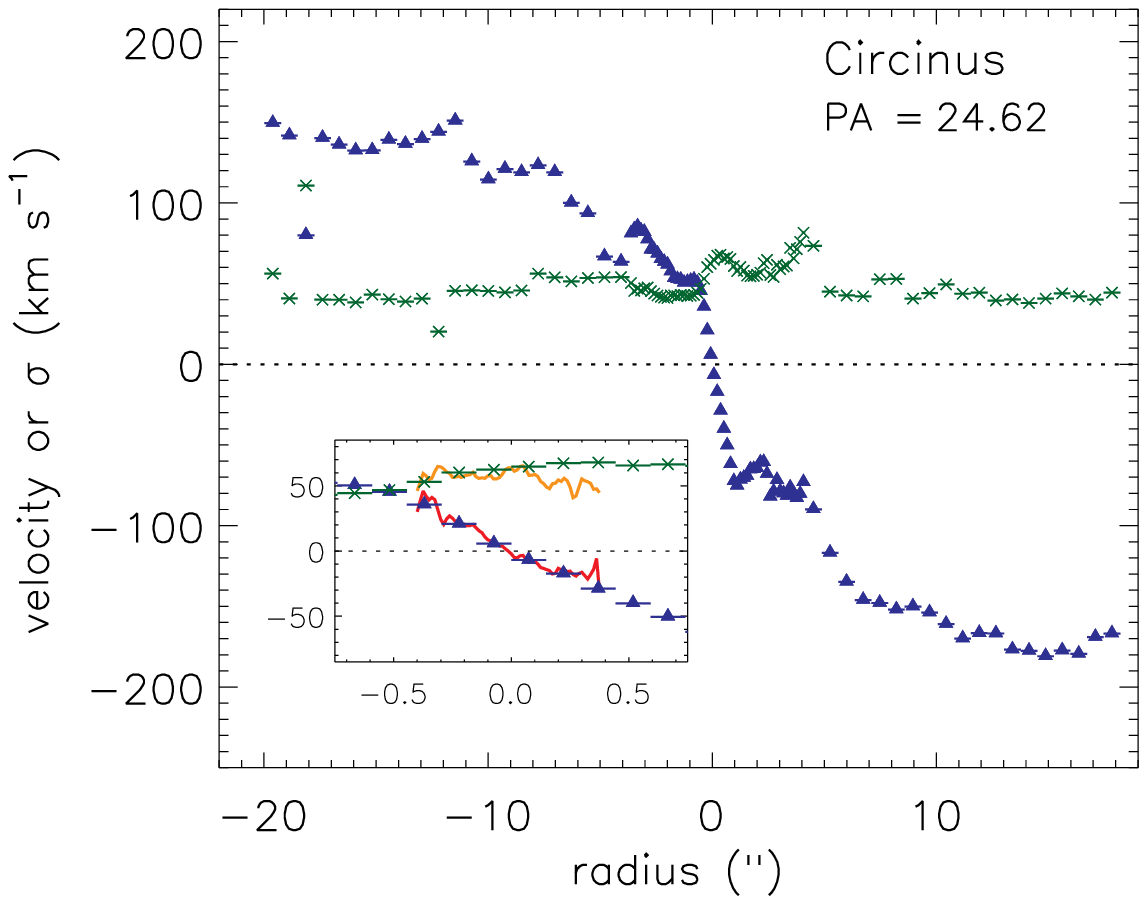}
\plotone{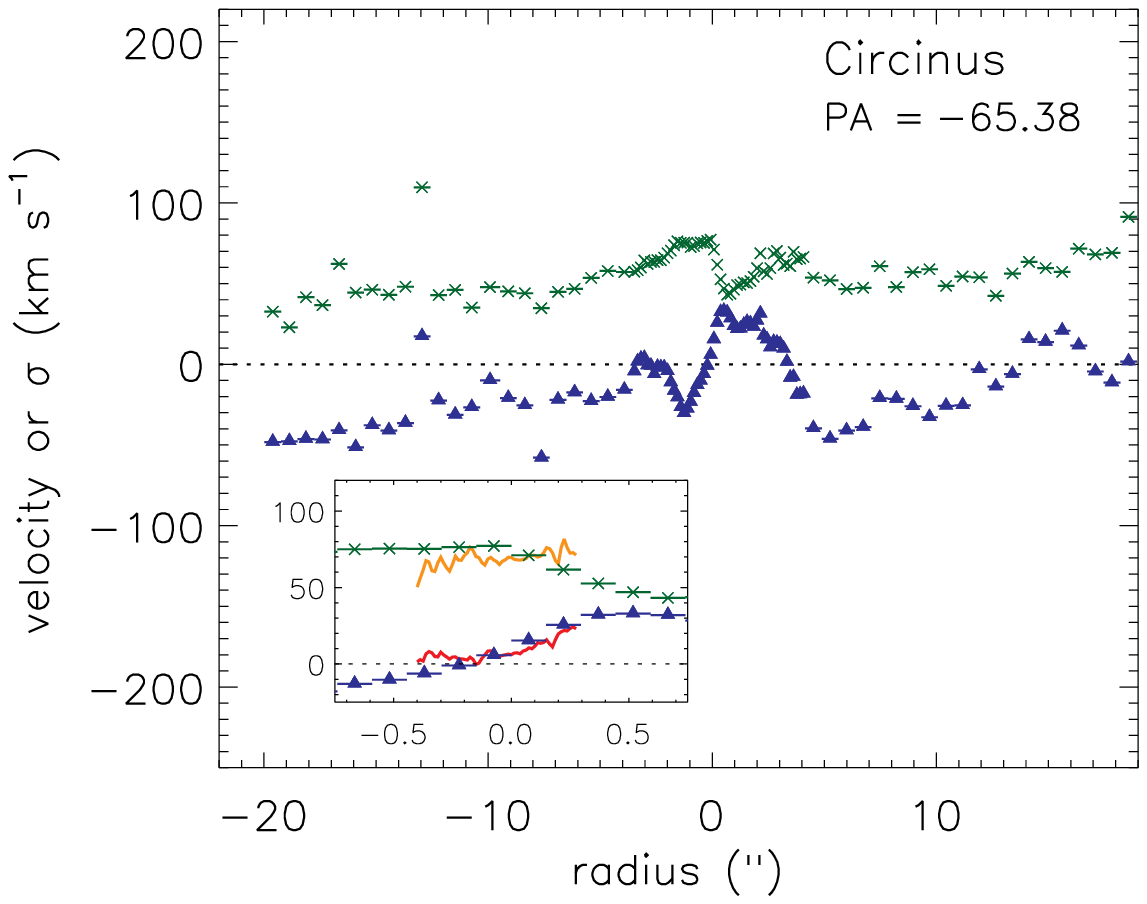} 
\caption[]{Observed \htwos\ velocity (triangles) and velocity dispersion (``x" data points) along the major (left column) and minor (right column) axes as measured with ISAAC.  The PA of the slit (East of North) and the galaxy name are as indicated at the top of each plot.  A comparison of the ISAAC data with the same region measured with SINFONI is shown in the smaller inset plot, where the higher resolution velocity and velocity dispersion are indicated by the solid curves. \label{fig:isaac}}
\end{figure*}

The flux distribution, rotational velocity, and velocity dispersion as a function of radius were also determined for each of the single slit ISAAC spectra.  This is achieved by fitting a Gaussian to the emission line profile in a spatial region determined by binning pixels until at least a 5\sig\ detection of the \htwo\ emission is reached.  A single Gaussian is found to be a suitable representation of the emission line profile in all three of the galaxies observed.  The \htwo\ velocity and \sig\ of the gas on the larger scales measured with ISAAC, as well as a comparison to the smaller scale kinematics measured with SINFONI, are shown in Fig. \ref{fig:isaac}.

We have chosen to exclude NGC 1068 from the AGN sample because it is the only galaxy in the sample for which the \htwo\ kinematics are dominated by noncircular motions rather than, as for all other galaxies in the sample, by rotation in the gravitational potential implied by the nuclear stellar kinematics (see $\S$ \ref{sec:kin_mass}).  The \htwo\ kinematics of NGC 1068 differ from circular rotation by as much as 100 \kms, and, as discussed in detail in \citet{mueller08}, suggest an inflow of material toward the central BH.  The high spatial resolution of the NGC 1068 observations can be ruled out as the reason for the complexity of the kinematics since the 6 pc resolution is comparable to that achieved with four other galaxies (FHWM $<$ 10 pc) and two of these have data with a resolution even better than that of NGC 1068.  Although we do not include NGC 1068 in our discussion of the general \htwo\ properties in AGNs, it is still presented along side the rest of the sample in all figures and tables for completeness.  For many of the \htwo\ properties discussed throughout this paper, the properties measured and derived for NGC 1068 are broadly consistent with those found for the rest of the sample.  However, with the advantage of 2-D spatial data, the kinematics clearly reveals that the nuclear properties of NGC 1068 are unique with respect to the other galaxies observed.

\begin{figure}[!ht]	
\epsscale{1.2}
\plotone{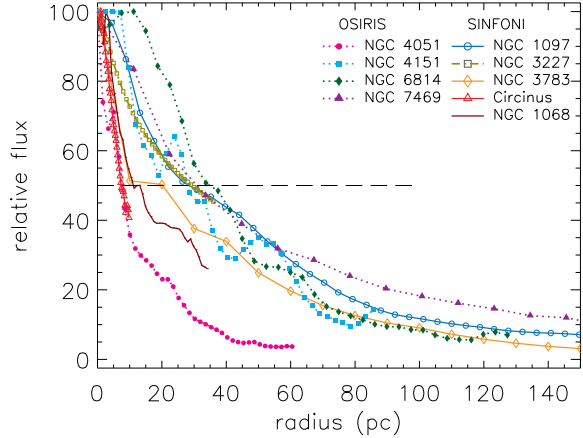}
\caption[]{Azimuthal averages of the H$_{2}$ flux distributions as indicated by the legend.  The horizontal dashed line indicates the HWHM.  For NGC 3227, the best fit S\'ersic function is shown (see $\S$ \ref{sec:obj-3227} for details).  Typically the standard deviation of the azimuthally averaged flux is 5\%. \label{fig:dist}}
\end{figure}

\begin{figure}[!ht]	
\epsscale{1.2}
\plotone{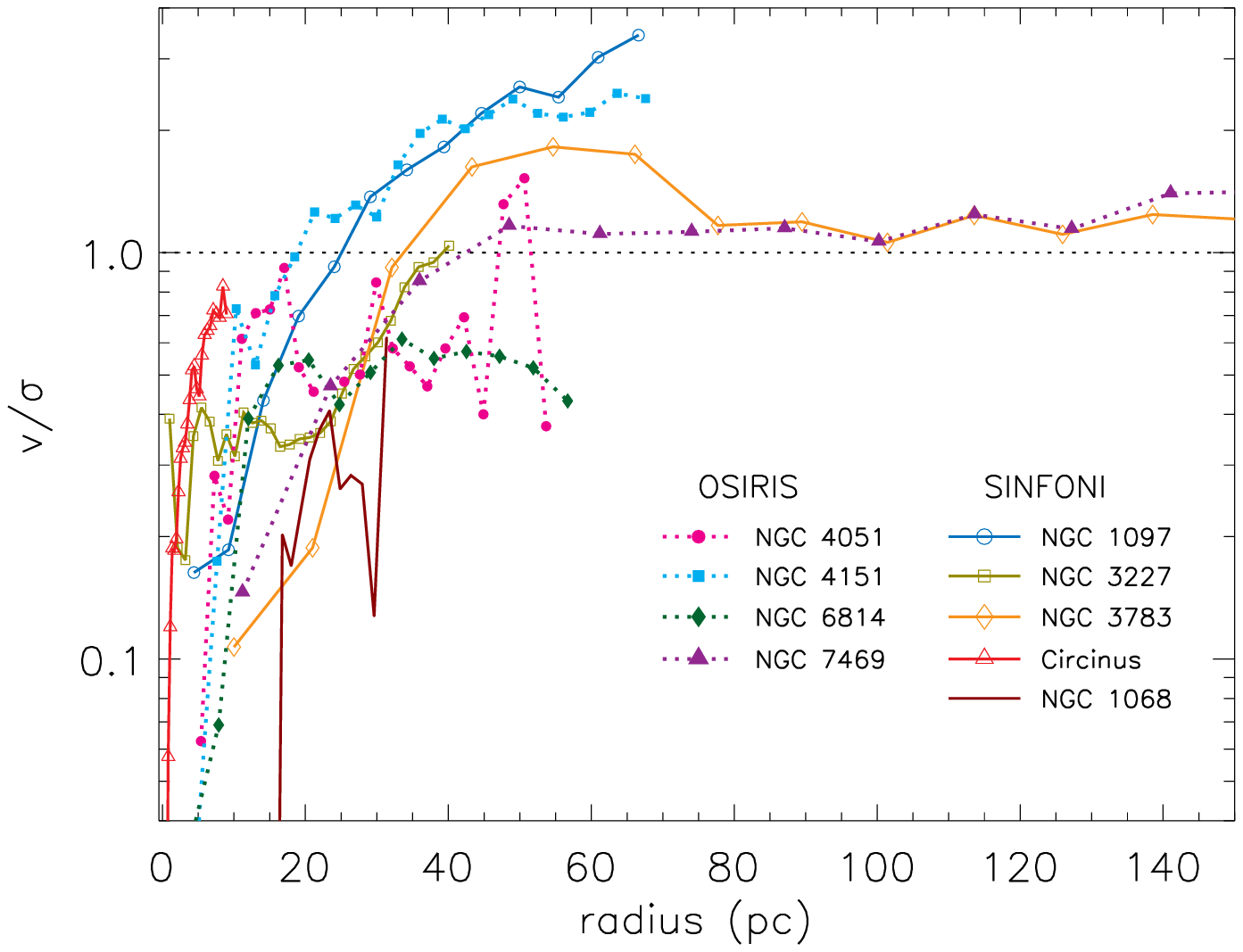}
\caption[]{Ratio of azimuthally averaged \vs\ as a function of radius for each galaxy as indicated by the legend.  The point where velocity is equal to $\sigma$ is indicated by the dotted horizontal line.  The typical standard deviation of the azimuthal average is less than 0.1. \label{fig:vs}}
\end{figure}


\subsection{Spatial Distribution}

As can be seen in the 2-D maps of the \htwo\ flux distribution in Figures \ref{fig:maps-sinfoni} and \ref{fig:maps-osiris}, there is a great deal of diversity in the nuclear \htwo\ flux distributions in the sample galaxies.  Although most of the galaxies have a relatively symmetric flux distribution, the 2-D maps reveal several nonsymmetric features, particularly in NGC 3227 and NGC 4151.  This should not come as a surprise given that \htwos\ is emitted from warm molecular gas (typically 1000-2000 K), and therefore the local environment will greatly influence the observed luminosity distribution.  For this reason it should be kept in mind that the flux distribution is not necessarily representative of the actual \htwo\ mass distribution.  In fact, particularly in the cases of NGC 3227 and NGC 4151, in addition to the brighter patches of emission, \htwo\ is also detected at a significant level throughout the measured FOV.  The \htwo\ flux distribution in the individual galaxies observed is discussed in more detail in the Appendix.

Despite the complexity of the observed flux distributions, the azimuthal average of the \htwo\ emission is consistent with a half-width at half-maximum (HWHM) less than 35 pc in all of the AGN, with a mean HWHM for the sample of $\sim$30 pc (Fig. \ref{fig:dist} and Table \ref{tab:fit}).  Of note is that D07 report a nuclear stellar component with similar size scales in the SINFONI subsample based on the excess stellar light beyond that expected from an extrapolation of either a smooth {\em r}$^{1/4}$ or an exponential function.  Fitting a generalized exponential (or S\'ersic function; \citealt{sersic68}) to the azimuthally averaged light profiles of the gas indicates disklike distributions, where the threshold of a disk profile is taken to be \n\ = 2 (\citealt{cresci05}) and \n\ = 1 is an exponential and \n\ = 4 is a de Vaucouleurs profile.  Table \ref{tab:fit} lists the best-fit \n\ values for each of the galaxies observed, which on average give \n\ = 1.6$\pm$0.4.

\input{tab2.tex}


\input{tab3.tex}

\input{tab4.tex}

\input{tab5.tex}


\subsection{Kinematics and Mass}
\label{sec:kin_mass}
In general, the nuclear \htwo\ kinematics in the sample galaxies exhibits ordered rotation with kinemetry fits that are consistent with co-planar circular rotation (i.e. higher orders of the harmonic expansion are within the velocity and \sig\ measurement errors).  Furthermore, kinemetry indicates that both the PA and inclination angle remain constant throughout the nuclear region (to within the measurement error of $\sim$10\deg), and residuals of fits with the PA and inclination angle held constant are consistent with the estimated kinematic errors ($\le$ 15 \kms).  This suggests that, in the galaxies observed, the nuclear gas has no significant radial motions (e.g. in/out flows) and that there is no detectable warp down to the smallest scales measured, typically $\sim$20 pc.  Although deviations of up to $\sim$30 \kms\ from co-planar disk rotation are seen in NGC 1097, NGC 3227, and NGC 4151 (see the Appendix for details), the kinematics in each case is still dominated by disk rotation.  We therefore conclude, based on the kinematics and spatial distribution of \htwo, that the properties of the nuclear molecular gas in this AGN sample are consistent with a disklike distribution with a radial scale on the order of a few tens of parsecs.   

The best-fit PA and inclination angles given by the kinemetry fits to the \htwo\ gas kinematics are listed Table \ref{tab:fit} (see the Appendix for a discussion of the individual galaxies).  In all galaxies where a comparison is possible, the best-fit parameters to the gas kinematics are consistent with those determined from the stellar kinematics measured using the CO bandheads detected in the same \kb\ data (except in NGC 1068, as discussed).  Values based on the stellar kinematics for the SINFONI subsample are those reported in D07, and those for the OSIRIS subsample are also determined via kinemetry and are reported for the first time here (the nuclear stellar properties of this subsample will be discussed in more detail in a future publication).  As stated, all azimuthally averaged properties are computed with the best-fit PA and inclination angle determined from the stellar kinematics, when possible.  However, there is no significant difference in the averages computed using the best-fit values determined from the \htwo\ kinematics.  On larger scales, others have also found that the orientation of the gas disk agrees with that of the stars.  For example, \citet{dumas07} find agreement between the gas (in this case [O III] and H$\beta$) and stellar disk orientation in the central kpc of a sample of local active and inactive galaxies.  Moreover, the similarity of the \htwo\ kinematics and distribution with that of the nuclear stellar component suggests that the gas and stars in the central tens of parsecs are spatially mixed. 

While the 2-D velocity fields of the nuclear \htwo\ gas indicate rotating disks, the velocity dispersion suggests that these disks are relatively thick.  On average, the rotational velocity to velocity dispersion ratio, \vs\ (where \vrot=$V_{obs}$/sin \inc), is 0.9$\pm$0.3 at a radius of 30 pc (Fig. \ref{fig:vs}).  This rather low value of \vs\ indicates that random motions play a significant role at these scales, despite the general rotation of the gas.  This conclusion is drawn from the fact that the kinetic energy of the random motion is of order 3\sig$^{2}$ (assuming the dispersion arises from macroscopic motions and that we are observing one dimension of an isotropic distribution), which is significant in comparison to the energy on the order of V$_{rot}^{2}$ from rotation.  Consequently, when determining a dynamical mass, the significant velocity dispersion of the gas must be considered.  To achieve this we estimate the dynamical mass by \mdyn\ = (V$_{rot}^{2}$ + 3\sig$^{2}$)$r$/$G$, where $r$ is the radius from the nucleus and $G$ is the gravitational constant.  The measured kinematics and estimated dynamical masses within the radii of 30pc and the \htwo\ light distribution HWHM are given in Table \ref{tab:kin} for each galaxy, and the mean \sig\ values within these radii are given in Table \ref{tab:meandisp}.  The significance and origin of the relatively high nuclear velocity dispersion observed in these galaxies will be discussed in $\S$ \ref{sec:disp}.

Comparing the 20\as\ FOV ISAAC single-slit spectroscopy with the higher spatial resolution kinematics measured with SINFONI indicates that the velocities measured in the central 0\as.5 are consistent with the general galactic rotation (Fig. \ref{fig:isaac}).  This supports, in at least those galaxies observed with ISAAC, the suggestion that the nuclear gas is distributed in a rotating disk.  In addition, the consistency of the nuclear kinematics with the large-scale disk orientation reported in the literature for the sample galaxies indicates that these nuclear disks are well aligned with the host galaxy.  In each of the three galaxies measured with ISAAC, the velocity dispersion increases within the central few arcseconds, with the outer disk \sig($r$ $>$ 5\as) = 43-48 \kms\ and the inner disk \sig($r$ $<$ 0\as.5) = 63-96 \kms.  These elevated \sig\ values are more than can be accounted for by beam smearing of the central velocity gradient, and therefore represent a real increase in \sig\ in comparison to the normal star-forming disk measured at radii outside of 5\as.  For comparison, Table \ref{tab:meandisp} gives the mean \sig\ measured for each of these galaxies within the outer disk ($r$ $>$ 5\as) and inner $r$ $<$ 0\as.5 from the seeing-limited single-slit ISAAC data, as well as that from the inner $r$ $<$ 0\as.5 of the integral field SINFONI data.
  
For those galaxies in which the cold molecular gas has been measured through millimeter interferometric CO observations on scales similar to those measured here, the kinematics of the relatively hot \htwo\ and cold gas are found to be qualitatively similar.  This indicates that nongravitational forces do not significantly influence the hot gas traced by \htwo\ and that it does indeed follow the mass distribution in the nuclear region.  The best-fit PA and inclination angles of the \htwo\ gas disks agree with those of the CO gas disks reported in the literature (see Table \ref{tab:co} for references), with differences of $\lesssim$10 \kms\ and thus within the measurement errors.  In addition, similar to the hot molecular gas, the velocity dispersion of the cold gas is significant with respect to its rotational velocity (Table \ref{tab:co}).  For example, in NGC 7469 the velocity dispersion measured (after taking into account beam smearing of the velocity gradient) is as high as 60 \kms\ compared to a rotational velocity of $\sim$100 \kms.  Although the velocity dispersion of the cold molecular gas is generally lower than that of the hotter \htwo\ gas, its relatively high value indicates that the cold gas, similar to the hot gas, has a significant component of random motion.

\subsection{Vertical Disk Height}
\label{sec:height}

With a velocity dispersion that is greater than the inclination-corrected rotational velocity out to radii of tens of parsecs (Fig. \ref{fig:vs}), a simple thin disk of \htwo\ gas can be ruled out.  As discussed in $\S$\ref{sec:kin_mass}, this is because, in this case, the kinetic energy of the random motion is greater than that from the ordered rotation.  In addition, the kinematics of the sample galaxies is consistent with co-planar circular rotation, suggesting that disk warps are not the cause of the high velocity dispersion.  Therefore, although the molecular gas exhibits bulk rotation, on these scales it must be in a more spherical-like distribution (e.g. a geometrically thick disk).  

As discussed in $\S$ \ref{sec:kin_mass}, the \sig\ measured in the extended stellar disks (r $>$ 5\as) of the galaxies measured with ISAAC indicates that processes associated with typical disk star formation are capable of producing an \htwo\ velocity dispersion of \sig\ $\sim$ 45 \kms\ (comparable to the \sig\ of the cold molecular gas; Table \ref{tab:co}).  An additional velocity dispersion of $\sim$ 30 \kms\ is then needed to produce the higher \htwo\ nuclear velocity dispersions observed in the AGN at a radius of 30 pc (this is found by subtracting, in quadrature, the \sig\ observed in the outer galactic disk).  A similar conclusion is reached when considering the mean \sig\ within a radius of 30 pc (Table \ref{tab:meandisp}).  The excess \htwo\ velocity dispersion is greater than 25 \kms\ in five of the eight galaxies, with values ranging from 26-87 \kms\ at a radius of 30 pc (9pc in Circinus).  In the remaining three galaxies (NGC 3783, NGC 4051, and NGC 6814) the measured \sig\ is comparable to the value seen in the outer star forming disks.  

Models have shown that strong \htwos\ emission can be produced via shocks of speeds 20-40 \kms\ (e.g. \citealt{hollenbach89}, \citealt{burton92}).  However, when the shock exceeds 40 \kms\ the molecular hydrogen is dissociated, and thus emission is no longer produced.  Even if the excitation of \htwo\ is dominated by shocks at the maximum speed of 40 \kms, and these shocks are randomly oriented, it is only possible to produce \sig\ $\sim$ 40 \kms\ (including instrumental dispersion).  It is, however, possible to produce a higher \htwo\ velocity dispersion with bow shocks, where the angle between the shock front and the propagation direction of the bow is highly oblique and the resulting shock does not dissociate the \htwo\ (examples include the Orion bullets and Herbig-Haro object HH99B; \citealt{tedds99} and \citealt{giannini08}, respectively).  In this scenario, the \htwos\ velocity dispersion measured when integrating over the molecular cloud corresponds to the cloud's bulk motion.  Therefore, the high gas velocity dispersion measured is likely to be the result of bulk motion of the molecular clouds.

Although it is not straightforward to derive the disk scale height from the nuclear velocity dispersions measured, we attempt to get a rough estimate via two methods.  The first is to assume that the gas and stars are self gravitating, in which case the scale height of the gas disk can be estimated with the relationship \z\ = \sig$^{2}$/2$\pi$$G$$\Sigma$, where $G$ is the gravitational constant and $\Sigma$ is the surface density of the disk assuming a gas mass determined from the \mdyn\ estimates derived in $\S$ \ref{sec:kin_mass} and the gas mass fraction of 10\% estimated in the following section, i.e. $\S$ \ref{sec:cd}.  Strictly speaking this relationship should only be applied to a thin isothermal disk and consequently the relationship breaks down at the smallest radii where \vs\ is significantly less than unity (r $\lesssim$ 20 pc).  Taking the measured \sig\ at a radius of 30 pc, where \vs\ $\sim$ 1, the implied scale height is on average \z\ = 40$\pm$6 pc for the galaxies observed (excluding Circinus, which was measured out to a radius of only 9 pc where the disk height is 13 pc).  At the HWHM radius of the light distribution, which is on average 23$\pm$11 pc, the average height is \z\ = 33$\pm$16 pc.  Estimates of the disk heights for the individual galaxies are given in Table \ref{tab:kin}.  

The second method used to determine the height of the gas disk assumes vertical hydrostatic equilibrium, which implies \vs\ $\sim$ $r$/\z.  For each galaxy, the estimated \z\ at a radius of 30 pc and at the HWHM radii are consistent, to within 10 pc, with that derived above under the assumption of a thin isothermal disk.  The average disk height given by this method is \z\ = 38$\pm$13 pc at a radius of 30 pc, again consistent with that derived above.  We therefore estimate that the heights of the molecular gas disks in the observed AGNs are on the order of tens of parsecs at the radii considered, with typical values of \z\ $\sim$ 30-40 pc at a radius of 30 pc.

\bigskip

\subsection{Column Density}
\label{sec:cd_both}

\input{tab6.tex}


\subsubsection{Column Density Derived from Dynamical Mass}
\label{sec:cd}

An order of magnitude estimate of the gas mass, and thus the column density, within the nuclear regions of the galaxies can be derived by assuming that the nuclear gas masses are a fixed fraction of the dynamical masses.  In addition, we make the simplifying assumption that the \htwo\ gas mass at a given radius is distributed uniformly.  The derived column densities are thus the average values within a given a radius, and provide a measure of the maximum extinction.  As will be shown in the following section, a significant fraction of the gas within the thick disks observed is in a clumpy distribution and the actual column densities are consequently lower than derived here.

\begin{figure}[!ht]	
\epsscale{1.15}
\plotone{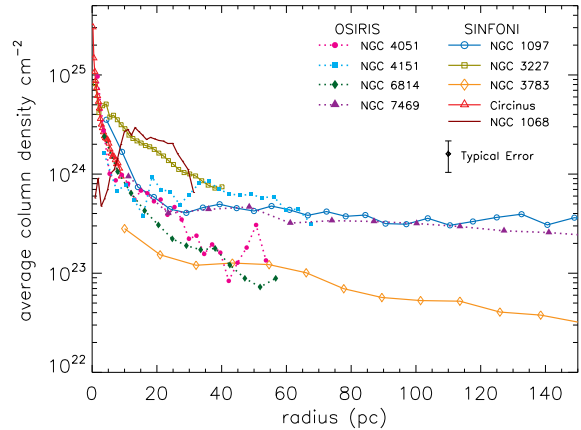}
\caption[]{Azimuthal average of the average gas column density for each galaxy as indicated by the legend.  The average column density is calculated assuming that the gas is uniformly distributed throughout the region considered and that the gas fraction is 10\% (see $\S$ \ref{sec:cd} for details).  As discussed in $\S$ \ref{sec:cd}, the molecular gas is actually in a clumpy distribution, and the densities plotted here represent only the averaged column densities within the given radius.  The error bar shows the typical 35\% error of the gas column density estimates.  \label{fig:cd}}
\end{figure}

This method of estimating the column densities requires that the gas masses are distributed as implied by \htwos\ and are not in a denser component such as the \h2o\ maser disks seen on parsec scales (e.g. \citealt{greenhill03}).  We argue that this is indeed the case based on evidence that the nuclear stellar component is spatially mixed with the thick gas disk observed ($\S$ \ref{sec:kin_mass}).  The thick stellar disk indicates that the cold component of gas from which the stars formed must also be in a similarly thick disk, which is supported by the high velocity dispersion measured in CO observations (e.g. \citealt{schinnerer00a}, \citealt{davies04b}; Table \ref{tab:co}).  Therefore, while evidence exists of dense thin disks on scales smaller than observed here, the distribution and kinematics of the stars and molecular gas (both hot and cold) suggests that, on scales of tens of parsecs, the mass is distributed in a thick disk consistent with the disk traced by \htwos.  

Although the gas mass fraction (\fg) is not accurately known in the nuclear regions of AGN, we attempt to constrain this value via several methods discussed below.  
\noindent 1.{\em Typical Galaxies \textemdash} Normal spiral galaxies, similar to the galaxies hosting the Seyfert AGN observed, have typical gas mass fractions of $\sim$10\%, with values ranging from 4\% for Sab to 25\% for Scd galaxies (e.g. \citealt{young91}).  Given the evidence reported by D07 for recent nuclear star formation in the SINFONI galaxy subsample, the gas mass fraction measured in starburst and ULIRG galaxies may also be applicable to the nuclear regions of these AGNs.  Local starburst and ULIRG galaxies are typically found to have \fg\ = 10-20\% (e.g. \citealt{young84}, \citealt{downes98}, \citealt{genzel01}, \citealt{greve05}, \citealt{tacconi06}).  Therefore, a gas mass fraction similar to that in spiral, starburst, or ULIRG galaxies suggests \fg\ $\sim$ 10\% with a range from 4\% up to 20-25\%. 

\noindent 2.{\em Cold Molecular Gas \textemdash } Measurements of the cold molecular gas within the nuclear region exist in the literature for four of the galaxies in this sample: NGC 1097, NGC 3227, NGC 1068, and NGC 7469.  The standard galactic CO-to-\htwo\ conversion factor is not assumed since evidence indicates that this conversion factor is lower in intense star forming environments such as the nuclei observed here (e.g. \citealt{downes98}, \citealt{papadopoulos00}).  We therefore use the conversion factor of (1.38$\pm0.46$)$\times$10$^{20}$ cm$^{-2}$ (K \kms)$^{-1}$ found by \citet{davies04b} for the central 800 pc in NGC 7469, which is 0.4-0.8 times the Galactic conversion measured by \citet{strong88}.  The total gas mass is then determined from the \htwo\ gas mass implied by the CO luminosity by assuming a mass correction for helium of 40\% and that the masses of H I and H II are negligible at these radii.  As can be seen in Table \ref{tab:co}, the gas fractions derived from the CO 2-1 integrated intensity vary significantly from source to source, with values of \fg\ = 1-29\%.  To date, there are no higher spatial resolution CO 2-1 observations of NGC 7469 than those of \citet{davies04b}, but over this relatively large region of 1.6 kpc, the independently determined \htwo\ mass (used to estimate the CO-to-\htwo\ conversion factor above) and the CO dynamical mass imply \fg\ = 58\%.

\noindent 3.{\em Sample Wide Average for Active Galaxies \textemdash } There now exist in the literature several studies of the cold molecular gas in the central hundreds of parsecs of galaxies of various activity levels, including Seyfert galaxies.  Taking a sample of five active galaxies that have been observed in CO 2-1 as part of the NUGA Survey (NGC 4321, NGC 4579, NGC 4826, NGC 6574, and NGC 6951) we use the reported gas and dynamical masses to derive a gas mass fraction (\citealt{lindtkrieg08}, \citealt{gb05}, \citealt{gb03}, \citealt{krips07}).  In several cases no dynamical mass is given in the literature at the small radii considered here, in which case it is estimated from the reported CO 2-1 kinematics.  Based on this sample we find a typical gas mass fraction of $\sim$10\%, with \fg\ = 1-15\% within the central $\sim$200 pc of four galaxies and \fg\ = 10\% out to a radius of 300 pc in the fifth galaxy.

\noindent 4.{\em \htwos\ Luminosity \textemdash } The gas mass can also be determined (to within a 1\sig\ uncertainty of a factor of about 2) from the luminosity of \htwos\ using the relationship reported by \citet{mueller06}, which has a conversion factor of 4000 \Msun/\Lsun\ (also see \citealt{dale05}).  A lower limit on the gas mass is estimated from the most extreme case in their sample, NGC 6240, which is known to be over luminous in \htwos~ (e.g. \citealt{sugai97}, \citealt{tecza00}), and has an order of magnitude lower conversion factor of 430 \Msun/\Lsun.  Based on this extreme conversion factor, the median gas mass fraction for the sample is \fg\ $\sim$ 13\%, with values ranging from 5-53\% (individual gas mass estimates are given in Table \ref{tab:mgas}).  In each of the individual galaxies, the gas mass fraction derived from the \htwo\ luminosity is consistent with that determined from the CO observations discussed above.  The typical conversion factor gives significantly higher gas masses and the lowest value in the sample is \fg\ = 47\%.  In some cases the typical conversion factor implies a gas mass fraction greater than 100\%, suggesting that these galaxies may also be over luminous in \htwo\ compared to the sample of galaxies studied by \citet{mueller06}, and that a conversion similar to that found for NGC 6240 is more appropriate.  We therefore deduce from this method that \fg\ is greater than a few percent, with no constraint on the upper limit. 

To summarize, typical galaxies have \fg\ $\sim$ 10\% (ranging from 4 to 25\%), CO measurements of active galaxies also suggest \fg\ $\sim$ 10\% (with a range from 1 to 15\%), CO measurements of three Seyferts in this sample on scales comparable to those measured here imply \fg\ = 1-29\%, and a conversion of the nuclear \htwo\ luminosity observed in the sample galaxies gives a lower limit on \fg\ of a few percent.  We therefore conclude that for the nuclear regions of the Seyfert galaxies observed that typically \fg\ $\sim$ 10\% with a range of \fg\ $\sim$ 1-20\%.  If the gas is uniformly distributed, then \fg\ $\sim$ 10\% gives an average column density within a radius of 30 pc of \nh\ $\sim$ (4.9$\pm$3.3) \tanh\ for the galaxy sample, with values ranging from 1.3 to 10.0 \tanh\ (Fig. \ref{fig:cd}).  The column densities within a radius of 30pc and the HWHM light radius are listed for each galaxy in Table \ref{tab:kin}.  Taking \fg\ $\sim$ 1\% still results in column density estimates greater than 10$^{22}$ cm$^{-2}$ in all galaxies, which is sufficient to obscure an AGN, and \fg\ $\sim$ 20\% gives an average \nh\ $\sim$ 10$^{24}$ cm$^{-2}$.  However, as will be shown in the following section, the gas is unlikely to be uniformly distributed within the nuclear region, and the actual clumpy distribution results in significantly less obscuration of the AGN (as well as the nuclear star formation).

\subsubsection{Column Density Derived from Extinction}
\label{sec:cd_ext}

\input{tab7.tex}

The column densities within the nuclear regions of the Seyfert galaxies can also be determined from the extinction of the nuclear stars, which can be constrained by fitting a reddened stellar template to the nuclear stellar continuum.  We use K-type stellar templates, but the template used does not significantly alter the derived column densities because in the \kb\ there is little difference in the intrinsic continuum slope for any reasonable stellar population.  The best-fit extinctions for the individual galaxies are given in Table \ref{tab:ext} assuming a model in which the gas and dust are mixed with the stars ($I_{obs}$=$I_{o}$(1-e$^{-\tau_{\lambda}}$)/$\tau_{\lambda}$, where $I_{o}$ and $I_{obs}$ are the intrinsic and observed luminosities, and $\tau_{\lambda}$ is the optical depth at a given wavelength).  Extinction estimates range from $A_{V}$ = 1 mag to 24 mag, with an average extinction of 12 mag for the sample.  This amount of extinction, assuming \nh = 1.9 $\times$ 10$^{21}$ cm$^{-2}$ mag$^{-1}$ $A_{V}$, implies an average column density of \nh\ = (2.4$\pm$1.6) $\times$ 10$^{22}$ cm$^{-2}$, with values ranging from 0.2 to 4.5 $\times$ 10$^{22}$ cm$^{-2}$ (see Table \ref{tab:ext}).  Also given in Table \ref{tab:ext} is the best-fit extinctions and implied column densities assuming screen extinction ($I_{obs}$=$I_{o}$e$^{-\tau_{\lambda}}$).  This model provides a lower limit since, as discussed in $\S$ \ref{sec:kin_mass}, the similarity of the kinematics and distribution of the stars and gas in these galaxies suggests they are mixed.  As a result a screen model is not expected to be applicable to the nuclear regions of these galaxies. 
 
The nuclear stellar continuum in the Seyfert galaxies is extracted from an annulus with inner and outer radii of 25 and 35 pc, and the region measured is therefore matched to the {\em r} $\sim$ 30 pc spatial scales of the nuclear gas disks.  Although the high spatial resolutions achieved with the AO typically gave PSFs with an HWHM of $<$ 0\as.15 (less than 16 pc at the average distance of the galaxies observed), in some galaxies the contamination within this annulus from the nonstellar AGN continuum in the PSF wings is nonnegligible.  To account for this contamination, the nuclear spectrum extracted from a region dominated by nonstellar continuum (within the FWHM of the PSF core; Table \ref{tab:obj}) is first subtracted from the annular spectrum.  The nuclear spectrum is scaled to match the annulus spectrum broad Br$\gamma$ emission, which is assumed to originate in the unresolved BLR.  The combination of the formal uncertainties given in Table \ref{tab:ext} and the uncertainty introduced by error in the correction for the AGN continuum contamination, as well as template mismatch, results in an overall accuracy of the extinction estimates of up to $\sim$50\% with typical values of $\sim$30\%.  Correction of the continuum for nonstellar contamination is highly uncertain in the two Seyfert 2 galaxies because there is no broad Br$\gamma$ emission to use for scaling of the nuclear spectrum.  A correction is attempted by scaling the nuclear spectrum by the flux level in the PSF wings at the radii measured, and these estimates are reported in Table \ref{tab:ext} along with the uncorrected estimates.  The corrected value for Circinus is used when determining the sample-wide properties, and, as discussed, NGC 1068 is excluded from the sample.  Therefore, the general conclusions of the stellar extinction in the sample are not affected by the additional uncertainty of the corrected estimates in the Seyfert 2 galaxies.  In addition, due to the greater distance of NGC 3783 and NGC 7469, an estimate of the extinction from {\em r} = 25-35 pc was not possible because the PSF HWHM already extends in both cases to $\gtrsim$20 pc resulting in significant nonstellar contamination of the continuum at {\em r} $\sim$ 30 pc.  Instead, the extinction is measured for these two galaxies in a larger annulus of {\em r} = 60-80 pc.  Therefore, for these two galaxies, the estimates are likely to represent a lower limit on the extinction at {\em r} $\sim$ 30 pc.  Also, for Circinus the extinction within {\em r} = 7-9 pc is given in Table \ref{tab:ext}, which is at the edge of the FOV. 

The relatively low extinction of the stellar continuum in these galaxies implies column densities that are over an order of magnitude lower on average than those derived within a similar spatial region in $\S$ \ref{sec:cd} under the assumption that the gas comprising a fixed fraction of the dynamical mass and is uniformly distributed (median \nh[extinction] = 0.04\nh[\htwo\ \fg=10\%]).  One way to reconcile this discrepancy is to assume a lower gas mass fraction.  However in all but one case (NGC 3783) a value of \fg\ $<$ 1\% is required to bring the column densities into agreement.  This value is lower than observed in any known Seyfert or typical spiral or starburst galaxy (see $\S$ \ref{sec:cd}).  A more likely scenario is that the gas within this nuclear region is in a clumpy distribution such that the nuclear stellar and central AGN emission is not heavily obscured despite the large gas masses present.  As will be discussed in more detail in the following section (and was mentioned in $\S$ \ref{sec:intro}), this clumpy distribution is also supported by theoretical models (e.g. \citealt{elitzur06b}), hydrodynamical simulations (e.g. \citealt{schartmann08b}), and observations of thermal dust emission (e.g. \citealt{tristram07}) and silicate absorption (e.g. \citealt{deo07}).     

\subsection{Relatively Transparent Thick Disks of Molecular Gas}
\label{sec:gas}

In summary, based on the kinematics of \htwos, the molecular gas within the central $\sim$100 pc of the Seyferts is in a rotating, thick (\z/{\em r} $\sim$ 1) disklike distribution that is spatially mixed with the nuclear stellar population.  This gas/stellar disk has spatial scales of less than 35 pc in all cases, with a typical HWHM of $\sim$30 pc.  Within the central $\sim$60 pc the dynamical mass is $\sim$10$^8$ \Msun, which suggests M$_{gas}$ $\sim$ 10$^7$ \Msun\ based on the typical gas mass fraction of 10\% (M$_{gas}$ = (0.1-2.0) $\times$ 10$^7$ \Msun\ assuming \fg\ = 1-20\%).  If a uniform gas distribution is assumed then this typical gas mass fraction implies an average column density at a radius of 30 pc of $\sim$ 5 \tanh.  However, the average column density implied by the extinction of the nuclear stellar continua is over an order of magnitude lower than this estimate (\nh\ $\sim$ 2 $\times$ 10$^{22}$ cm$^{-2}$), suggesting that the nuclear gas in these galaxies is actually in a clumpy distribution.  The disklike distribution of gas on $\sim$10 pc scales in Seyfert galaxies is therefore relatively transparent and does not heavily obscure the nuclear stellar population or the central AGN.

In addition to theoretical models and hydrodynamical simulations indicating that the gas on $\sim$10 pc scales is distributed in discrete clouds rather than evenly throughout the nuclear region (e.g. \citealt{elitzur06b}, \citealt{schartmann08b}), additional support for this conclusion comes from spectroscopic observations of 10 \mic\ silicate in Seyfert galaxies.  It is understood that significant silicate absorption requires the radiation source be deeply embedded in dust that is both optically and geometrically thick (e.g. \citealt{levenson07}).  Strong silicate absorption therefore suggests that a fairly uniform thick shell of gas and dust surrounds a compact starburst or AGN.  On the other hand a medium that consists of clouds illuminated from the outside will not produce silicate absorption since there is no strong temperature gradient across the clouds leading to a relatively constant dust temperature.  \citet{hao07} and \citet{deo07} found that only a small fraction of Seyferts have deep silicate absorption (seven out of 61 Seyfert galaxies, six of which are type 2s).  The lack of silicate absorption in most Seyfert galaxies (\citealt{shi06}, \citealt{spoon07}, \citealt{uddin07}, \citealt{thompson07}) is interpreted to mean that the gas and dust is clumpy.  Moreover, \citet{deo07} find that the few Seyfert with deep silicate absorption are either highly inclined or merging systems, suggesting that the spectral feature arises in a cool dusty region on scales of tens of parsecs or greater.

\citet{marr93} find that in our own Galaxy the molecular gas in the central tens of parsecs is confined to virial clumps with sizes of order 0.1 pc and densities of 10$^6$ cm$^{-3}$.  Similarly, \citet{christopher05} find numerous dense cores with typical diameters of 0.25 pc, densities of 3-4 $\times$ 10$^7$ cm$^{-3}$, and masses of $\sim$ 2 $\times$ 10$^4$ \Msun.  With such high densities, these clouds are stable against tidal disruption and have long lifetimes ($\sim$10$^7$ yrs).  In addition to these virial estimates, similar cloud densities are also derived from multitransition modeling (\citealt{jackson93}) and are also found for clouds in the central 30 pc of Seyferts using analytical modeling (\citealt{vollmer08}).  If we assume that all of the molecular gas in the observed Seyfert galaxies is composed of clouds with properties similar to those seen in the Galactic Center (0.25 pc diameter, $\sim$ 2 $\times$ 10$^4$ \Msun), then there must be about 500 clouds within a radius of 30 pc for a gas mass of $\sim$ 10$^7$ \Msun.  This results in only a $\sim$1\% chance that a given line of sight will be blocked by a cloud, and considering the full range of \fg\ values of 1-20\%, the chance of obscuration by a cloud is 0.1-2\%.  These results are qualitatively in agreement with the small fraction of Seyferts that show deep silicate absorption.  However, this is an extreme case in which all of the gas is assumed to be within dense clumps.  Since the reddened stellar continuum in these galaxies implies significant extinction, it is likely that in addition to dense clouds of gas there is a nonnegligible fraction of the nuclear gas in a diffuse component.

\begin{figure}[!t]	
\epsscale{1.2}
\plotone{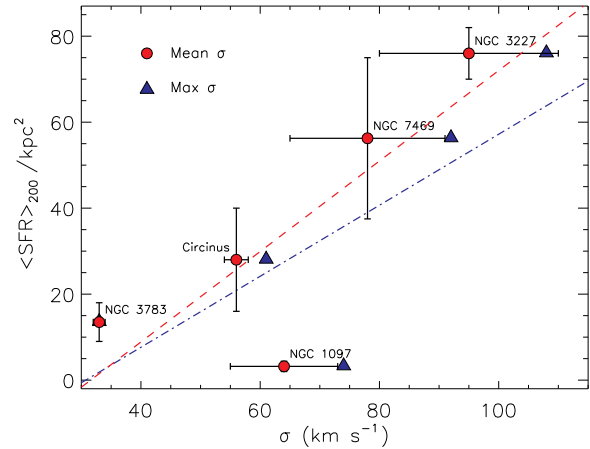}
\caption[]{\htwo\ velocity dispersion vs. the density of the time-averaged SFR (averaged over the time elapsed since the beginning of the star-forming episode), as determined by D07 and adjusted so all galaxies are averaged over 200 Myr.  For each galaxy, a circle shows the mean dispersion within the radius measured by D07 and a triangle to the right of each data point indicates the maximum dispersion within this radius.  The horizontal error bars show the standard deviation of the velocity dispersion and does not take into account the typically 5-15 \kms\ measurement errors.  Fits to the mean and max velocity dispersions with respect to the time-averaged SFRs are shown by the dashed and dash-dotted lines, respectively.  NGC 1068, which is excluded from the sample (see $\S$ \ref{sec:prop}), has a time-averaged SFR of 163 \Msun\ kpc$^{-2}$ over 200 Myr, placing it outside of the region plotted.  \label{fig:ss}}
\end{figure}

\section{Correlation of the Molecular Gas and Star Formation Properties}
\label{sec:sb-agn}

We find a correlation between the \htwos\ velocity dispersion and the SFR density in the sense that galaxies with greater star formation have higher velocity dispersion, implying a thicker gas disk.  This relationship can be seen in Fig. \ref{fig:ss}, where the time-averaged SFRs determined by D07 are all normalized to 200 Myr.  Plotted for each galaxy are both the mean and maximum \sig\ of \htwo\ within the region considered by D07 (0\as.25-0\as.4, or 8-128 pc, depending on the galaxy; see Table \ref{tab:meandisp}).  Linear fits to both of these quantities with respect to the SFR density indicate that there is a significant correlation, with linear correlation coefficients of $\sim$0.8.  The significance of this correlation and the suggested link between star formation and height of the gas disk will be investigated further in the following section, i.e. $\S$ \ref{sec:disp}.

\begin{figure}[!ht]	
\epsscale{1.2}
\plotone{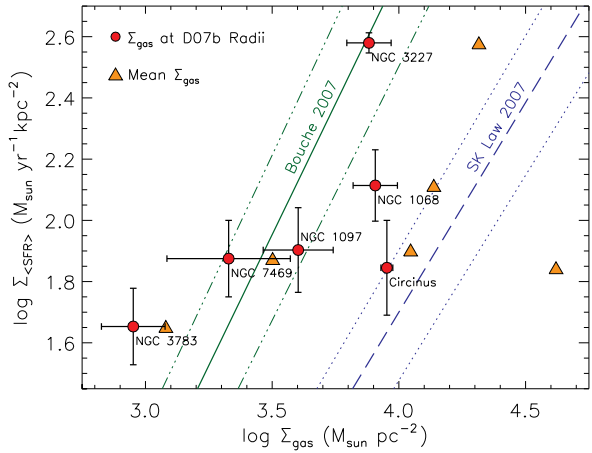}
\caption[]{Gas surface density vs. the SFR per unit area (averaged over the time since the beginning of the star-forming episode, see D07 for details) for the SINFONI subsample galaxies compared to the Schmidt-Kennicutt relation (\citealt{kennicutt07}; dashed line with the formal error of the fit indicated by dotted lines) and the adjusted relation reported by \citet{bouche07} (solid line with the fit error indicated by the dash-dotted lines).  The gas surface density at the radii measured by D07 are shown by the circles, and to the right of each data point the mean surface density within this radius is indicated.  The data shown assume \fg\ = 10\%.  If \fg\ = 1\% is assumed, then all galaxies have gas surface densities that are lower at a given SFR per unit area than either relation.  The scatter of the galaxies observed is similar to that seen in other samples (see, for example, Fig. 3 of \citealt{bouche07}).  \label{fig:ks}}
\end{figure}

A correlation is also found between the surface density of the molecular gas and the SFR per unit area that is consistent with known scaling relations.  As shown in Fig. \ref{fig:ks}, the observed AGN for which SFRs have been determined generally agree with both the Schmidt-Kennicutt relation, which assumes a constant CO-to-\htwo\ conversion typical of normal galaxies (\citealt{kennicutt98}, \citealt{kennicutt07}), and with the modified relation of \citealt{bouche07}, which accounts for the decreased CO-to-\htwo\ conversion found in denser and more active systems.  Based on the time-averaged SFRs (D07) shown in Fig. \ref{fig:ks}, the AGN are in better agreement with the modified relation.  Although the current SFRs, which are typically considered in this relation, are not well constrained in the observed AGNs, they are likely to be at least an order of magnitude lower than the time-averaged SFRs (D07).  Applying this correction, the AGNs are then found to lie closest to the original Schmidt-Kennicutt relation of \citealt{kennicutt07}.  Furthermore, a gas mass fraction of 10\% is assumed to derive the gas surface densities shown in Fig. \ref{fig:ks}.  If a gas mass fraction of only 1\% is assumed, which, as discussed in $\S$ \ref{sec:cd}, is also a plausible value for the nuclear regions of these galaxies, then all galaxies have a lower gas surface density for a given SFR per unit area than either relation.  The correlation of the gas surface density with the SFR density, and general agreement with the relations observed in other galaxies, suggests that the high SFRs observed in these AGNs should not come as a surprise given the high gas surface densities in the nuclear regions.  

\section{Achieving the Implied Disk Height}
\label{sec:disp}

As discussed in $\S$ \ref{sec:height}, the high velocity dispersion with respect to the rotational velocity implies a geometrically thick gas disk.  However, the individual molecular clouds of the thick disk would cool via supersonic shocks from collisions on a relatively short timescale (comparable to the orbital period of a few million years), thus energy must continually be injected into the region in order to maintain this vertical structure.  Many theories have been proposed in the literature to address how the vertical structure of the obscuring medium, or ``torus", can be maintained.  The feasibility of these mechanisms in maintaining the vertical structure on scales of tens of parsecs in the observed Seyfert galaxies is now considered.

As discussed in $\S$ \ref{sec:kin_mass}, kinematically the molecular hydrogen in the galaxies observed is consistent with a rotating disk and there is no evidence of a significant warp (e.g. twisting of the kinematics major axis PA and/or inclination angle) in the galaxies down to scales of $\sim$20 pc, nor of significant radial motion such as an outflow.  Therefore, for the galaxies in this sample, models of disk winds (e.g. \citealt{elitzur06}) and warped disks (\citealt{nay05}, \citealt{caproni06}) are ruled out as the dominant source of the high velocity dispersion at the radii considered here, and are thus not viable mechanisms to explain the implied vertical height of the gas disk.  It has, however, been suggested that there is evidence of warped gas disks on these scales in other AGN (e.g. \citealt{krajnovic07}, \citealt{neumayer07}), and, based on \h2o\ maser observations, on smaller $\lesssim$ 1 pc scales (e.g. \citealt{herrnstein96}, \citealt{greenhill03}).  In addition, a disk wind may still be present on these scales but if the surface brightness is too low then it would be undetected.  In this case, the disk wind would not be a significant component of the \htwo\ emission detected and would thus not appreciably impact the \htwo\ distribution nor kinematics measured. 

Based on stellar continuum intensity of the galaxies also measured with ISAAC, the nuclear starburst has a significantly higher SFR in comparison to the star formation of the outer disk (r $>$ 5\as), where a velocity dispersion of only $\sim$45 \kms\ is measured (see $\S$ \ref{sec:kin_mass}).  The contribution from star formation to the velocity dispersion, and therefore to the disk height, is thus likely to be greater in the nuclear region, especially since the similarity of the distribution and kinematics of the molecular gas and stars suggest they are spatially mixed.  The nuclear star formation can contribute to the \htwo\ disk height through at least three different mechanisms: stellar radiation pressure, supernovae, and stellar winds.  

The model of Thompson et al. (2005; see also \citealt{ballantyne08}) suggests that stellar radiation pressure can result in disk heights comparable to the radial distance, provided the SFR is sufficiently high.  In their optically thin model SFR scales as SFR $\sim f_{gas}^{2} \sigma^{4}$, and in their optically thick model SFR $\sim f_{gas} \sigma^{2}$, where $f_{gas}$ is the gas fraction and \sig\ is the gas velocity dispersion.  Assuming a gas fraction of 10\% (see $\S$ \ref{sec:cd}), the time-averaged SFRs reported in D07 are high enough to achieve the disk heights implied by the \sig\ at 30 pc in both of these models.  For example, in NGC 3227 the velocity dispersion of 98 \kms\ requires a SFR of at least 0.25 \Msun\ yr$^{-1}$ to be maintained via radiation pressure from star formation (based on the optically thin model), and the time-averaged star formation is estimated to be 1.2 \Msun\ yr$^{-1}$, almost five times higher than required.  Considering that the SFRs at their peak were at least a factor of a few up to as much as an order of magnitude greater than the time-averaged SFRs (D07), it is feasible, based on these models, that the disk height is generated via stellar radiation pressure.  If lower gas mass fractions are considered then the required SFRs decrease and this mechanism becomes even more likely to play a role in maintaining the disk height.  However, although D07 find that the nuclear star formation is recent, it is no longer active, and depending on the evolution of the star formation with time (which is not known) the current star formation could be as much as an order of magnitude lower than the time-averaged SFR (D07).  Therefore, although this mechanism may have played a critical role in generating the disk height during the peak of the star formation and for some time afterward, it is unclear if the current SFRs can maintain the estimated disk heights.

Another possible way by which star formation can contribute to the height of the gas disk is with energy input from supernovae.  \citet{wada02} have shown that with turbulence due to supernovae it is possible to achieve disk heights of tens of parsecs.  However, the supernovae rate required to reach these heights is $\sim$1 yr$^{-1}$, which is at least one, and as much as four, orders of magnitude higher than the supernovae rate predicted for the nuclear starbursts in the SINFONI subsample (D07).  The last mechanism related to star formation considered for explaining the vertical structure of the gas disk is stellar winds.  \citet{nay07} model the obscuring properties of stellar winds from a starburst within an accretion disk and find that a disk height of only a few parsecs can be achieved even with a significant outflow rate of 1-100 \Msun\ yr$^{-1}$.  Although in principle both of these mechanisms, supernovae and stellar winds, are likely to play some role in maintaining the \htwo\ disk height, neither method significantly contributes to the disk height of tens of parsecs implied by the high velocity dispersion.

An alternative energy source capable of supporting a geometrically thick gas disk is radiation pressure from the AGN emission.  \citet{pier92} show qualitatively that disk heights of \z=1-10$r_{in}$ can occur when $L_{BH}$/$L_{Edd}$=0.01-0.1, where $r_{in}$ is the inner radius of the torus as determined by the dust sublimation temperature.  In the galaxies observed $r_{in}$ $\sim$ 0.2 pc, implying that this mechanism can only achieve disk heights on the order of parsecs, with the greatest height of 4 pc achieved in NGC 7469.  The results of the more detailed study of \citet{krolik07} also suggest that for the observed galaxies the maximum disk height is on the order of parsecs, and that at a radius of $\sim$30 pc there is no significant disk height achieved via AGN radiation pressure.  Therefore, although AGN radiation pressure will inevitably play a role in the vertical structure of the disk, it is not enough on its own to explain the height of the disk observed at the radii considered.

An alternative to the models discussed above is production of the disk height through energy input from mass accretion into the nuclear region.  This mechanism has been shown by \citet{vollmer08} to be capable of producing disk heights on the order of tens of parsecs with predicted inflow rates of 0.07-4.27 \Msun\ yr$^{-1}$ for the SINFONI subsample of galaxies (their Table 3).  In this scenario an initial, short, but massive, infall of gas creates a turbulent disk in which significant star formation occurs.  After the resulting supernovae explosions clear the intercloud medium, there remains a massive, geometrically thick, collisional disk of dense gas clouds.  Based on the observed properties, the Seyferts in the SINFONI subsample are thought to be in this second phase, which has a lifetime of at least 100 Myr.  In this model, it is assumed that the physical properties of the gas on these scales are determined chiefly by the mass accretion rate from scales of $\sim$100 pc.  The disk height of the collisional disk is then determined by the cloud sizes, the transparency of the gas, and the cloud volume filling factor, each of which can be estimated from the observed \htwo\ kinematics and an assumed cloud mass of 10 \Msun\ and a sound speed of 1.5 \kms\ within the cloud.  While the mass inflow rate is significant during this phase, it is slowly decreasing, and once it drops sufficiently the collisional disk transitions to become geometrically thin.  

In conclusion, on a scale of tens of parsecs there is no evidence that outflows or warps contribute significantly to the \htwos\ kinematics of the sample galaxies, which would tend to exclude models invoking such mechanisms to obscure the AGN at these radii.  Although radiation pressure from AGN, stellar winds, and supernovae, are all likely to contribute to the \htwo\ disk height at some level, none are capable of producing the disk heights at the radii considered here.  At the peak of the star formation in the nuclear region, it is possible that radiation pressure from the stars can thicken the gas disk to the heights implied by the kinematics. However, without knowing the evolution of the star formation with time it is not possible to say if the current star formation is capable of maintaining the vertical structure.  Based on a model in which the gas properties on scales of tens of parsecs are dictated by the rate of inflow, the disk heights in the observed Seyferts may be maintained by energy input from the accretion of material into the nuclear region.  Only with improved constraints on the physical structure of the interstellar medium on these scales from future higher spatial resolution observations, and increased sample sizes, will it be possible to quantify the relative contributions of the various mechanisms discussed to maintain the disks heights implied in the Seyfert galaxies observed.

\section{Relevance to AGN}
\label{sec:AGN}

\subsection{Are We Seeing an Obscuring Torus?}
In half of the sample galaxies (four of the eight), the extinction of the nuclear star formation derived in $\S$ \ref{sec:cd_ext} implies gas column densities greater than the 10$^{22}$ cm$^{-2}$ required to obscure the optical and UV emission from the AGN and its associated BLR (\citealt{risaliti99}, \citealt{treister04}).  Of note is that the Seyfert 2 galaxy Circinus (and NGC 1068, which has been excluded from the analysis of \htwo; see $\S$ \ref{sec:prop}) falls within this group of obscured AGN.  In another two galaxies, the derived column densities are \nh\ $\sim$ 10$^{22}$ cm$^{-2}$, but considering the estimated errors it is uncertain if the densities are enough to obscure AGN emission.  The column densities of the remaining two galaxies are below the threshold for obscuring the AGN.  The additional clumpy component of dense gas implied by the high gas masses derived from the \htwo\ kinematics ($\S$ \ref{sec:cd_both}) is, as discussed in $\S$ \ref{sec:gas}, unlikely to significantly contribute to obscuration of the AGN since only a few percent of the lines of sight will encounter a gas clump.  We therefore conclude that in approximately half of the Seyfert galaxies observed the column density along lines of sight that pass through the molecular gas detected on scales of tens of parsecs will be great enough to obscure the central AGN.  

As discussed in $\S$ \ref{sec:height} and $\S$ \ref{sec:disp}, the high velocity dispersion and radial profile of the molecular gas on the scales observed suggests that the molecular gas disk in Seyfert galaxies is vertically thick due to bulk motion of the gas, and thus some fraction of the lines of sight will pass through this disk.  Observations of the space density of Seyfert 1s (unobscured) versus Seyfert 2s (obscured) suggest that the half-opening angle of the torus predicted by popular AGN unification schemes is 39\deg-48\deg\ (\citealt{osterbrock93}, \citealt{rush96}, \citealt{schmitt01}, \citealt{hao05}), implying that \z/$r$ $\sim$ 0.9-1.3.  The disk heights derived in $\S$\ref{sec:height} are consistent with this prediction with an estimated average \z/$r$ of 1.3 $\pm$ 0.2.  This fraction is unchanged if only those galaxies for which \nh\ $>$ 10$^{22}$ cm$^{-2}$ are considered.  Therefore, if the molecular disks observed on scales of tens of parsecs are capable of obscuring AGN emission (as half of the galaxies observed suggest), then the derived disk heights imply the fraction of Seyferts observed through the thick obscuring disk is consistent with the observed space densities of Seyfert 1s and 2s.

Since the observed properties of the nuclear molecular gas suggest a rotating, geometrically thick, clumpy, gas disk, we conclude that in Seyfert galaxies the gas on scales of tens of parsecs is likely to be associated with the smaller scale torus invoked in AGN unification models.  This clumpy disklike structure is consistent with the radii predicted by models of clumpy tori (e.g. \citealt{cameron93}, \citealt{schartmann05}, \citealt{honig06}, \citealt {fritz06}, \citealt{schartmann08a}, \citealt{schartmann08b}).  Given the estimated column densities, it is also possible that the gas on these scales plays a role in obscuring the AGN.  While there is evidence that obscuring material exists at smaller scales than the $\sim$20 pc scales probed here (e.g. \citealt{greenhill03}, \citealt{jaffe04}, \citealt{tristram07}), we would argue that the molecular gas measured reveals the larger scale extended structure of the nuclear obscuring material.  Although it is most likely that the smaller scale torus component is largely responsible for obscuring the AGN, and other functions such as collimation of ionization cones, the contribution of the extended structure to AGN obscuration can be significant.  Furthermore, clumpy torus models predict similar additional larger scale structures.  For example, \citet{schartmann08b} find that a clumpy torus consisting of a dense compact ($\sim$5 pc) component embedded in a more extended and diffuse component ($\sim$50 pc) is required to match the observed spectral energy distributions of Seyfert galaxies.  In addition, the alignment of the clumpy disklike structure with the large-scale orientation of the host galaxy disk is consistent with evidence that Seyfert galaxies are likely to fuel their AGN via secular processes rather than by the major mergers that are thought to trigger the fueling of higher luminosity quasars (e.g. \citealt{derobertis98}, \citealt{gb05}). 

\subsection{Starburst-AGN connection}
As shown by a detailed analysis of the SINFONI subsample of Seyfert galaxies in D07, as well as by other studies of mostly Seyfert 2 galaxies (e.g. \citealt{gonzalez01}, \citealt{fernandes04}, \citealt{riffel07}), recent nuclear star formation is associated with the AGN activity.  Since the star formation measured on scales similar to the \htwo\ distribution is likely to have been Eddington limited (D07), it is inevitably short-lived.  This implies that the star formation may be episodic in nature with multiple short bursts, the frequency of which would be dependent on the variability of the inflow of gas to the nuclear region (central $\sim$100 pc).  With the typical velocity, velocity dispersion, and M$_{gas}$ values observed (e.g. 75 \kms, 50 \kms, and 10$^7$ \Msun, respectively), an inflow rate of $\sim$10 \Msun\ yr$^{-1}$ is expected (\citealt{forster06}).  This is reasonable for a short time period, but is certainly not likely to be sustainable.  

The correlation of the gas velocity dispersion with star formation that is shown in Fig. \ref{fig:ss} suggests that the height of the molecular gas disk is related to the state of the nuclear star formation.  As discussed in $\S$ \ref{sec:disp}, the disk height may be maintained by stellar radiation pressure (\citealt{thompson05}).  It is also possible that the nuclear star formation and disk height are both dictated by the inflow rate into the nuclear region (\citealt{vollmer08}) and that stellar radiation pressure is a secondary effect.  This later scenario is consistent with the clumpy nature of the gas ($\S$ \ref{sec:gas}), since on these scales the gas is predicted to be in a geometrically thick, collisional disk that is dominated by compact clouds rather than a diffuse medium, and is thus transparent.  Only immediately after a short, massive inflow of material into the central $\sim$100 pc would the structure be opaque, and this phase lasts for only $\sim$10 Myr.  \citet{garciaburillo08} suggest a possible triggering mechanism for this inflow event via the dynamical decoupling of a nuclear bar with respect to an outer bar, which results in significant gas inflow.  In addition, they find in numerical simulations that gravitational torques combined with viscosity can result in recurrent episodes of inflow. 

If the disk height is related to the state of the nuclear star formation and/or the inflow rate to the central $\sim$100 pc, we may expect that the thick molecular gas disks detected in the observed Seyferts is a dynamic structure.  Each burst of gas inflow or star formation could increase the \sig\ of the gas, resulting in a thickening of the disk.  In contrast, when the inflow rate decreases and nuclear star formation is in a low intensity state, the \sig\ of the gas would quickly decrease, allowing the gas to settle into a thin disk.  In cases where the gas disk on these scales is capable of obscuring the AGN, as is the case in at least half of the observed Seyferts, a high state of inflow or star formation would result in an obscuring torus-like structure on scales of tens of parsecs and in a low state no obscuring medium on these scales would be present.  In this case, the fraction of Seyfert 1s versus 2s would depend not only on orientation but also on the nuclear inflow rate and the level of star formation.  This would imply that on average type 2 Seyferts have a higher rate of inflow into the nuclear region and/or a higher SFR per unit area out to radii of tens of parsecs.  Unfortunately, it is not possible to rigorously test either of these predictions with the data currently available.  The star formation properties on scales of tens of parsecs have been investigated in only a small number of Seyferts (e.g. D07), and the inflow rate into the nuclear region has been determined in even fewer AGNs, the majority of which are at most low luminosity Seyfert galaxies (e.g. \citealt{garciaburillo08}).  {\em HST} imaging studies have however, found that the central kpc of Seyfert 2 galaxies exhibit a more disturbed morphology (more twisting of the isophotes) than is seen in Seyfert 1 galaxies, which can be interpreted as a sign of a greater inflow rate (\citealt{hunt04}).  It is important to note that obscuration of the AGN is also likely to occur on scales smaller than those probed here.  Therefore even if a state of low inflow and/or star formation leads to a thin disk on scales of tens of parsecs, and thus less probability of obscuration of the AGN, it is still possible for Seyfert 2-like properties to be observed. 

\section{Conclusions}
\label{sec:conc}

With a mean spatial resolution of $\sim$20 pc (ranging from 4 to 36 pc) we have mapped the 2-D distribution and kinematics of \htwo\ in the central $\sim$100 pc of a sample of local AGN, most of which are Seyfert 1 galaxies (i.e. have a direct view of the nuclear region).  From this sample we conclude that, with few exceptions (notably NGC 1068), the molecular gas is in a rotating disklike distribution that is geometrically thick due to bulk random motion.  The HWHM of this disk is $\sim$30 pc and the kinematics, as well as distribution, of the molecular gas is consistent with that of the stars in the nuclear region, suggesting that the stars and gas are spatially mixed on these scales.

The disk height implied by the velocity dispersion of the gas is on the order of the disk radius (\z\ $\sim$ {\em r}).  The gas column density is determined using the typical gas mass fraction of \fg\ $\sim$ 10\% suggested by four independent estimates.  This value implies an average \nh = (4.9$\pm$3.3) \tanh\ for the sample of galaxies, with values ranging from 1.3 to 10.0 \tanh.  However, the estimated stellar extinction within a similar region implies an average gas column density of only \nh\ = (2.4$\pm$1.6) $\times$ 10$^{22}$ cm$^{-2}$.  The molecular gas on these scales is therefore in a clumpy distribution rather than uniformly dispersed.  Although in some Seyfert galaxies the column densities derived by the stellar extinction are great enough to obscure an AGN, in approximately half of the galaxies the gas on these scales is transparent despite the $\sim$10$^7$ \Msun\ of gas present within the central $\sim$60 pc.  We conclude that the molecular gas on the $\sim$20 pc scales observed in the sample Seyfert galaxies is in a generally rotating, geometrically thick clumpy disk that is associated with the smaller scale obscuring torus invoked in popular AGN unification models, but is not in all cases obscuring the AGN. 

Given the time-averaged SFRs and the estimated peak SFRs for the SINFONI subsample of galaxies, the vertical height of the gas disk can be produced through radiation pressure from the nuclear star formation occurring within the nuclear region.  However, it is uncertain if the current SFRs are able to maintain the full disk height implied by the observed velocity dispersion.  Radiation pressure from the AGN, stellar winds, and supernovae are also likely to play a role, albeit at a less significant level.  There is no evidence that outflows or disk warps are a general mechanism for obscuring the AGN in these Seyfert galaxies, as there is no indication of either in the kinematics measured down to the smallest scales.  Alternatively, the disk heights may be maintained via converted gravitational energy as a result of gas accretion into the nuclear region, as suggested by \citet{vollmer08}.  The kinematics of the nuclear gas disks within the central $\sim$100 pc are consistent with the larger scale orientations of the galaxy disks in the sample galaxies, and, for those galaxies for which single-slit ISAAC spectra were obtained, the small-scale kinematics are also consistent with the rotation of the galaxies on kpc scales.

Since the vertical structure of the gas disk may be dependent on the inflow rate and/or intensity of the nuclear star formation, we propose that the thick disks observed may be a dynamical structure.  Significant nuclear gas inflow or star formation will form a thick disk, while with low levels of inflow and/or star formation the gas will quickly cool into a thin disk.  This scenario is supported by the correlation of the time-averaged SFR per unit area with the \htwo\ velocity dispersion, and by the consistency of the nuclear star formation and gas densities with ``Schmidt-Kennicutt'' relations observed in other galaxies.  For those galaxies in which the gas on scales of tens of parsecs plays a role in obscuring the AGN (approximately half of the observed galaxies), then this scenario suggests the nuclear star formation and/or gas inflow rate will in part determine the classification of a Seyfert galaxy as type 1 or 2.


\acknowledgments
We are grateful to the staff of ESO and the W. M. Keck Observatory for their support.  The authors wish to recognize and acknowledge the very significant cultural role and reverence that the summit of Mauna Kea has always had within the indigenous Hawaiian community.  We are most fortunate to have the opportunity to conduct observations from this mountain.  E. K. S. H. wishes to acknowledge support from the American Astronomical Society's Small Research Grant Program.  The authors thank H. Engel for valuable discussions on the science presented, and we are also grateful to the anonymous referee for useful comments on the manuscript.

{\it Facilities:} 
\facility{Keck:II (OSIRIS)}, 
\facility{VLT:Yepun (SINFONI)},
\facility{VLT:Antu (ISAAC)}.

\appendix
\section{Description of Hot Molecular Gas in Individual Objects}
\label{sec:obj}

\subsection{NGC 1097}
NGC 1097 has an \htwo\ flux distribution that is centrally concentrated with an elongation in a roughly north-south direction and a peak flux coincident with both the \kb\ continuum and nonstellar emission.  The distribution has an HWHM of 28.5 pc, which is resolved based on a spatial resolution determined from the nonstellar emission of 0\as.25, or 22 pc at the distance of NGC 1097.  A S\'ersic fit to the azimuthal average of the flux distribution gives a disklike fit with \n=1.5.  The smoothness of the \htwo\ velocity field indicates there is no significant warp in the gas disk down to the smallest scales measured.  The best-fit kinematic parameters to the gas velocity field are consistent with that of the stars, with a best fit PA of -52\deg\ and an inclination of 42\deg\ given by the gas and PA of -49\deg\ and an inclination of 43\deg\ by the stars.  The kinemetry fit to the gas velocity field reveal deviations from co-planar disk rotation of the order of 30 \kms\ (greater than the $\sim$10 \kms\ error of the velocity measurements) and outline a spiral structure.  An investigation of the detailed kinematics in NGC 1097 will be presented in a future paper.

As reported by D07, the central stellar velocity dispersion decreases significantly, displaying what has become known as a ``sigma-drop".  Within the central 1\as\ the dispersion drops to 100 \kms\ compared to a relatively constant value of 150 \kms\ beyond this radius.  Based on this sigma-drop, and evidence of recent star formation, the presence of a dynamically cold nuclear disk is suggested.  The kinematics of the gas supports this in that the central \sig\ and rotational velocity change within the central 1\as\ to more closely match that measured for the stars.  The gas \sig\ increases from an average of 45 \kms\ outside of $r$ = 0\as.5 to an average of 63 \kms\ and a maximum of 74 \kms\ at the nucleus (Fig \ref{fig:isaac} and Table \ref{tab:meandisp}).  The offset of about 10 \kms\ higher \sig\ measured by ISAAC compared to that measured with SINFONI is due to blurring of the steep velocity gradient ($\sim$300 \kms\ in the central 2\as) in the lower spatial resolution seeing-limited ISAAC data.  

The dynamical mass derived from the gas kinematics is \mdyn\ = 9.5\tMsun\ at 30 pc, and is similar to the HWHM radius of 28.5 pc.  This mass is lower by a factor of 1.4 than the dynamical mass determined from the stellar kinematics (D07), but is consistent when the velocity and \sig\ errors of $\pm$10 \kms\ are considered.  The kinematics within this region is therefore dominated by the BH mass, which is estimated to be within the range of 4-12 \tMsun\ (\citealt{lewis06}).  Although the \htwo\ velocity dispersion is not as significant in comparison to the rotational velocity as is seen in other galaxies in the sample (\vs\ reaches unity at a radius of 24 pc, only slightly greater that the spatial resolution), the disk height at a radius of 30 pc is still estimated to be comparable to the radial scale.  The gas column density based on the \mdyn\ and the adopted \fg\ = 10\% is above 10$^{23}$ cm$^{-2}$ out to over 150 pc, with \nh\ = 4.2 \tanh\ at 30 pc.  Based on the cold molecular gas measured in CO (\citealt{hsieh08}), the gas mass fraction may be as low as 1.3\% (Table \ref{tab:co}), which lowers the column density to \nh\ = 5.5 \tcnh\ at 30 pc.  The column density implied by the extinction of the stellar continuum based on the mixed model is 1.7 \tcnh\, which is just above the 10$^{22}$ cm$^{-2}$ limit of obscuring the AGN.

\subsection{NGC 3227}
\label{sec:obj-3227}
\begin{figure}[!b]	
\epsscale{0.5}
\plotone{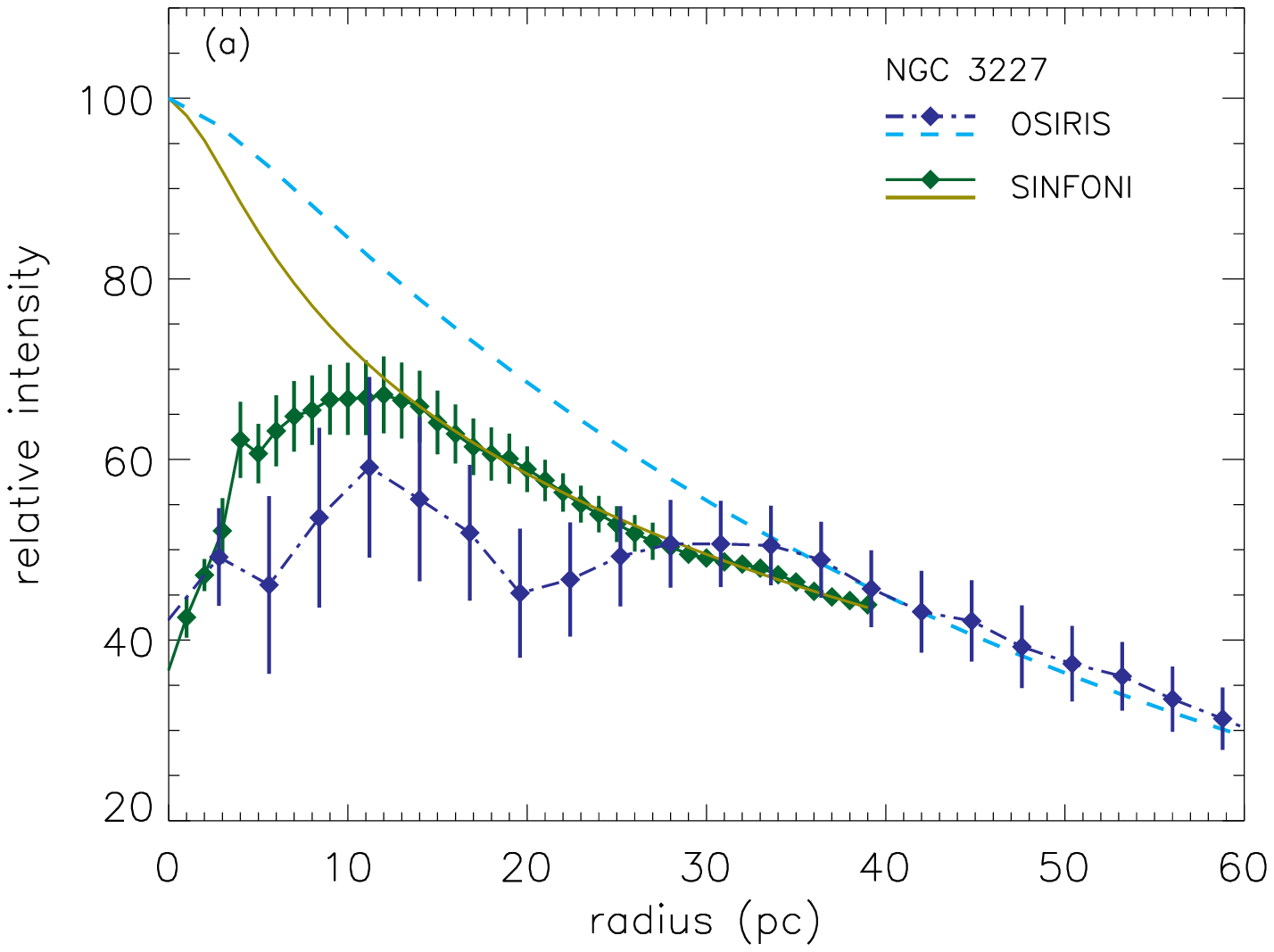}
\plotone{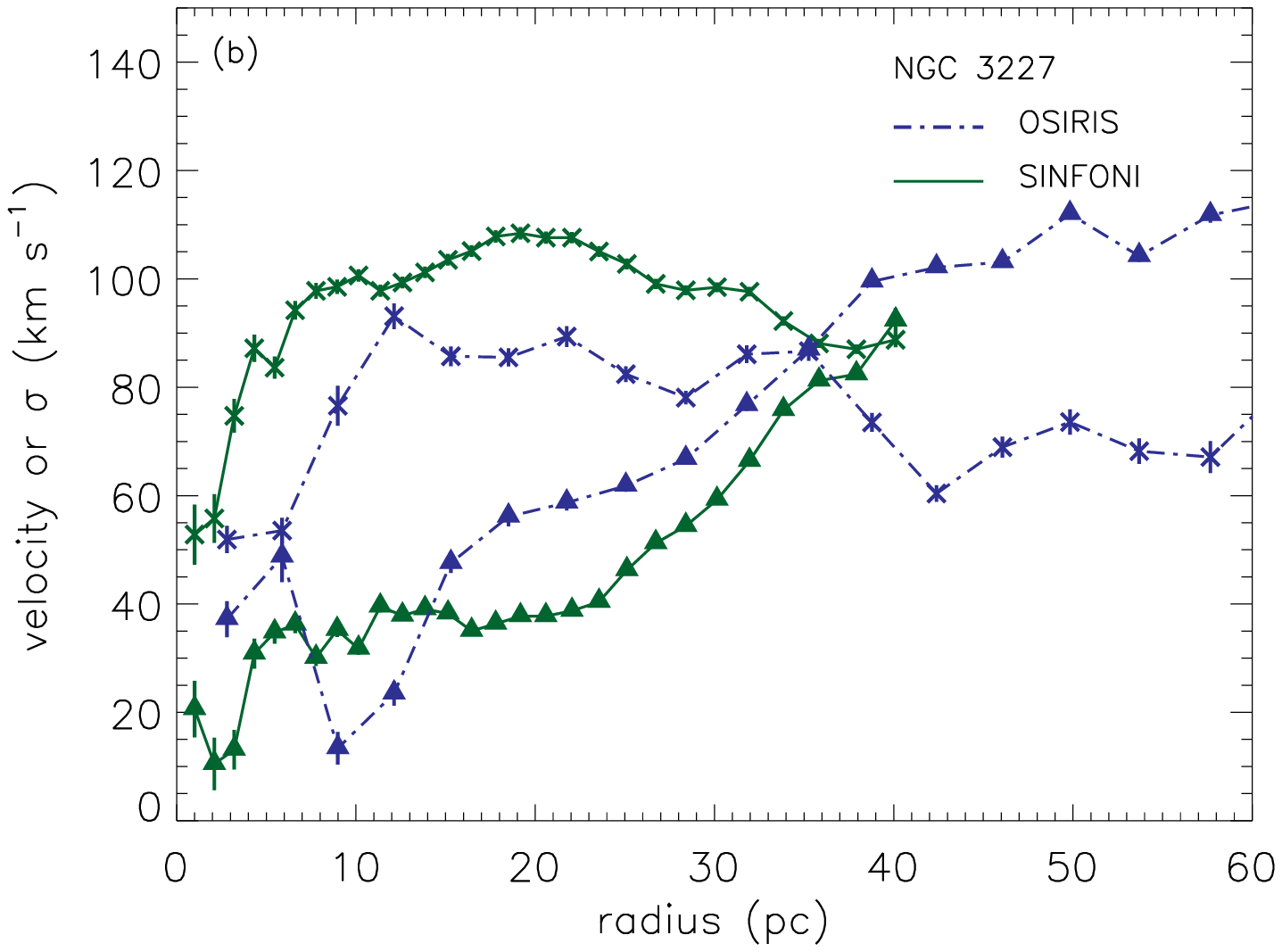}
\caption[]{Comparison of the azimuthally averaged (a) flux distribution and (b) velocity (triangles) and velocity dispersion (``x" symbols) in NGC 3227 measured with SINFONI and OSIRIS.  As indicated by the legend, the solid lines are for SINFONI measured data and dash-dotted lines are those measurements made with OSIRIS.  The flux distribution has been normalized to the peak of the extrapolated S\'ersic fits, which for the SINFONI (solid curve) and OSIRIS (dashed curve) data were fit beyond a radius of 10 pc (0\as.13) and 31 pc (0\as.39), respectively (see $\S$ \ref{sec:obj-3227} for details). \label{fig:comp3227}}
\end{figure}

The flux distribution of the hot molecular gas in NGC 3227 is relatively complex compared to the other galaxies in the sample.  The strongest \htwos\ emission is not found at the location of the nucleus, but is instead distributed along a PA of -45\deg\ with peaks of emission 0\as.24 northwest and 0\as.35 and 0\as.81 southeast of the nucleus.  \citet{schinnerer00a} also report that strongest emission from the cold molecular gas, measured by CO on a larger scale of 0\as.6, is not coincident with the nucleus.  The azimuthal average of this complex emission structure peaks at a radius of about 0\as.13.  The azimuthal average of the OSIRIS data differs from that measured with SINFONI because of the FOV coverage (Fig. \ref{fig:comp3227}).  The OSIRIS data only covers east of the nucleus, and therefore does not include the northwest emission knot, which largely is responsible for the off-center emission peak of the full FOV azimuthal average measured with SINFONI.  For this reason, the properties derived from the SINFONI data are taken to be more representative of the full nuclear region in NGC 3227.  

With an azimuthal average it is of course not possible to represent the complex emission structure seen in NGC 3227, and a S\'ersic fit to the averaged distribution at the radii of the nonsymmetric emission is thus of limited value.  We therefore fit a S\'ersic function to the azimuthally averaged flux distribution beyond the radii effected by the knots of emission, and the HWHM of the observed distribution can then be estimated from an extrapolation of this fit.  For the SINFNOI data a fit is preformed to the fall in emission beyond the peak at 0\as.13 (10 pc), and a best fit is found with $n$=1.76.  The OSIRIS data are fit outside a radius of 0\as.39 (30 pc), which is beyond the majority of the bright emission knots and most likely represents the more extended emission.  The best fit in this case is found with a $n$=1.16 S\'ersic function.  The HWHMs of the extrapolated S\'ersic fits are 29.3 pc and 34.9 pc for the SINFONI and OSIRIS data, respectively.  For comparison, the HWHM of the azimuthally averaged observed distribution measured with OSIRIS is 60.9 pc with respect to the nucleus and 28.1 pc with respect to the peak of the distribution.  These HWHM estimates, both the observed values and those determined from the S\'ersic fits, are well resolved based on the spatial resolution given by the broad Br$\gamma$ emission, which for the SINFONI data is 0\as.085 (7 pc) and for the OSIRIS data is 0\as.07 (6 pc).  

Despite the complexity of the \htwo\ flux distribution, the kinematics of the molecular gas is fairly regular.  The best fit inclination and PAs of the gas disk determined from the SINFONI and OSIRIS data are consistent, with \inc\ = 50\deg-60\deg\ and PA = -30 to -45\deg.  These disk parameters are in agreement with the best fit to the stellar kinematics of \inc\ = 55\deg-60\deg\ and PA = -45\deg, as well as with the large-scale galactic disk (\citealt{devau91}).  As discussed in \citealt{davies06}, the majority of the gas is in uniform rotation, with only the weak emission along the minor axis showing signs of nonrotational motions, possibly an outflow.  The kinemetry fit shows deviations from co-planar disk rotation along the minor axis on the order of $\sim$30 \kms\ (compared to the $\sim$10 \kms\ error of the velocity measurements).  The velocity dispersion is rather high, with an average value of $\sim$86 \kms\ within {\em r} = 0\as.5 (94 \kms\ measured with SINFONI and 77 \kms\ measured with OSIRIS).  The velocity field, as well as the velocity dispersion, is comparable to that measured for the stars in the nuclear region, with a maximum difference of 25 \kms\ along the minor axis.  A comparison of the azimuthally averaged velocity and velocity dispersion as measured with SINFONI and OSIRIS is shown in Fig. \ref{fig:comp3227}.

The gas kinematics indicate that \mdyn\ = 20 \tMsun\ at a radius of 30 pc.  Modeling of the \htwos\ kinematics suggests that NGC 3227 has a BH of 2.0$^{+1.0}_{-0.4}$ \tMsun\ (\citealt{hicks08}), and modeling of the stellar kinematics gives a mass estimate of 0.7-2.0 \tMsun\ (\citealt{davies06}), therefore the BH is not a significant fraction of the dynamical mass measured.  The \htwo\ velocity dispersion is greater than the inclination corrected rotational speed out beyond 30 pc, suggesting that the gas is in a thick disk $\sim$45 pc high at 30 pc from the nucleus.  Although measurements of the cold molecular gas detected with CO (2-1) show that \vs\ greater than one outside of 1\as, even the cold gas has a significant velocity dispersion of 60 \kms\ inside of this radius (Table \ref{tab:co}; \citealt{schinnerer00a}).  Assuming \fg\ = 10\% (which is consistent with estimates derived from CO observations; Table \ref{tab:co}), and that this gas is uniformly distributed, the average gas column density inside of 30 pc is estimated to be $\sim$ 9 \tanh.  Based on the stellar extinction a mixed model suggests that the column density is 1.9 \tcnh, which is well over an order of magnitude lower than that derived from the dynamical mass, suggesting that the majority of the nuclear gas is in a clumpy distribution. 

A more detailed analysis of the stellar and gas kinematics in NGC 3227 is presented in \citet{davies06} and \citet{hicks08}.

\subsection{NGC 3783}
The \htwos\ flux distribution in NGC 3783 is centrally concentrated and the strongest emission is consistent with the nucleus location.  The observations of NGC 3783 have a spatial resolution (based on both the nonstellar emission and the broad Br$\gamma$ emission) of 0\as.18, which is equivalent to 36 pc at the galaxy distance.  A S\'ersic fit to the azimuthally averaged light distribution gives a best fit with \n\ = 1.85 and the azimuthally averaged light profile has an observed HWHM of 20.2 pc.

The \htwo\ kinematics can reliably be measured out to only about 1\as\ (equivalent to 200 pc).  The velocity field is relatively regular, with a best-fit PA of -15\deg, and inclination of 35\deg.  Over the central 0\as.5 less than 4\% of the \kb\ continuum is stellar (D07), which prevents a reliable measurement of the stellar kinematics, and thus no comparison with the gas kinematics is possible.  A comparison to the \kb\ photometry (\citealt{mulchaey97}) indicates that the gas kinematics is consistent with the large scale isophotes, suggesting that there is no warp in the gas disk.

At a radius of 30 pc, a dynamical mass of \mdyn\ = 2.8 \tMsun\ is derived from the gas kinematics, and at the observed HWHM of 20.2 pc \mdyn\ = 1.6 \tMsun.  Based on the mass estimated from reverberation mapping (\citealt{peterson04}) the central BH has a mass comparable to the dynamical mass within a radius of 30 pc and therefore it likely dominates the kinematics.  The disk height is estimated to be similar to the radial scale, with a vertical scale of 40 pc at a 30 pc radius.  The gas column density in NGC 3783 (assuming \fg\ = 10\%) is a bit lower than estimated for the other galaxies in the sample, but is still above 10$^{23}$ cm$^{-2}$ out to a radius of $\sim$70 pc and is 1.3 \tanh\ at 30 pc.  The stellar extinction implies \nh\ = 5.3 \tcnh, which is still high enough to obscure the AGN.

\subsection{NGC 4051}
In NGC 4051, the \htwo\ flux distribution is symmetric and centrally concentrated at a location coincident with the nonstellar emission.  The observed azimuthally averaged HWHM is 7.6 pc, which is just barely resolved based on the spatial resolution determined from the nonstellar emission of 0\as.12, or 6 pc at the distance of NGC 4015.  The best fit to the azimuthally averaged flux distribution is a \n=1.8 S\'ersic function. 

The measured velocity field is consistent with uniform disk rotation without any indication of a warp down to the smallest scales measured, but the limited FOV and relatively low signal-to-noise ratio of these data do not constrain the kinematics as well as in the other galaxies.  The best fit to the \htwo\ kinematics suggests a disk with \inc\ = 50\deg\ and PA = 5\deg.  At a radius of 30 pc the \vs\ ratio is 0.83 and a disk height of \z\ = 41 pc is derived from the kinematics.  The dynamical mass within this radius is 5.0 \tMsun\, and, based on the reverberation mapped \mbh\ estimate of 0.2 \tMsun\ (\citealt{peterson04}), the \mbh\ does not dominate the kinematics on these scales.  The derived gas column density within $r$ = 30 pc (assuming \fg\ = 10\%) is then 2.2 \tanh\ and remains above 10$^{23}$ cm$^{-2}$ within the measured FOV.  Unlike all other galaxies in the sample, there is very little extinction of the nuclear stars with a mixed extinction model predicting at most $A_{V}$ = 1 mag. 

\subsection{NGC 4151}
\label{sec:obj-4151}
The \htwo\ flux distribution in the nuclear region of NGC 4151 is relatively complex due to knots of emission along a PA of roughly 90\deg.  As noted by \citet{hicks08}, several of these knots are coincident with radio emission associated with a nuclear jet (e.g. \citealt{mundell03}).  Underlying these knots of emission is a component of extended \htwo\ emission that is centrally peaked at the location of the nonstellar emission.  We do not attempt to examine the bright knots of emission further in this paper, and exclude them from the azimuthal average by only considering flux below 25\% of the peak.  Furthermore, \htwo\ is not well detected in the central 0\as.2, and this region is binned to obtain a reliable measure of the central flux and kinematics.  Unfortunately, a treatment of the bright emission knots such as that carried out for NGC 3227 is not possible because the knots extent throughout the measured FOV.  The azimuthally averaged flux distribution of the extended \htwo\ emission has an observed HWHM of 28.1 pc and is best fit with an \n\ = 1.3 S\'ersic function.  If instead all of the emission is considered in the azimuthal average then there are two additional peaks at $\sim$0\as.4 and 0\as.7 due to knots of emission on either side of the nucleus at these radii.  In Fig. \ref{fig:rad-osiris} this radial profile (smaller diamonds with dashed error bars) can be compared with that of the diffuse emission (solid diamonds and error bars).  Based on the nonstellar emission, the spatial resolution of these NGC 4151 data is 0\as.11, or 7.5 pc at the distance of NGC 4151.

The underlying \htwo\ emission has a velocity field suggestive of uniform disk rotation with a best fit of PA = 10\deg\ and \inc\ = 45\deg, which is consistent with the kinematics of the stars in the nuclear region and roughly consistent with the large scale galactic rotation (\citealt{devau91}).  Coincident with the knots of bright emission mentioned above are patches of velocity inconsistent with co-planar circular rotation, with residuals on the order of $\sim$30 \kms.  This deviation from rotation is not surprising if indeed the knots of emission are associated with the radio jet.  The velocity dispersion throughout the measured FOV is $\sim$55 \kms, and is greater than the inclination corrected rotational velocity out to a radius of about 20 pc, beyond which \vs\ rises steadily above unity.  The kinematics at this radius implies a disk height of \z\ = 33 pc and a dynamical mass of 14.0 \tMsun, suggesting that the central BH does not dominate the kinematics at this radius based on the estimated BH mass (\citealt{hicks08}, \citealt{peterson04}).  The gas column density remains above 10$^{23}$ cm$^{-2}$ out to the edge of the FOV and is estimated to be 6.2 \tanh\ at a radius of 30 pc, assuming \fg\ = 10\%.  A column density of an order of magnitude less is predicted by fitting a mixed extinction model to the stellar continuum.  The extinction suggests that \nh\ = 2.8 \tcnh, which is still high enough to obscure the AGN.

\subsection{NGC 6814}
The \htwo\ emission detected in NGC 6814 is nearly symmetric and peaks at the AGN location of the nonstellar emission.  With a spatial resolution of 0\as.17 (18 pc), the observed HWHM of the azimuthally averaged distribution is measured to be 34.8 pc.  Fitting a S\'ersic function indicates a disklike distribution with \n\ = 1.04.

Like the other galaxies in the sample, NGC 6814 exhibits regular rotation indicative of uniform disk rotation.  The best fit to the gas kinematics suggests a disk with \inc\ = -40\deg\ and PA = 55\deg, which is consistent with results from fitting the stellar kinematics.  The velocity dispersion is greater than the inclination corrected rotational velocity out to the point where a reliable fit can no longer be obtained at about 0\as.5.  At a radius of 30 pc, the disk height based on the kinematics is estimated to be comparable to the radial distance with \z\ = 46 pc.  The dynamical mass at this radius is 4.2 \tMsun, a quarter of which is estimated to be mass attributable to the central BH (\citealt{wandel02}).  The derived gas column density is greater than 10$^{23}$ cm$^{-2}$ to a radius of 40 pc, with \nh = 1.9 \tanh\ at 30 pc assuming \fg\ = 10\%.  The stellar extinction in this galaxy is relatively low, with an implied column density of only \nh\ = 5.7 $\times$ 10$^{21}$ cm$^{-2}$.  This estimate is well over an order of magnitude lower than derived from the dynamical mass, suggesting that the majority of the gas in this galaxy, on scales of tens of parsecs, is in a clumpy distribution.

\subsection{NGC 7469}
\begin{figure}[!b]	
\epsscale{0.45}
\plotone{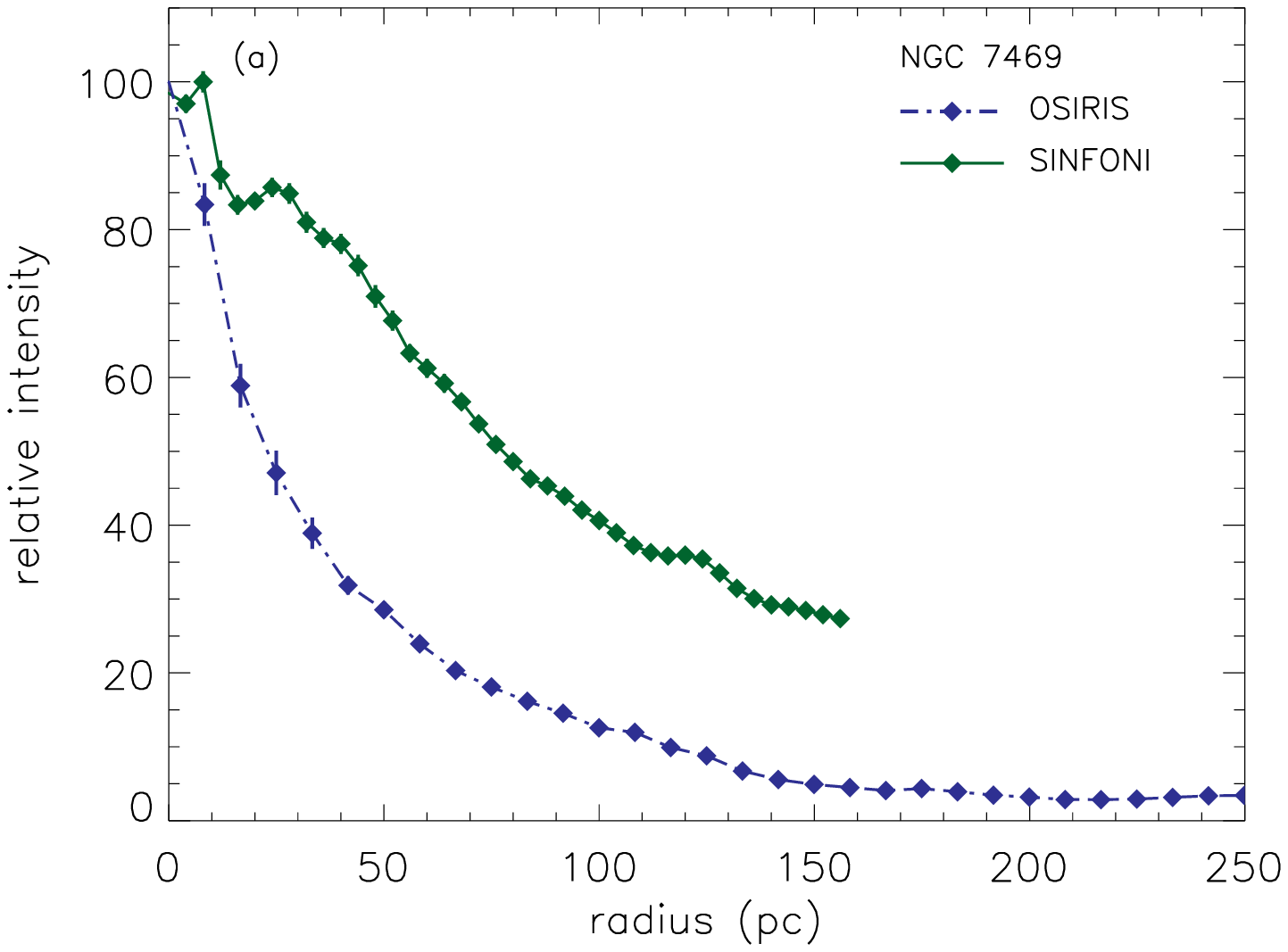}
\plotone{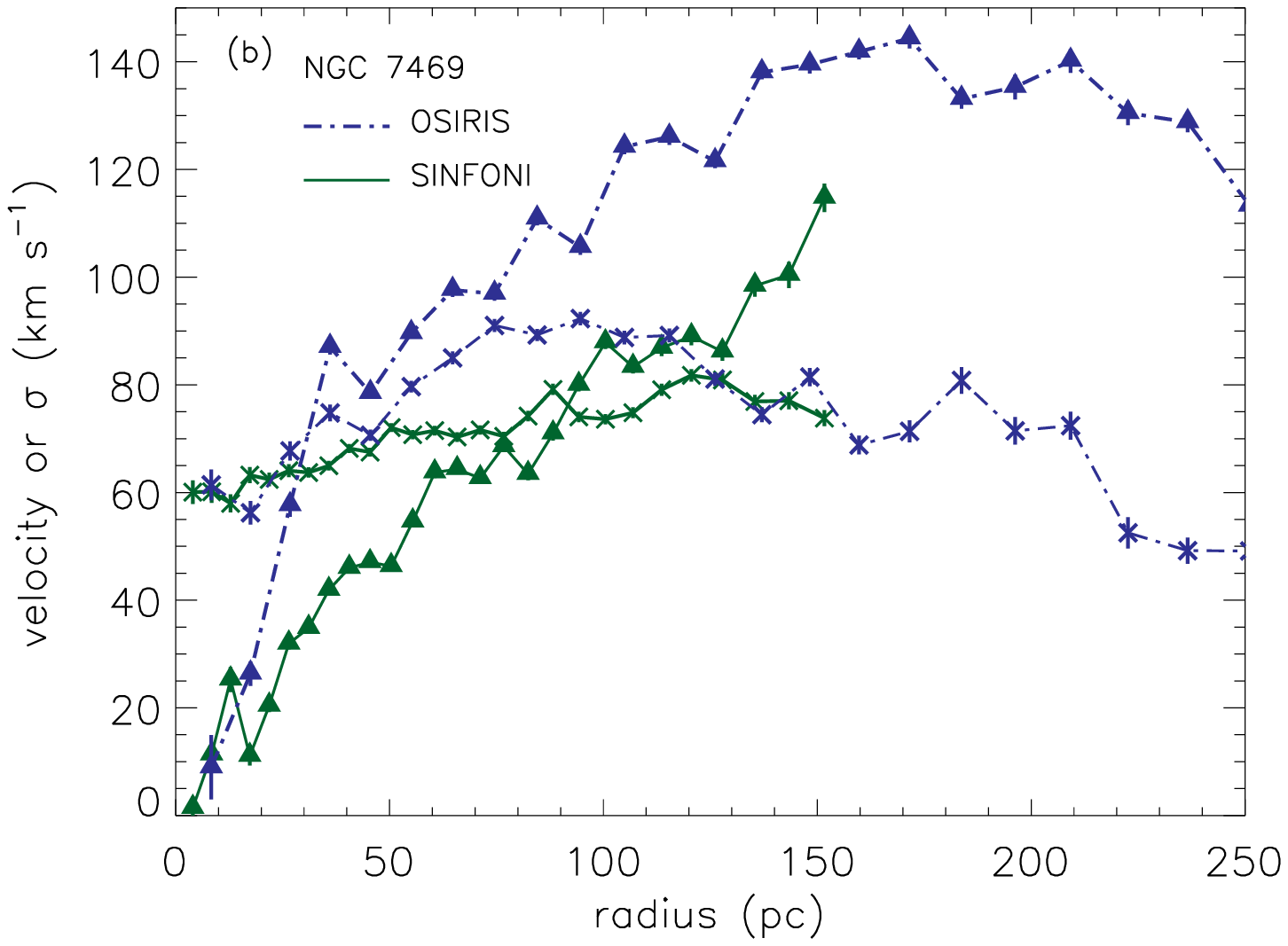}
\caption[]{Comparison of the azimuthally averaged (a) flux distribution and (b) velocity (triangles) and velocity dispersion (``x" symbols) in NGC 7469 measured with SINFONI and OSIRIS.  As indicated by the legend, the solid lines are for SINFONI measured data and dash-dotted lines are those measurements made with OSIRIS.  \label{fig:comp7469}}
\end{figure}

The \htwos\ flux distribution in NGC 7469 is very symmetric with a peak coincident with both the nonstellar and stellar emission.  The spatial resolution of the SINFONI data is estimated from the nonstellar continuum to be 0\as.15, which is 48 pc at the distance of NGC 7469.  The azimuthal average of the SINFONI observed distribution has an HWHM of 77.4 pc, and a fit to the azimuthal average finds a S\'ersic \n\ = 1.3 distribution.  The OSIRIS data has a higher spatial resolution of 0\as.11, equivalent to 35 pc, and for this reason the properties measured from the OSIRIS have data are taken to be more accurate, although, with the exception of the observed flux HWHM, the measured and derived properties from both data sets are consistent.  The observed HWHM of the azimuthally averaged flux distribution measured with OSIRIS is 30.8 pc and a best fit to the flux profile is found with a S\'ersic \n\ = 1.98 distribution.  A comparison of the azimuthally averaged flux distribution measured with OSIRIS and SINFONI, as well as that of the velocity and velocity dispersion, is shown in Fig. \ref{fig:comp7469}.

The \htwo\ velocity field is suggestive of uniform rotation, with no indication of a significant warp in the gas disk down to the smallest scales measured.  The gas kinematics is best fit with \inc\ = 50\deg\ and PA = -40\deg\ in the SINFONI data, in comparison to \inc\ = 30\deg\ and PA = -48\deg\ in the higher spatial resolution OSIRIS data.  These results just barely agree to within the expected $\pm$10\deg\ errors of the measurements, and both agree with the fit to the nuclear stellar kinematics and with the larger scale gas disk measured in CO (\citealt{davies04b}).  The velocity dispersion is about 60 \kms\ in the center and rises as high as 89 \kms\ at $r$ = 0\as.5.  This dispersion in velocity is greater than the (inclination corrected) rotational velocity in the central 0\as.29 and \vs\ remains near unity out to at least 1\as.  The kinematics implies a disk height of about 43 pc at a radius of 30 pc.  The dynamical mass derived from the gas kinematics gives \mdyn\ = 10.1 \tMsun\ within $r$ = 30 pc, and assuming \fg\ = 10\%, the derived gas column density within 30 pc is 4.2 \tanh.  Based on CO observations of the cold molecular gas (Table \ref{tab:co}; \citealt{davies04b}), the gas mass fraction is likely to have a significantly higher value of $\sim$60\%, suggesting an average gas column density of 2.4 \tbnh\ out to a radius of 30 pc.  A mixed extinction model also suggests column densities high enough to obscure the AGN, with \nh\ $\sim$ 5 \tcnh.  This column density is still an order of magnitude or more lower than implied by the dynamical mass, suggesting that a significant fraction of the nuclear gas is distributed in clumps rather than uniformly.

\subsection{Circinus}
The \htwos\ emission in Circinus is symmetric with a peak coincident with that of the \kb\ continuum.  The HWHM of the azimuthally averaged light profile is 7.4 pc, observed with a spatial resolution of 0\as.22, or 4.2 pc at the galaxy distance.  The light profile is best fit with a \n\ = 1.1 S\'ersic function.  The \htwo\ velocity field suggests uniform rotation, with the exception of a region at the western edge of the FOV at 9 pc, which is most likely associated with the ionization cone.  A gas disk is best fit to the kinematics with a PA of 30\deg\ and an inclination of 55\deg.  Within the errors, these parameters agree with the large scale optical photometry (PA = 20\deg\ and inclination of 65\deg; \citealt{freeman77}), suggesting that there is no warp down to at least a few parsecs (there is evidence of a warp on much smaller scales of 0.1-0.4 pc; \citealt{greenhill03}).

The dynamical mass is estimated to be \mdyn\ = 1.9\tMsun\ within 9 pc, and \mdyn\ = 1.8\tMsun\ within the HWHM radius of 7.4 pc.  Based on this \mdyn\ and assuming a gas fraction of 10\%, the gas column density within 9 pc is 9.4 \tanh, and increases to 1.3 \tbnh\ within the HWHM radius.  The stellar continuum is heavily obscured in this galaxy, consistent with its classification as a Seyfert 2.  A mixed extinction model implies column densities of \nh\ = 4.3 \tcnh, which is high enough to obscure the AGN.  This estimate is still over an order of magnitude lower than that derived from the dynamical mass, suggesting a clumpy medium.  The velocity dispersion is relatively constant throughout the FOV measured (\sig\ $\sim$ 56$\pm$3 \kms\ for r$<$0\as.5; Table \ref{tab:meandisp}) and the rotational velocity to \sig\ ratio is relatively low with \vs\ only reaching 0.7 at the edge of the FOV.  At this radius, the disk height is estimated to be 13 pc, and a height of 11 pc is estimated at the HWHM radius. 

A detailed analysis of the nuclear region of Circinus, including the \htwo\ properties as measured with SINFONI, is presented in \citet{mueller06}.  

\subsection{NGC 1068}
NGC 1068 has the most complex \htwos\ properties of any of the galaxies in the sample and is discussed in detail in \citet{mueller08}.  Within the central 1\as, there are two knots of \htwo\ emission, one coincident with the nucleus and another to the north, and there is no detectable \htwo\ emission in the southwest.  An azimuthal average of this distribution gives an observed HWHM of 13.2 pc and a S\'ersic fit finds \n\ = 2.09, although, like in NGC 3227, these values obviously do not reflect the two dimensional nature of the nuclear \htwo\ distribution.  The nuclear stars are heavily obscured in this galaxy, and even after correcting for contamination from the nonstellar continuum, the extinction is so great that it is beyond the saturation limit of a mixed extinction model (which is at an optical depth of a few).  The spatial resolution based on the nonstellar emission is 0\as.08, equivalent to 5 pc at the distance of NGC 1068.

Noncircular motions dominate the \htwo\ kinematics, and \citet{mueller08} argue that streaming of the gas toward the nucleus is suggested.  This should therefore be kept in mind when evaluating the azimuthally averaged rotational velocity and \sig\ shown in Fig. \ref{fig:rad-sinfoni}.  Because the \htwo\ properties in NGC 1068 so greatly differ from the disk rotation seen in the stars, unlike the consistency seen in the other sample galaxies, this galaxy is not included in the analysis of general \htwo\ properties in AGNs.  This galaxy therefore represents an exception to the general conclusions presented.

\end{document}

%% file: tab1.tex
\begin{deluxetable*}{llccllcccc}
\tabletypesize{\scriptsize}
\tablecaption{Summary of AGN data included in this study \label{tab:obj}} 
\tablewidth{0pt}
\tablehead{
\colhead{} &
\colhead{} &
\colhead{} &
\colhead{} & 
\multicolumn{4}{c}{IFU AO Observations} &
\multicolumn{2}{c}{ISAAC Observations} \\
\cline{5-8} \\
\colhead{Object} &
\colhead{Classification\tablenotemark{a}} &
\colhead{D(Mpc)\tablenotemark{a}} &
\colhead{pc/\as} &
\colhead{Band} & 
\colhead{Res.\tablenotemark{b} (\arcsec)} &
\colhead{Res.\tablenotemark{b} (pc)} &
\colhead{Date} &
\colhead{PAs (\deg)} &
\colhead{Date} \\
}

\startdata
                           &                   &       &        & {\em SINFONI} &      &      &          &        &           \\
NGC\,1097                  & LINER, Sy\,1      &    18 & \phn88 & {\em H+K } & 0.25 & 22.0 & 2005 Oct & -33,57 & 2006 Sep  \\
NGC\,3227\tablenotemark{c} & Sy\,1             &    17 & \phn80 & {\em K } & 0.09 & \phn7.2  & 2004 Dec & -30,60 & 2006 Jan \\
NGC\,3783                  & Sy\,1             &    42 &    200 & {\em K } & 0.18 & 36.0   & 2005 Mar & \nodata & \nodata \\
NGC\,7469                  & Sy\,1             &    66 &    320 & {\em K } & 0.15 & 48.0   & 2004 Jul & \nodata & \nodata \\
Circinus\tablenotemark{c}  & Sy\,2             & \phn4 & \phn20 & {\em K } & 0.22 & \phn4.4   & 2004 Jul & -65,25 & 2006 Jan,Feb \\
NGC\,1068\tablenotemark{c} & Sy\,2             &    14 & \phn70 & {\em H+K } & 0.08 & \phn5.6 & 2005 Oct, 2006 Nov & \nodata & \nodata \\
\cline{1-10}
                           &                   &       &        & {\em OSIRIS}  &      &      &          &        &           \\
NGC\,3227                  & Sy\,1             &    17 & \phn80 & {\em K } & 0.07 &\phn5.6   & 2006 Mar, 2008 Jan & \nodata & \nodata \\
NGC\,4051                  & Sy\,1             &    10 & \phn48 & {\em K } & 0.12 & \phn5.8   & 2006 Apr, 2008 Jan & \nodata & \nodata \\
NGC\,4151                  & Sy\,1             &    14 & \phn68 & {\em K } & 0.11 &\phn7.5   & 2006 Mar & \nodata & \nodata \\
NGC\,6814                  & Sy\,1             &    22 &    107 & {\em K } & 0.17 & 18.2   & 2006 Apr, Sep & \nodata & \nodata \\
NGC\,7469                  & Sy\,1             &    67 &    320 & {\em K } & 0.11 & 35.2  & 2006 Sep & \nodata & \nodata \\
\enddata

\tablenotetext{a}{Classification and distance (D) are from D07 and the NASA/IPAC Extragalactic Database.}

\tablenotetext{b}{Spatial resolution (FWHM) estimated from the data itself, using the methods described in $\S$ 2.2.}

\tablenotetext{c}{References to detailed studies of individual objects: 
Circinus \citep{mueller06},
NGC 3227 \citep{davies06},
NGC 1068 \citep{mueller08}.}

\end{deluxetable*}


%% file: tab2.tex
\begin{deluxetable}{lccccccccc}
\tabletypesize{\tiny}
\tablecaption{Best Fit Kinematic Parameter Values\label{tab:fit}} 
\tablewidth{0pt}
\tablehead{
\colhead{} &
\multicolumn{2}{c}{Stellar\tablenotemark{a}} &
\colhead{} &
\multicolumn{2}{c}{H$_{2}$ Gas} &
\colhead{} &
\colhead{} &
\colhead{} \\
\cline{2-3}
\cline{5-6} \\
\colhead{Galaxy} & 
\colhead{PA} & 
\colhead{$\it i$} &
\colhead{} &
\colhead{PA} &
\colhead{i} &
\colhead{} &
\colhead{S\'ersic} &
\colhead{HWHM\tablenotemark{c}} \\
 \colhead{} & 
\colhead{} & 
\colhead{(\deg)} &
\colhead{} &
\colhead{(\deg)} &
\colhead{(\deg)} &
\colhead{} &
\colhead{n} &
\colhead{(pc)} \\
}
\startdata

NGC 1097  &  -49  &  43  &  &  -52  &  42  & & 1.52 & 28.5 \\
NGC 3227  &  -45  &  55  &  &  -30  &  50  & & 1.76\tablenotemark{b} & 29.3 \\
          &  -45  &  60  &  &  -45  &  60  & & 1.16\tablenotemark{b} & 34.9 \\
NGC 3783  &  -18  &  27  &  &  -15  &  35  & & 1.85 & 20.2 \\
NGC 4051  &  -10  &  55  &  &   \phn\phn5  &  50  & & 1.83 & \phn7.6 \\
NGC 4151  &   \phn-5  &  50  &  &   \phn10  &  45  & & 1.25 & 28.1 \\
NGC 6814  &  -40  &  55  &  &  -40  &  55  & & 1.04 & 34.8 \\
NGC 7469  &  -52  &  45  &  &  -40  &  50  & & 1.32 & 77.4 \\ 
          &  -50  &  45  &  &  -48  &  30  & & 1.98 & 30.8\\ 
Circinus  &   \phn20  &  65  &  &   30  &  55  & & 1.13 & \phn7.4 \\  
NGC 1068  &  \phn85   &  40  &  &  -16  &  50  & & 2.09 & 13.2 \\

\enddata

\tablecomments{PA is given as degrees East of North.  For NGC 3227 and NGC 7469 the first and second lines are based on SINFONI and OSIRIS data, respectively.  As discussed in $\S$ \ref{sec:prop}, NGC 1068 is excluded from the AGN sample.}

\tablenotetext{a}{For NGC 3783 and Circinus no stellar kinematics could be measured from the \kb\ data presented and instead listed here are fits to the large scale \kb\ (\citealt{mulchaey97}) and optical (\citealt{freeman77}) photometry, respectively.  In addition, for NGC 7469 the best-fit parameters based on CO 2-1 molecular gas (assuming the gas ring at r=2\as.3 is circular; \citealt{davies04b}) are given on the first line instead of the SINFONI measured best-fit stellar parameters (which agree with those derived from the OSIRIS data shown on the second line).}

\tablenotetext{b}{See $\S$ \ref{sec:obj-3227} for details of the S\'ersic fit to NGC 3227.}

\tablenotetext{c}{Measurement of direct HWHM radius from the observed azimuthal average, except for NGC 3227, where the HWHM is that of the S\'ersic function fit as described in $\S$ \ref{sec:obj-3227}.}


\end{deluxetable} 


%% file: tab3.tex
\begin{deluxetable*}{lccccccccccccccc}
\tabletypesize{\scriptsize}
\tablecaption{Quantities Derived from the \htwos~ Kinematics \label{tab:kin}} 
\tablewidth{0pt}
\tablehead{
\colhead{} &
\multicolumn{6}{c}{At r=30pc}  &
\colhead{} &
\multicolumn{6}{c}{At HWHM Radius} \\
\cline{2-7}
\cline{9-15} \\
\colhead{Galaxy} & 
\colhead{v} & 
\colhead{\sig} &
\colhead{v/\sig} &
\colhead{\z} &
\colhead{M$_{dyn}$}  &
\colhead{n$_{H}$ (\fg=0.1)} & 
\colhead{} &
\colhead{r} &
\colhead{v} & 
\colhead{\sig} &
\colhead{v/\sig} &
\colhead{\z} &
\colhead{M$_{dyn}$}  &
\colhead{n$_{H}$ (\fg=0.1)} \\ 
\colhead{} & 
\colhead{\kms} &  
\colhead{\kms} &
\colhead{} &   
\colhead{pc} &
\colhead{10$^{7}$ \Msun} &
\colhead{10$^{23}$ cm$^{-2}$} &
\colhead{} &
\colhead{pc} & 
\colhead{\kms} &  
\colhead{\kms} &
\colhead{} &   
\colhead{pc} &
\colhead{10$^{7}$ \Msun} &
\colhead{10$^{23}$ cm$^{-2}$} \\
}
\startdata

NGC 1097 & 73 & \phn52 & 1.42 & 30 & \phn9.5 & \phn4.2 & & 28.5  & 68    & \phn52 & 1.32  & 30    & \phn8.4 & \phn4.1 \\ 
NGC 3227 & 59 & \phn98 & 0.60 & 45 & 22.6    & 10.0    & & 29.3  &  57    & \phn98 & 0.58  & 44    & 22.0    & 10.2    \\
         & 71 & \phn82 & 0.87 & 40 & 17.7    & \phn7.8 & & 34.9  &  86    & \phn87 & 0.99  & 44    & 24.3    & \phn7.9 \\
NGC 3783 & 26 & \phn33 & 0.78 & 40 & \phn2.8 & \phn1.3 & & 20.2  &  \phn6 & \phn34 & 0.18  & 33    & \phn1.6 & \phn1.6 \\
NGC 4051 & 37 & \phn44 & 0.83 & 41 & \phn5.0 & \phn2.2 & & \phn7.6  &  13    & \phn49 & 0.26  & 13    & \phn1.3 & \phn8.9 \\
NGC 4151 & 82 & \phn67 & 1.23 & 33 & 14.0    & \phn6.2 & & 28.1  &  76    & \phn59 & 1.28  & 30    & 10.8    & \phn5.4 \\
NGC 6814 & 23 & \phn43 & 0.53 & 46 & \phn4.2 & \phn1.9 & & 34.8  &  26    & \phn47 & 0.55  & 57    & \phn6.3 & \phn1.8 \\
NGC 7469 & 34 & \phn64 & 0.54 & 46 & \phn9.3 & \phn4.2 & & 77.4  &  68    & \phn71 & 0.96  & 99    & 35.4    & \phn2.4 \\
         & 43 & \phn62 & 0.69 & 43 & 10.1    & \phn4.2 & & 30.8  &  45    & \phn63 & 0.71  & 43    & 10.7    & \phn4.2 \\
Circinus\tablenotemark{a} & 36 & \phn51 & 0.70 & 13 & \phn1.9 & \phn9.4 & & \phn7.4  &  38    & \phn54 & 0.71  & 11    & \phn1.8 & 12.9    \\
NGC 1068 & 21 & 102    & 0.21 & 49 & 22.5    & 10.1    & & 13.2  &  \phn8 &    118 & 0.07  & 22    & 12.8    & 29.2    \\
\enddata

\tablecomments{The typical uncertainty on the velocity and velocity dispersion measurements is 5-15 \kms.  For NGC 3227 and NGC 7469 the first and second lines are based on SINFONI and OSIRIS data, respectively.  As discussed in $\S$ \ref{sec:prop}, NGC 1068 is excluded from the AGN sample.  The disk scale height, \z, is derived as described in $\S$ \ref{sec:height}, using the relationship \z~ = \sig$^{2}$/2$\pi$G$\Sigma$.  \mdyn~ and \nh~ are derived assuming the typical 10\% gas mass fraction - see $\S$3.2-3.4 for details and discussion.}

\tablenotetext{a}{Quantities measured at 9 pc (the edge of the measured FOV) rather than 30 pc.}

\end{deluxetable*} 


%% file: tab4.tex
\begin{deluxetable*}{lccccccccccc}
\tabletypesize{\scriptsize}
\tablecaption{Mean \htwos\ Velocity Dispersion \label{tab:meandisp}} 
\tablewidth{0pt}
\tablehead{
\colhead{} &
\multicolumn{7}{c}{SINFONI \& OSIRIS} &
\colhead{} &
\multicolumn{3}{c}{ISAAC} \\
\cline{2-8}
\cline{10-12} \\
\colhead{Galaxy} &
\colhead{$<$\sig$>$$_{30pc}$} &
\colhead{$<$\sig$>$$_{HWHM}$} &
\colhead{$<$\sig$>$$_{0\as.5}$} &
\colhead{} &
\colhead{r} &
\colhead{$<$\sig$>$$_{D07b}$} &
\colhead{\sig$_{max}$} &
\colhead{} &
\colhead{$<$\sig$>$$_{20\as}$} &
\colhead{$<$\sig$>$$_{>5\as}$} &
\colhead{$<$\sig$>$$_{0\as.5}$} \\
\colhead{} &
\colhead{\kms} &
\colhead{\kms} &
\colhead{\kms} &
\colhead{} &
\colhead{pc} &
\colhead{\kms} &
\colhead{\kms} &
\colhead{} & 
\colhead{\kms} &
\colhead{\kms} & 
\colhead{\kms} \\
}
\startdata

NGC 1097 & 61$\pm$\phn9 & 62$\pm$10 & 60$\pm$\phn9 & & \phn22    & 64$\pm$\phn9 & \phn74 & &  58$\pm$10     & 47$\pm$11   & 71$\pm$\phn1\\
NGC 3227 & 95$\pm$15 & 95$\pm$15 & 94$\pm$14 & & \phn32    & 95$\pm$15    & 108    & & 74$\pm$15     & 43$\pm$23   & 96$\pm$\phn5\\
         & 78$\pm$14 & 79$\pm$14 & 77$\pm$14 & &    & 79$\pm$14  & \phn93  & &   &   &   \\
NGC 3783 & 33$\pm$\phn1 & 33$\pm$\phn1 & 34$\pm$\phn2 & & \phn60    & 33$\pm$\phn1 & \phn33 & & \nodata & \nodata & \nodata \\
NGC 4051 & 55$\pm$\phn8 & 56$\pm$12 & 56$\pm$\phn8 & & \nodata  & \nodata & \nodata & & \nodata & \nodata &  \nodata \\
NGC 4151 & 56$\pm$10 & 54$\pm$10 & 57$\pm$11 & & \nodata  & \nodata &  \nodata & & \nodata & \nodata & \nodata \\
NGC 6814 & 48$\pm$\phn6 & 48$\pm$\phn5 & 45$\pm$\phn6 & & \nodata & \nodata & \nodata & & \nodata & \nodata &  \nodata \\
NGC 7469 & 62$\pm$\phn2 & 67$\pm$\phn5 & 71$\pm$\phn7 & & 128 & 70$\pm$\phn7 & \phn82 & & \nodata & \nodata & \nodata \\  
         & 62$\pm$\phn6 & 62$\pm$\phn6 & 79$\pm$12 & &  &   78$\pm$13     &  \phn92      & &         &         &         \\
Circinus\tablenotemark{a} & 56$\pm$\phn3 & 57$\pm$\phn2 & 56$\pm$\phn3 & & \phn\phn8 & 56$\pm$\phn2 & \phn61 & &  54$\pm$\phn9  & 48$\pm$10   & 63$\pm$\phn4 \\
NGC 1068 & 88$\pm$42 & 58$\pm$40 & 88$\pm$42 & & \phn30    & 88$\pm$42    & 134    & & \nodata & \nodata & \nodata \\

\enddata

\tablecomments{A uniformly weighted mean within a radius given by the subscript, except where it is indicated to be outside of the stated radius.  For NGC 3227 and NGC 7469 the first and second lines are based on SINFONI and OSIRIS data, respectively.  As discussed in $\S$ \ref{sec:prop}, NGC 1068 is excluded from the AGN sample.}

\tablenotetext{a}{Quantities measured at 9 pc (the edge of the measured FOV) rather than 30 pc.}

\end{deluxetable*} 


%% file: tab5.tex
\begin{deluxetable*}{lrrcccccc}
\tabletypesize{\small}
\tablecaption{Summary of CO Measurements \label{tab:co}} 
\tablewidth{0pt}
\tablehead{
\colhead{Galaxy} &
\multicolumn{2}{c}{Radius} &
\colhead{\mgas} &
\colhead{\mdyn} & 
\colhead{\fg} &
\colhead{\sig} &
\colhead{v/\sig} &
\colhead{References} \\
\colhead{} &
\colhead{\as} &
\colhead{pc} &
\colhead{10$^7$ \Msun} &
\colhead{10$^7$ \Msun} &
\colhead{\%} &
\colhead{\kms} &
\colhead{} &
\colhead{} \\
}

\startdata

NGC 1097 & 1.8 & 158 & 2.6 & 207 & 1.3$\pm$0.2 & 50 & 5  & 1 \\
NGC 3227 & 0.3 & \phn24 & 0.8$\pm$0.3 & 6.4 & 13$\pm$4 & 60 & 0.8  & 2 \\
NGC 7469 & 2.5 & 800 & 380 & 650 & 58 & 60 & 0.6 & 3 \\
NGC 1068 & 0.5 & \phn35 & 2.3-4.0 & 13.6 & 17-29 & 30  & 1.2  & 4 \\


\enddata


\tablecomments{References for CO measurements and derived \mgas\ and \mdyn\ are as follows: (1) \citealt{hsieh08}, (2) \citealt{schinnerer00a}, (3) \citealt{davies04b}, (4) \citealt{schinnerer00b}, \citealt{mueller08}, D07b.}

\end{deluxetable*}


%% file: tab6.tex
\begin{deluxetable}{lcccc}
\tabletypesize{\small}
\tablecaption{Gas Mass Estimates as Derived in \S \ref{sec:cd}\label{tab:mgas}} 
\tablewidth{0pt}
\tablehead{
\colhead{} &
\colhead{} &
\colhead{\fg=10\%} &
\multicolumn{2}{c}{L$_{H2}$ Conversion\tablenotemark{a}} \\
\cline{4-5} \\
\colhead{Galaxy} &
\colhead{M$_{dyn}$} &
\colhead{M$_{gas}$} &
\colhead{M$_{gas}$} &
\colhead{\fg} \\
\colhead{} &
\colhead{10$^7$ \Msun} &
\colhead{10$^7$ \Msun} &
\colhead{10$^7$ \Msun} &
\colhead{\%} \\
}
\startdata

NGC 1097 & \phn9.5 & 1.0 & 0.5-4.5 &  5-47 \\
NGC 3227 & 22.6    & 2.3 & 2.4-22.2 &  11-98 \\
         & 17.7    & 1.8 & 2.2-20.2 &  $\geq$12\\
NGC 3783 & \phn2.8 & 0.3 & 1.1-10.6 &  $\geq$39 \\
NGC 4051 & \phn5.0 & 0.5 & 0.7-6.6 &  $\geq$14 \\
NGC 4151 & 14.0    & 1.4 & 1.2-10.8 &  9-77 \\
NGC 6814 & \phn4.2 & 0.4 & 0.6-5.5 &  $\geq$14 \\
NGC 7469 & \phn9.3 & 0.9 & 4.9-45.5 &  $\geq$53 \\
         & 10.1    & 1.0 & 4.7-43.3 &  $\geq$47 \\
Circinus & \phn1.9 & 0.2 & 0.2-1.5 &  11-79 \\
NGC 1068 & 22.5    & 2.3 & 4.2-39.3 &  $\geq$19 \\

\enddata

\tablecomments{Dynamical masses are based on the \htwo\ kinematics as discussed in $\S$\ref{sec:kin_mass}.  All masses are derived for a radius of 30 pc, with the exception of Circinus, for which the radius is 9 pc.   For NGC 3227 and NGC 7469 the first and second lines are based on SINFONI and OSIRIS data, respectively.  As discussed in $\S$ \ref{sec:prop}, NGC 1068 is excluded from the AGN sample.}
\tablenotetext{a}{For cases where the typical conversion from \htwo\ luminosity to M$_{gas}$ results in a mass higher than M$_{dyn}$, only the \fg\ lower limit derived from the more extreme conversion factor is given (see the discussion in $\S$\ref{sec:cd}).}

\end{deluxetable} 


%% file: tab7.tex
\begin{deluxetable*}{lcccccc}
\tabletypesize{\footnotesize}
\tablecaption{Best-Fit Extinction\label{tab:ext}} 
\tablewidth{0pt}
\tablehead{
\colhead{} &
\colhead{} &
\multicolumn{2}{c}{Screen} &
\colhead{} &
\multicolumn{2}{c}{Mixed}  \\
\cline{3-4} 
\cline{6-7} \\
\colhead{Galaxy} &
\colhead{r} & 
\colhead{A$_{v}$} & 
\colhead{n$_{H}$} &
\colhead{} &
\colhead{A$_{v}$} & 
\colhead{n$_{H}$}  \\ 
\colhead{} &
\colhead{pc} & 
\colhead{mags} & 
\colhead{10$^{21}$ cm$^{-2}$} &
\colhead{} &
\colhead{mags} & 
\colhead{10$^{21}$ cm$^{-2}$}  \\ 
}
\startdata

NGC 1097  & 25-35  & 3.8      & 7.2      &  & 8.8       & 16.7      \\
          &        & (0.5)    & (1.0)    &  & (1.5)     & (2.8)     \\
NGC 3227  & 25-35  & 4.1      & 7.7      &  &  9.7      & 18.5      \\
          &        & (0.3)    & (0.7)    &  & (1.0)     & (1.9)     \\
          & 25-35  & 4.3      & 8.1      &  & 10.2      & 19.3      \\
          &        & (0.4)    & (0.7)    &  & (1.1)     & (2.0)     \\
NGC 3783  & 60-80  & 8.5      & 16.1     &  & 27.7      & 52.7      \\
          &        & (0.1)    & (0.2)    &  & (0.6)     & (1.1)     \\
NGC 4051  & 25-35  & 0.0      & 0.0      &  & 1.0       & 1.8       \\
          &        & (0.4)    & (0.7)    &  & (1.1)     & (2.1)     \\
NGC 4151  & 25-35  & 5.6      & 10.7     &  & 14.5      & 27.6      \\
          &        & (0.6)    & (1.2)    &  & (2.2)     & (4.2)     \\
NGC 6814  & 25-35  & 1.4      & 2.7      &  & 3.0       & 5.7       \\
          &        & (0.4)    & (0.8)    &  & (0.9)     & (1.7)     \\
NGC 7469  & 60-80  & 9.5      & 18.0     &  & 37.1      & 70.5      \\
          &        & (0.1)    & (0.2)    &  & (1.4)     & (2.7)     \\
          & 60-80  & 6.5      & 12.4     &  & 17.7      & 33.6      \\
          &        & (0.2)    & (0.4)    &  & (0.8)     & (1.5)     \\
Circinus  & 7-9    & 7.5      & 14.2     &  & 22.5      & 42.7      \\
          &        & (0.3)    & (0.5)    &  & (1.6)     & (3.1)     \\
NGC 1068  & 25-35  & 12.8     & 24.3     &  & 121$^{\dag}$ & 230$^{\dag}$ \\
          &        & (0.4)    & (0.7)    &  & (\nodata) & (\nodata) \\
\enddata

\tablecomments{Values in parenthesis are the error of the measurement immediately above.  For NGC 3227 and NGC 7469 the first and second set of lines are based on SINFONI and OSIRIS data, respectively.  For the Seyfert 2 galaxies Circinus and NGC 1068 the correction for potential non-stellar contamination is uncertain (see text for details).  The values marked with \dag\ for NGC 1068 are beyond the saturation limit of the mixed extinction model, which is at an optical depth of a few.  As discussed in $\S$ \ref{sec:prop}, NGC 1068 is excluded from the AGN sample.}

\end{deluxetable*}